\documentclass[aps,prx,reprint,floatfix,superscriptaddress,footinbib,twocolumn]{revtex4}
\usepackage{xcolor}
\usepackage{microtype}
\usepackage[bookmarks=true,
bookmarksnumbered=false, % true means bookmarks in
bookmarksopen=false, % true means only level 1
colorlinks=true,
linkcolor=blue]{hyperref}
\definecolor{webblue}{rgb}{0, 0, 0.5} % less intense blue
\hypersetup{colorlinks=true,urlcolor=blue,citecolor=blue}
\usepackage{graphicx}
\usepackage{subfigure}
\usepackage{url}
\usepackage{hyperref}
\usepackage{inconsolata}
\usepackage{amsmath}
\usepackage[capitalize]{cleveref}
\usepackage{braket}
\usepackage{outlines}
\usepackage{enumitem}
\usepackage{soul}
\usepackage{color}

\usepackage{bm} 
\DeclareMathOperator{\Tr}{Tr}
\usepackage[utf8]{inputenc}
\usepackage{siunitx,booktabs}
\usepackage{multirow}
\usepackage{amssymb}

\usepackage{relsize}

\usepackage{soul}
\usepackage{tabularx}

\usepackage{amsmath}
\usepackage{amssymb}
\usepackage{bbm}
\usepackage{braket}
\usepackage{xcolor}

\usepackage{comment}
\usepackage{pifont}
\newcommand{\cmark}{\ding{51}}%
\newcommand{\xmark}{\ding{55}}%

\begin{document}

\title{Chiral excitonic order from twofold van Hove singularities in kagome metals}

\author{Harley D. Scammell}

\email{h.scammell@unsw.edu.au}
\affiliation{School of Physics, University of New South Wales, Sydney 2052, Australia}
\affiliation{Australian Research Council Centre of Excellence in Future Low-Energy Electronics Technologies, University of New South Wales, Sydney 2052, Australia}
\author{Julian Ingham}
\email{jingham@bu.edu}
\affiliation{Physics Department, Boston University, Commonwealth Avenue, Boston, MA 02215, USA}
\author{Tommy Li}
\affiliation{Dahlem Center for Complex Quantum Systems and Fachbereich Physik, Freie Universit\"{a}t Berlin, Arnimallee 14, 14195 Berlin, Germany}
\author{Oleg P. Sushkov}
\affiliation{School of Physics, University of New South Wales, Sydney 2052, Australia}
\affiliation{Australian Research Council Centre of Excellence in Future Low-Energy Electronics Technologies, University of New South Wales, Sydney 2052, Australia}

\date{\today}

\begin{abstract}

Recent experiments on kagome metals AV$_3$Sb$_5$ (A=K,Rb,Cs) identify twofold van Hove singularities (TvHS)  with opposite concavity near the Fermi energy, generating two approximately hexagonal Fermi surfaces -- one electron-like and the other hole-like. Here we propose that a TvHS generates a novel time-reversal symmetry breaking excitonic order -- arising due to bound pairs of electrons and holes located at opposite concavity van Hove singularities. We introduce a minimal model for the TvHS and investigate interaction induced many-body instabilities via the perturbative renormalisation group technique and a free energy analysis. Specialising to parameters appropriate for the kagome metals AV$_3$Sb$_5$, we construct a phase diagram comprising chiral excitons, charge density wave and a region of coexistence. We propose this as an explanation of a diverse range of experimental observations in AV$_3$Sb$_5$. Notably, the chiral excitonic state gives rise to a quantum anomalous Hall conductance, providing an appealing interpretation of the observed anomalous Hall effect in kagome metals. Possible alternative realisations of the TvHS mechanism in bilayer materials are also discussed. We suggest that TvHS open up interesting possibilities for correlated phases, enriching the set of competing ground states to include excitonic order.

\end{abstract}

\maketitle

\section{Introduction}
\label{intro}
Kagome systems have been a major focus of theoretical and experimental investigation; due to their ability to realise Dirac points, flat bands and van Hove singularities, they have been predicted to host a range of novel correlated phases of matter \cite{Johnston1990,Tan2011,Sun2011,Green2010,Yu2012,Kiesel2013,Wen2010,Kiesel2012,Li2022}. Recently, a new class of materials AV$_3$Sb$_5$ (A=K,Rb,Cs) have attracted a great deal of attention due to their demonstration of unconventional superconductivity alongside competing density wave order, spatially modulated superconducting order and possible signatures of Majorana states in superconducting vortices \cite{Ortiz2019,Ortiz2020,Ortiz2021,Yin2021,Xu2021,Gupta2021, Duan2021, Yang2021, Jiang2021,Kenney2021, Li2021b, Zhao2021, Li2021c, Shumiya2021, Mielke2021b, Miao2021, Ni2021, Chen2021, Liang2021, Zhu2021, Chen2021b, Du2021b, Zhang2021, Du2021c, Tsirlin2021, Qian2021, Liu2021, Oey2021, Yang2021b, Kang2021, Hu2021, Xu2022, Denner2021,lin2021complex,Ortiz2021b,Wu2021,Park2021,Christensen2021}. Unusually, the materials exhibit time-reversal symmetry breaking with an anomalous Hall conductivity in spite of the absence of magnetic ordering; the origins and relationship between superconductivity, competing order, and the anomalous Hall effect remain an open question.

The materials consist of a stack of two dimensional layers -- a kagome lattice of vanadium and antimony alternating with a hexagonal lattice of antimony and triangular lattice of the alkali metal K/Rb/Cs -- with electrical transport predominantly in-plane, as demonstrated by the large ratio between the out-of- and in- plane resistivity $R_c/R_{ab}\approx 600$. The Fermi surface of these materials consists of several distinct contours, including nearly circular contours centered at the $\Gamma$ and $K$ points  as well as two approximately hexagonal contours \cite{Ortiz2021b}. Systems with hexagonal Fermi surfaces, corresponding to saddle points in the electronic dispersion, have been predicted to give rise to chiral superconductivity and competing density wave order, due to the effects of Fermi surface nesting \cite{Nandkishore2012}. However, ARPES and DFT results reveal that the hexagonal Fermi surfaces in the vanadium metals exhibit an unusual feature -- two-fold van Hove singularities (TvHS), for which the saddle points at each Fermi surface possess opposite concavity, resulting in one electron-like Fermi surface and one hole-like Fermi surface \cite{Kang2021}.

\begin{figure*}[t]
\includegraphics[width=18cm]{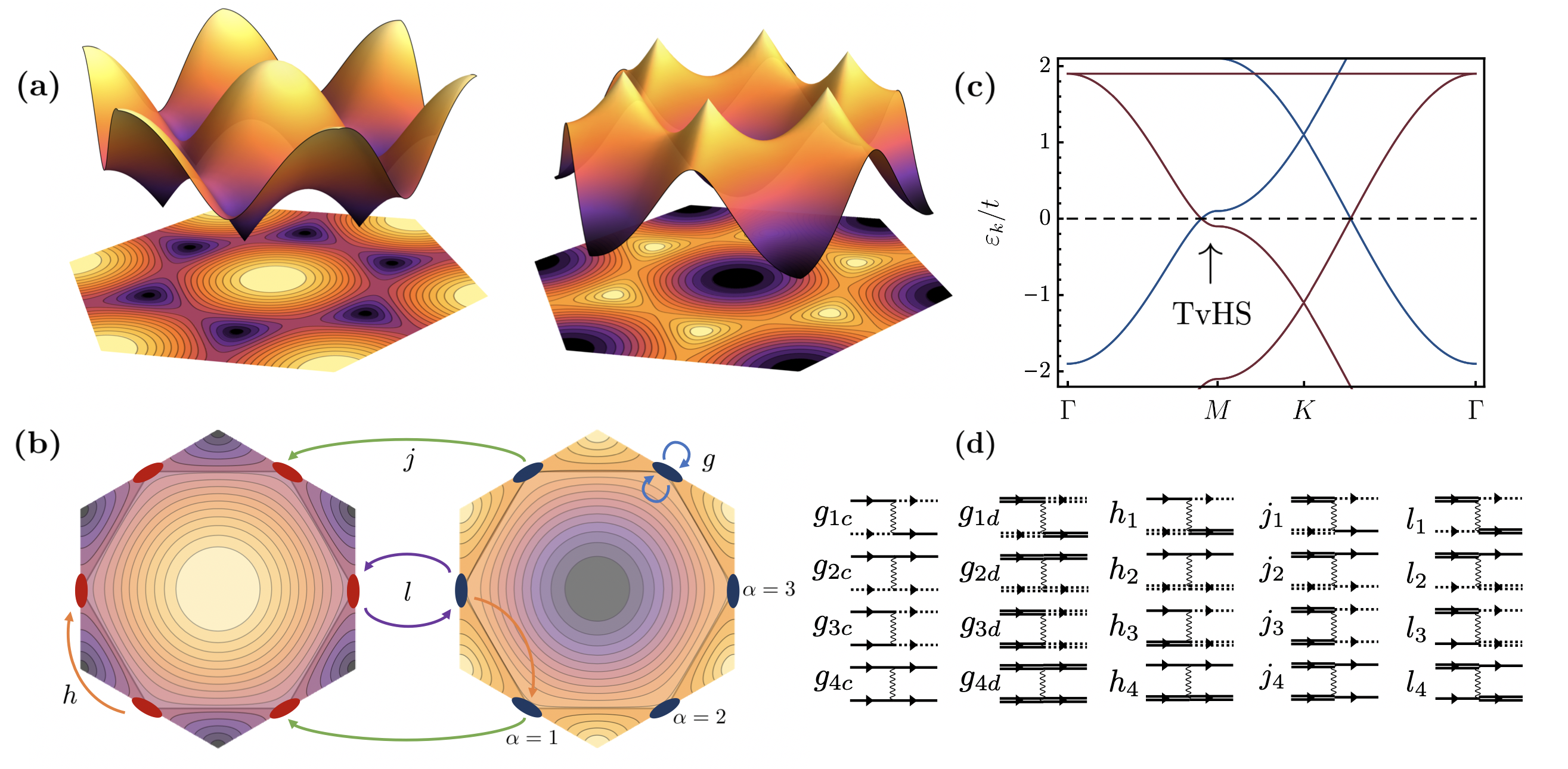}
\caption{Theoretical model: (a) Dispersion plots for a hexagonal tight-binding model, featuring saddle-points with opposite concavity. Dark (light) colours represent negative (positive) energy states relative to the saddle point. (b)  Band structure for \eqref{dft_tb}, with $t^c=t^d=t$, $\epsilon^c/t=2.1$, $\epsilon^d/t=-0.1$, demonstrating positive and negative concavity van Hove singularities near the Fermi level. Bands arising from different orbitals are coloured blue and red, respectively contributing an {\it m-} and {\it p}-type vHS near the Fermi level.  (c) Representative interaction processes from each of the classes $V_g, V_h, V_j, V_l$ \eqref{V}, accounting for scattering processes on or between the patches $\alpha=1,2,3,$ and flavours (left and right hexagons), illustrated on a contour plot of the energy in the first Brillouin zone. The opposite concavities of the saddle points can be seen by the opposite sign of the energies inside the Fermi surface. (d) Feynman diagrams representing the full set of allowed scattering processes. Double/single lines denote fermions from the two distinct Fermi surfaces, while dashed/solid lines represent fermions at different patches.} 
\label{f:Model}
\end{figure*}
%%%%%%%%%%%%%%%%% 

We argue that doping a system to a TvHS has an ineluctable influence on the low-energy physics. A single vHS results in a tendency towards density wave ordering and superconductivity. The appearance of TvHS introduces an additional tendency towards excitonic order -- corresponding to a condensation of electron-hole pairs -- owing to the coupling between an electron-like and hole-like Fermi surface. We introduce a low energy model which incorporates the TvHS -- featuring an electron-like and hole-like Fermi surface, each doped near their respective vHS. To understand the interplay and competition between the various many-body instabilities, we employ the perturbative renormalisation group (RG) method to determine the dominant ground state order \cite{Maiti2013,Schulz1987,Dzyaloshinskii1987,Furukawa1998,Nandkishore2012}, complemented by a Landau-Ginzburg free energy analysis of competing ground states. 
A chiral excitonic order naturally emerges, which breaks time-reversal symmetry and exhibits a quantum anomalous Hall effect. The chiral excitonic state appears as a generic weak coupling instability, but explicit modeling for AV$_3$Sb$_5$ suggests these materials exist in an intermediate coupling regime; guided by \textit{ab initio} results we generate a phase diagram featuring charge density wave order, chiral excitonic order and a region of coexistence. We suggest that the phenomenology encompassed by the TvHS model accounts for key features observed in the vanadium-based kagome metals, and could further motivate TvHS engineering in van der Waals heterostructures and bilayer materials.

%In Section \ref{model} we introduce our model, and describe key aspects of the RG method. In Section \ref{results} we discuss the solution to the RG equations, generating a phase diagram, and discuss aspects of the excitonic phase. In Section \ref{disc} we discuss our findings, and draw links to existing and future experiments. 

\section{Results}
\label{model}

\noindent {\bf Tight-binding Hamiltonians with TvHS.}
A TvHS consists of two Fermi surfaces with opposite concavity vHS so that one surface is electron-like and the other hole-like -- e.g. arising from doping near the $M$-point of a 2D hexagonal Brillouin zone as shown in Fig \ref{f:Model}a. The opposing concavities of the respective saddle points can be seen from the colour plot in Fig. \ref{f:Model}b; going from outside to inside the hexagonal Fermi surface, the energy changes sign, but the sign change is opposite for the two Fermi surfaces. The two Fermi surfaces may arise due to two hexagonal (honeycomb or kagome) bilayers, or a single layer with two sets of orbitals -- the latter case is the origin of the TvHS in vanadium metals AV$_3$Sb$_5$. 

To be explicit, we will introduce a particular lattice model realisation of a TvHS. A tight-binding model of a kagome monolayer with two sets of orbitals that has been used to describe AV$_3$Sb$_5$ is given by
\begin{align}
\label{dft_tb}
H_{tb} &=-\sum_{\braket{{i},{j}},\nu} t^\nu a_{i,\nu}^\dag a_{j,\nu} - \sum_{{i},\nu} \epsilon^\nu a_{i,\nu}^\dag a_{i,\nu},
\end{align}
where $a^\dag_{i,\nu}$ creates fermions on site $i$ and in orbital $\nu=c,d$. The differing orbital potentials, $\epsilon^c-\epsilon^d\approx t^c+t^d$ shift the energies of the two bands, aligning their valence and conduction bands and resulting in a TvHS. 

The bandstructure of a TvHS can be realised in both honeycomb and kagome systems (we discuss alternative tight-binding models in the Supplementary Material). However, the orbital content of the wavefunction at the $M$-points is qualitatively different in these two cases. For honeycomb, with two sublattices, the wavefunction at the $M$-points has equal support on both sublattices. Meanwhile, in kagome systems the wavefunction near the $M$-points exhibits the so-called `sublattice interference effect' \cite{Kiesel2012}: at a given $M$-point, the conduction band wavefunctions have support only on one sublattice and are referred to as $p$-type (owing to their `pure' sublattice composition) while the valence band wavefunctions have support on the other two sublattices and are referred to as $m$-type (due to their `mixed' sublattice composition). The sublattice  structure has important consequences when considering interaction effects, as we discuss below.

\  

\noindent {\bf Patch model.}
The problem of interaction driven instabilities on a single hexagonal Fermi surface (i.e. single vHS) has been previously studied using a three patch model \cite{Nandkishore2012}, whereby the full Brillouin zone is restricted to three momentum space patches near the vHS at the $M$-points, since they dominate the density of states. Following this approach, we define a three patch model and further introduce a flavour degree of freedom to account for the two opposite concavity Fermi surfaces -- fermions of one flavour (created by $c^\dag$) are electron-like, while fermions of the other flavour (created by $d^\dag$) are hole-like,
\begin{align}
\label{H0}
 H_0 & = \sum_{\bm k, \alpha}^\Lambda   (\varepsilon^c_{\bm k,\alpha}-\mu) c_{\bm k,\alpha}^\dag c_{\bm k,\alpha} -( \varepsilon^d_{\bm k,\alpha}+\mu)d_{\bm k,\alpha}^\dag d_{\bm k,\alpha}.
\end{align}
The patch index $\alpha=1,2,3$ indicates a fermionic excitation within a cut-off $\Lambda$ of the momentum $\bm{M}_\alpha$. Setting $\mu=0$ corresponds to doping exactly to the TvHS. The opposite concavity is encoded by the relative minus sign between the $c$ and $d$ dispersions. The TvHS should be contrasted with the problem of fixed concavity vHS with multiple flavours \cite{lin2019chiral} -- the opposite concavity of the two vHS plays a fundamental role in the interaction physics. The patch dispersion take the saddle point form $\varepsilon^{\nu}_{\bm k,1}= \tfrac{1}{2}t^{\nu}(k_x^2+\sqrt{3}k_xk_y)$, $ \varepsilon^{\nu}_{\bm k,2} = \tfrac{1}{4}t^{\nu}(-k_x^2+3k_y^2)$, $\varepsilon^{\nu}_{\bm k,3}=\tfrac{1}{2}t^{\nu}(k_x^2-\sqrt{3}k_xk_y)$, where $t^{\nu}$ is a characteristic energy scale, and equals the nearest neighbour hopping of the $\nu$-fermions ($\nu=c,d$) in the simple tight-binding model. Fermions at patches $\alpha\neq \beta$ are connected by the nesting vector $\bm Q_{\alpha\beta}=\bm M_{\alpha}-\bm M_{\beta}$, for which $\varepsilon^{\nu}_{\bm k + \bm Q_{\alpha\beta},\beta}\approx-\varepsilon^{\nu}_{\bm k,\alpha}$. 

Making contact with {\it ab initio} results for AV$_3$Sb$_5$, the $c$- ($p$-type) and $d$- ($m$-type) fermions arise from the vanadium $d_{yz}$ and $d_{xz}$ orbitals respectively, and have $t^c\approx 0.5$ eV, $t^d\approx 1$ eV \cite{Wu2021}. In the patch model, this sets $ t^c/t^d\equiv\kappa= 2$. For completeness we will analyse both $\kappa=1$ and $\kappa=2$. It is known from ARPES that the $c$-band vHS is near-perfectly nested, while the $d$-band vHS exhibits quartic corrections \cite{Kang2021}. Close to the $M$-point these corrections are subdominant to the quadratic part of the dispersion, and hence only influence the ultraviolet behaviour of the theory, near the cut-off $\Lambda \approx$ 0.5 eV. Since our analysis probes infrared scales far below $\Lambda$, it is well-justified to ignore the quartic corrections.

Below we will analyse three distinct cases: (i) honeycomb systems, for which the sublattice support on the two flavour vHS is the same, in the particle-hole symmetric limit $\kappa=1$; (ii) kagome systems, in which the two flavour vHS have different sublattice support i.e. $m$- and $p$-type, with $\kappa=1$; (iii) kagome systems with $\kappa=2$, which we have argued to describe kagome metals AV$_3$Sb$_5$.

\begin{table}[tb]
\begin{center}
\caption{Estimates of the bare coupling values in AV$_3$Sb$_5$. Projecting the pure and mixed sublattice form factors onto the cRPA results of \cite{Wu2021} results in the below values, where the intra-orbital, inter-orbital, Hund's, pair hopping, and nearest neighbour repulsions are $U=1$-$2$ eV, with $U'=0.8U$, $J=J'=0.1U$ and $V=0.3U$.} 
\label{couplings_estimate}
\vspace{0.1cm}
\begin{ruledtabular}
 \begin{tabular} {lllllllllllllllllllll} \\[-3.5mm]
 & $g_{i,c}$  & $g_{i,d}$ & $h_{i}$ & $j_{i}$ & $l_i$ \\[1mm] \hline \vspace{-0.2cm} \\ \vspace{0.2cm} 
$i=1$ & $0$ & $\tfrac{1}{4}(U+V)$ & $0$ & $0$ & $\tfrac{1}{2} J$ \\   \vspace{0.2cm}
$i=2$ & $V$ & $\tfrac{1}{4}U+V$ & $\tfrac{1}{2}U'+V$ & $0$ & $0$ \\   \vspace{0.2cm}
$i=3$ & $0$ & $\tfrac{1}{4}(U+V)$ & $0$ & $\tfrac{1}{2}J'$ & $0$ \\   \vspace{0.2cm}
$i=4$ & $U+V$ & $\tfrac{1}{2}U+V$ & $V$ & $0$ & $0$ \\[-1mm]
 \end{tabular}
\end{ruledtabular}
\end{center}
\end{table}

\  

\noindent {\bf Interactions.}
We now consider the possible couplings between the fermions. Due to the large density of states near the TvHS the Coulomb repulsion is expected to be strongly screened and we therefore model the interactions as short-ranged. The most general set of interactions between patches/flavours allowed by momentum conservation are 
\begin{align}
\label{V}
V&= \frac{1}{2}\sum_{\alpha,\beta}\left[V_{g,\nu} + V_h + V_j +V_l \right]
\end{align}
where $V_{g,\nu}$ are intraflavour couplings, $V_{h}$ are  interflavour density-density couplings, $V_{j}$ are flavour pair hopping,  and $V_{l}$ are flavour exchange couplings, resulting in 20 independent interactions. A schematic illustration of the $g,h,j,l$ couplings, as well as their representation in terms of Feynman diagrams, is shown in Fig \ref{f:Model}b and d. Additional details are found in the Supplementary Material.

In the kagome case, projecting the sublattice wavefunctions onto the Coulomb interaction results in different intraflavour couplings depending on whether the flavour has pure or mixed sublattice structure, a manifestation of the sublattice interference effect in  kagome patch models. We have therefore allowed for different couplings $V_{g,\nu}$ on each flavour. Performing this projection explicitly and using the calculations of \cite{Wu2021} gives the estimates of the bare coupling values shown in Table \ref{couplings_estimate}. {The values taken from \cite{Wu2021} are defined at the lattice scale; using these as input to our effective theory neglects the renormalisation flow between the lattice scale and $\Lambda$.} The sublattice interference effect has crucial consequences for the bare couplings: for instance, on a $p$-type vHS, the wavefunctions at different patches are orthogonal, and so the interpatch Coulomb repulsion is suppressed, resulting in $g_{1,c}=g_{3,c}=0$.
Thus, any attractive interactions present in the system, for e.g. due to phonons, immediately result in attractive couplings. 
%It has previously been noted in the literature that negative couplings are essential to obtain CDW phases in single vHS patch models \cite{Park2021}. 

In the honeycomb case, the orbital form factors are the same for both flavours and so we expect $V_{g,c} \approx V_{g,d}$. This reduces the number of independent coupling constants from 20 to 16. We shall present results for both models below.

\  

\noindent {\bf Instabilities.} Considering the interacting Hamiltonian,
\begin{align}
\label{Hint}
H&=H_0 + V,
\end{align}
we determine which instabilities arise within the framework of RG. The instability of the metallic phase and onset of an ordered ground state is signalled by the susceptibility of the associated order parameter: the strongest ordering tendencies are those with most divergent susceptibility. In the case of a nested Fermi surface, a density wave instability arises because the nesting condition $\varepsilon_{\bm p} \approx - \varepsilon_{\bm p+\bm Q}$, implies the total energy of a particle with momentum $\bm p$ and hole with momentum $\bm p + \bm Q$ is approximately zero. Similarly, the energy of an electron and a hole at the TvHS is $\varepsilon^c_{\bm p}+\varepsilon^d_{\bm p}\approx 0$. Without including interactions, it costs zero energy to create either of these particle-hole states, and hence, for an arbitrarily small attraction between particles and holes the system becomes unstable to lowering its energy by spontaneously creating many such pairs, analogous to the usual superconducting instability.  The RG method provides an unbiased approach to study competing orders on an equal footing, by resumming the logarithmically divergent corrections to the bare couplings and determining which ordering tendency dominates \cite{Maiti2013,Schulz1987,Dzyaloshinskii1987,Furukawa1998,Nandkishore2012}.

In Table \ref{order_params}, we enumerate the ordered states which naturally arise in the TvHS model, i.e. those with nesting tendencies. The first three -- CDW, SDW and SC  -- occur in the case of a single vHS. The next three -- singlet and triplet excitonic order, as well as interflavour pair density wave (PDW) -- are new instabilities introduced by the TvHS.

\begin{table}[tb]
\begin{center}
\caption{The leading ordered states. Notation: $\alpha,\beta$ index patch, $\sigma_\nu$ act on flavour, indexed by Latin characters $\nu=c,d$ with $\sigma_c = \tfrac{1}{2}(\sigma_0+\sigma_z)$, $\sigma_d = \tfrac{1}{2}(\sigma_0-\sigma_z)$, $\sigma_\pm = \tfrac{1}{2}(\sigma_x\pm i\sigma_y)$,  and $\vec{s}$ is the vector of Pauli matrices acting on spin. The final two columns indicate if the given ordered state arises in the presence of a single nested vHS and/or TvHS.} 
\label{order_params}
\vspace{0.1cm}
\begin{ruledtabular}
 \begin{tabular} {lllllllllllllllllllll} \\[-3.5mm]
 & Structure  & vHS & TvHS  \\[1mm] \hline \vspace{-0.2cm} \\ \vspace{0.2cm}
CDW & $\mathcal C_{\alpha\beta \nu}=\langle \psi^\dag_\alpha \sigma_\nu \psi_{\beta} \rangle$ & \ \cmark & \  \cmark  \\   \vspace{0.2cm}
SDW & $\mathcal S_{\alpha\beta \nu} = \langle \psi^\dag_\alpha \sigma_\nu \vec{s} \psi_{\beta} \rangle$ & \  \cmark & \  \cmark\\   \vspace{0.2cm}
SC & $\Delta_{\alpha \nu} = \langle \psi_\alpha \sigma_\nu \psi_\alpha \rangle$ & \  \cmark & \  \cmark  \\   \vspace{0.2cm}
Singlet exciton & $\Phi^{C}_{\alpha \pm} =\langle \psi^\dag_\alpha \sigma_\pm \psi_\alpha \rangle$ & \ \xmark & \ \cmark  \\ \vspace{0.2cm}
Triplet exciton & $\Phi^{S}_{\alpha \pm} =\langle \psi^\dag_\alpha \sigma_\pm \vec{s}\psi_\alpha \rangle$ & \  \xmark & \  \cmark  \\ \vspace{0.2cm}
PDW & ${\cal P}_{\alpha\beta \pm} = \langle \psi^\dag_\alpha \sigma_\pm \psi^\dag_{{\beta}} \rangle$ &  \ \xmark & \  \cmark  \\
 \end{tabular}
\end{ruledtabular}
\end{center}
\end{table}

\  

\noindent {\bf RG analysis.} We turn now to the RG treatment which identifies the leading instabilities, i.e. the dominant ground states in Table \ref{order_params}. Firstly, we compute the leading log$^2$ corrections to the bare couplings defined in \eqref{V}. The equations define how the couplings evolve with the RG time $t$ which is a proxy for the energy scale; here $t\to\infty$ corresponds to taking $T\to0$. The full RG equations for our model are lengthy, since they involve twenty independent interaction constants (Fig. \ref{f:Model}d), so we state their general form here and reserve explicit expressions for the Supplementary Material. The RG equations describing the flow of the couplings $g_i$, $h_i$, $j_i$, $l_i$ (where $i=1,2,3,4$) take the form
\begin{align}
\label{intRG}
   \notag  \tfrac{\partial}{\partial t} g_{i,\nu}  &= \beta_{g_{i,\nu}}(g,j,h,l), &&  \tfrac{\partial}{\partial t}  h_i = \beta_{h_i}(g,j,h,l),\\
    \tfrac{\partial}{\partial t}  j_i &= \beta_{j_i}(g,j,h,l), && \tfrac{\partial}{\partial t} l_i  = \beta_{l_i}(g,j,h,l),
\end{align}
where $\beta_{g_{i,\nu}}$, $\beta_{h_i}$, $\beta_{j_i}$, $\beta_{l_i}$ are functions of all twenty couplings.  Secondly, we compute the leading log$^2$ corrections to the order parameters, which generates the linear set of gap equations,
\begin{align}
\label{gap_eqn}
\tfrac{\partial}{\partial t} \mathcal{O}_i &= \sum_j \mathcal{V}_{ij}(g,j,h,l) \ \mathcal{O}_j
\end{align}
where $\mathcal{O}_i=\{\mathcal S_{\alpha\beta \nu}, \mathcal C_{\alpha\beta \nu}, \Delta_{\alpha \nu}, {\cal P}_{\alpha\beta \pm},\Phi^{C}_{\alpha \pm},\Phi^{S}_{\alpha \pm}\}$. Diagonalising the gap equation matrix $\mathcal{V}_{ij}$ and integrating over the RG time $t$, one identifies the leading eigenvalue $\lambda_i(t)$ which diverges fastest with $t$. The associated eigenvector is the order parameter with the largest critical temperature $T_c =\Lambda e^{-1/(\nu_0 \lambda_i)^{1/2}}$, and is therefore the dominant order at $T\lesssim T_c$. Multiple orders of comparable $T_c$ may arise, in which case one must compute the Landau-Ginzburg free energy to ascertain whether such phases compete or coexist.

 %%%%%%%%%%%%%%%%%
\begin{figure}[t]
\includegraphics[width=8cm]{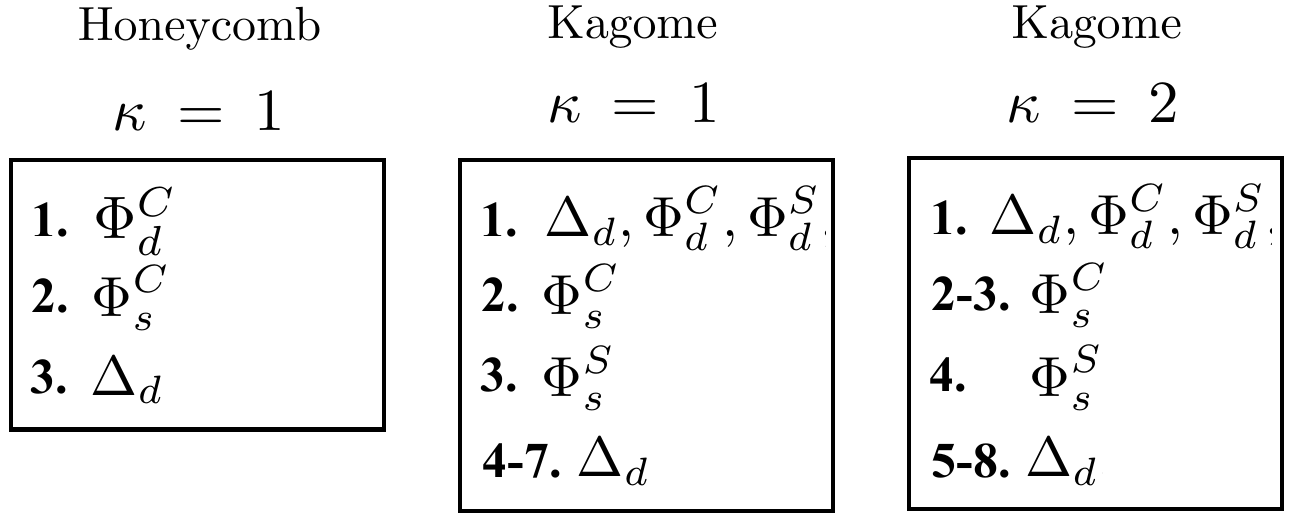}
 \caption{RG Fixed rays: RG fixed rays for the TvHS patch model in honeycomb and kagome systems. The parameter $\kappa$ measures particle-hole symmetry, c.f. discussion after Eq. \eqref{H0};  $\kappa=2$ is appropriate to AV$_3$Sb$_5$ \cite{Wu2021}. }
\label{f:FPs}
\end{figure}
%%%%%%%%%%%%%%%%%  

%\section{Results}
%\label{results}
\  

\noindent {\bf Dominant instabilities.}
A subset of the couplings diverge with increasing RG time $t\rightarrow \infty$. In this limit, the diverging couplings tend towards fixed constant ratios of each other referred to as {\it fixed rays}. The relative magnitudes of the couplings determine which ground state dominates. All possible choices of bare initial coupling values flow to one of these possible sets of ratios in the deep infrared, which therefore represent universal properties of the model. We now present the set of fixed rays possible in our TvHS patch model (for a derivation see the Supplementary Material). Despite the large number of interaction terms there turn out to be only a small set of fixed rays, which exhaustively characterise the possible ground states in the weak coupling regime. We summarise for three different cases: 
\begin{enumerate}
\item Honeycomb systems with $\kappa=1$ have three fixed rays: comprising chiral superconductivity $\Delta_d$, chiral $d$-wave excitons $\Phi^C_d$, and $s$-wave excitons $\Phi^{C}_s$.

\item Kagome systems with $\kappa=1$ have seven fixed rays: comprising chiral superconductivity $\Delta_d$, chiral $d$-wave excitons $\Phi^{C/S}_d$, and $s$-wave excitons.

\item Kagome systems with $\kappa=2$ have eight fixed rays: comprising chiral superconductivity $\Delta_d$, chiral $d$-wave excitons $\Phi^{C/S}_d$, and $s$-wave excitons $\Phi^{C/S}_s$.  

\end{enumerate}
Crucially, in all cases $d$-wave excitons emerge at a fixed trajectory, demonstrating the naturalness of excitonic order. As in the case of single vHS \cite{Nandkishore2012}, we find that $d$-wave superconductivity is also a natural instability of the TvHS model. 

For arbitrarily small initial couplings, the fixed rays are reached at long RG times, which corresponds to the deep infrared. However, the initial couplings could be sufficiently large that an instability occurs before the fixed ray is reached. In such a case it is appropriate to instead explicitly compute the flow from a specific set of initial conditions, and examine when an instability is reached. Given the significant magnitude of the bare values of the couplings in AV$_3$Sb$_5$ (Table \ref{couplings_estimate}) we believe that such an analysis is more appropriate when comparing with experiment, and is presented below.

 %%%%%%%%%%%%%%%%%
\begin{figure}[t!]
\includegraphics[width=8.5cm]{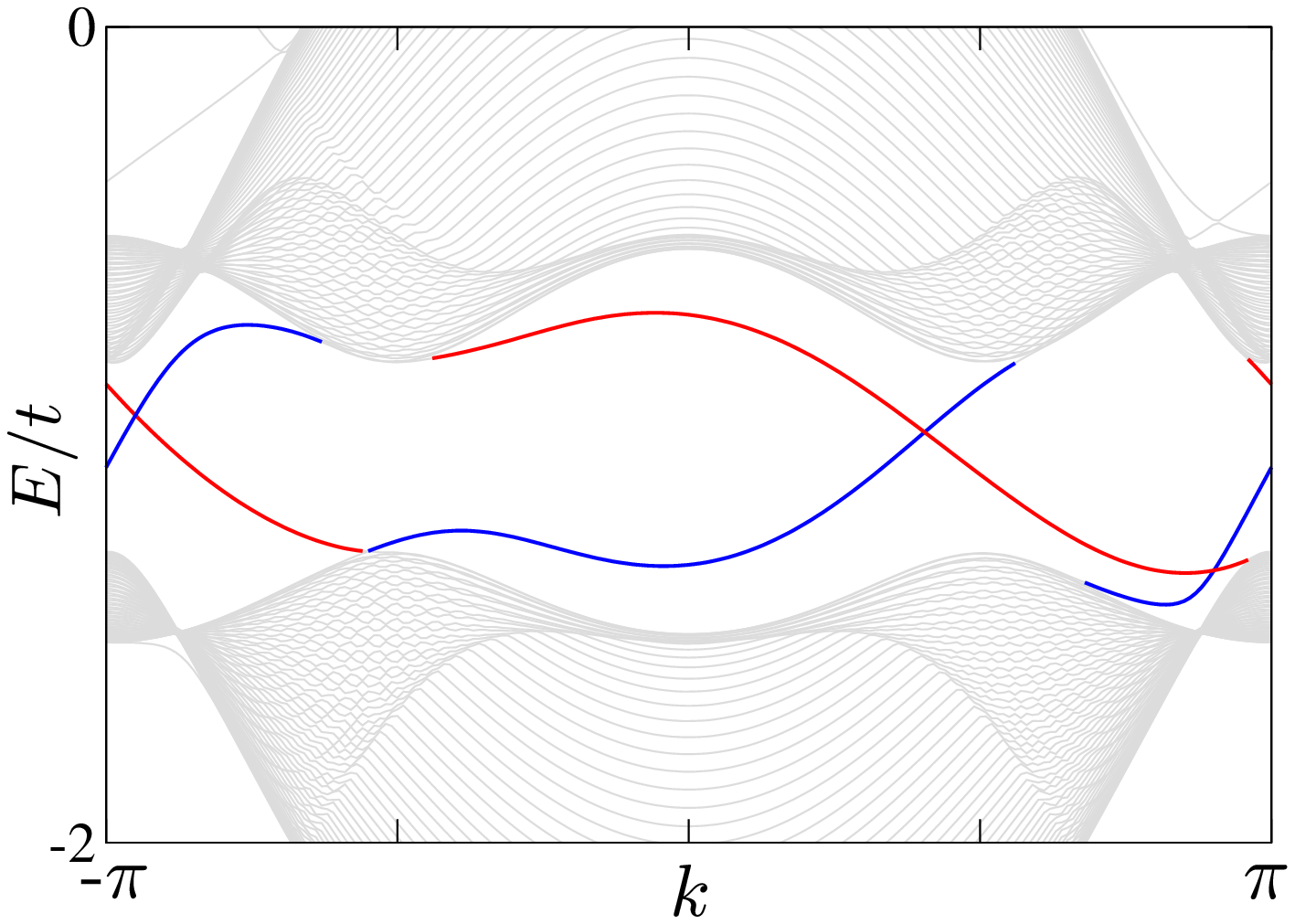}
\vspace{-0.5cm}
 \caption{Edge modes:  The 1D dispersion of a $d+id$ excitonic insulator in a two-orbital kagome system, for an infinite ribbon of width 120 unit cells.
 The edge states propagating along the top/bottom of the ribbon are plotted in red/blue.
}
\label{f:edge}
\end{figure}
%%%%%%%%%%%%%%%%% 

\  

\noindent {\bf Properties of the chiral excitonic condensate.}
We focus attention on some key properties of the excitonic phases which appear. The two $d$-wave excitonic order parameter structures which appear are
\begin{align}
    \notag \Phi^C_{\pm,a}&=\Phi_a e^{\pm i\phi_{a}}\tfrac{1}{\sqrt{6}}\left(1,-2,1\right), \\ \label{Xsol} \Phi^C_{\pm,b}&=\Phi_b e^{\pm i\phi_{b}}\tfrac{1}{\sqrt{2}}\left(1,0,-1\right)
\end{align}
Here $\Phi_a,\Phi_b$ are real scalars, $\phi_{a},\phi_{b}$ are distinct U(1) phases, and the row vectors enumerate patch indices, therefore encoding the spatial structure of the two order parameters. Continuing to the full Fermi surface, the spatial vectors schematically behave as  $\sim\cos(2\theta_{\bm k})$, $\sin(2\theta_{\bm k})$, with $\theta_{\bm k}$ the momentum angle. Similar $d$-wave eigenvectors appear for the superconducting states $\Delta$, which are the two-flavour analogues of the  superconducting states found in \cite{Nandkishore2012}.

Near the critical temperature, the Landau-Ginzburg free energy is found to be
\begin{align}
\label{F_Phi}
    &\mathcal{F}_\Phi = \mathcal{F}_0 + (\tfrac{1}{2\lambda_\Phi} - a_\Phi)(|\Phi_a|^2+|\Phi_b|^2)\\
   \notag &+ c_\Phi\left(\Phi_a^4 +\Phi_b^4 + \tfrac{4}{3}\Phi_a^2\Phi_b^2 + \tfrac{2}{3} \Phi_a^2{\Phi}_b^2\cos(2(\phi_a-\phi_b))\right)
\end{align}
where $\mathcal{F}_0$ is the free energy of the free fermions, with expansion coefficients $a_\Phi,c_\Phi>0$. The free energy is minimised by coexisting order parameters, with $\Phi_a=\Phi_b=\Phi_0$ and $\phi_a-\phi_b=\pi/2$ (mod $\pi$). The coexisting states form a single order parameter of the form $\Phi=\Phi_0 e^{\pm i\theta_\alpha}$, $\theta_\alpha = \{a,b,c\}$. Continuing around the Fermi surface, the combined order parameter becomes $\Phi\sim \Phi_{0,\bm k} (\cos(2\theta_{\bm k}) \pm i\sin(2\theta_{\bm k}))$, which is a chiral $d\pm id$ order. The chirality $\pm$ is spontaneously selected by the ground state, which therefore breaks time-reversal symmetry.

 %%%%%%%%%%%%%%%%%
\begin{figure*}[t!]
\includegraphics[width=15cm]{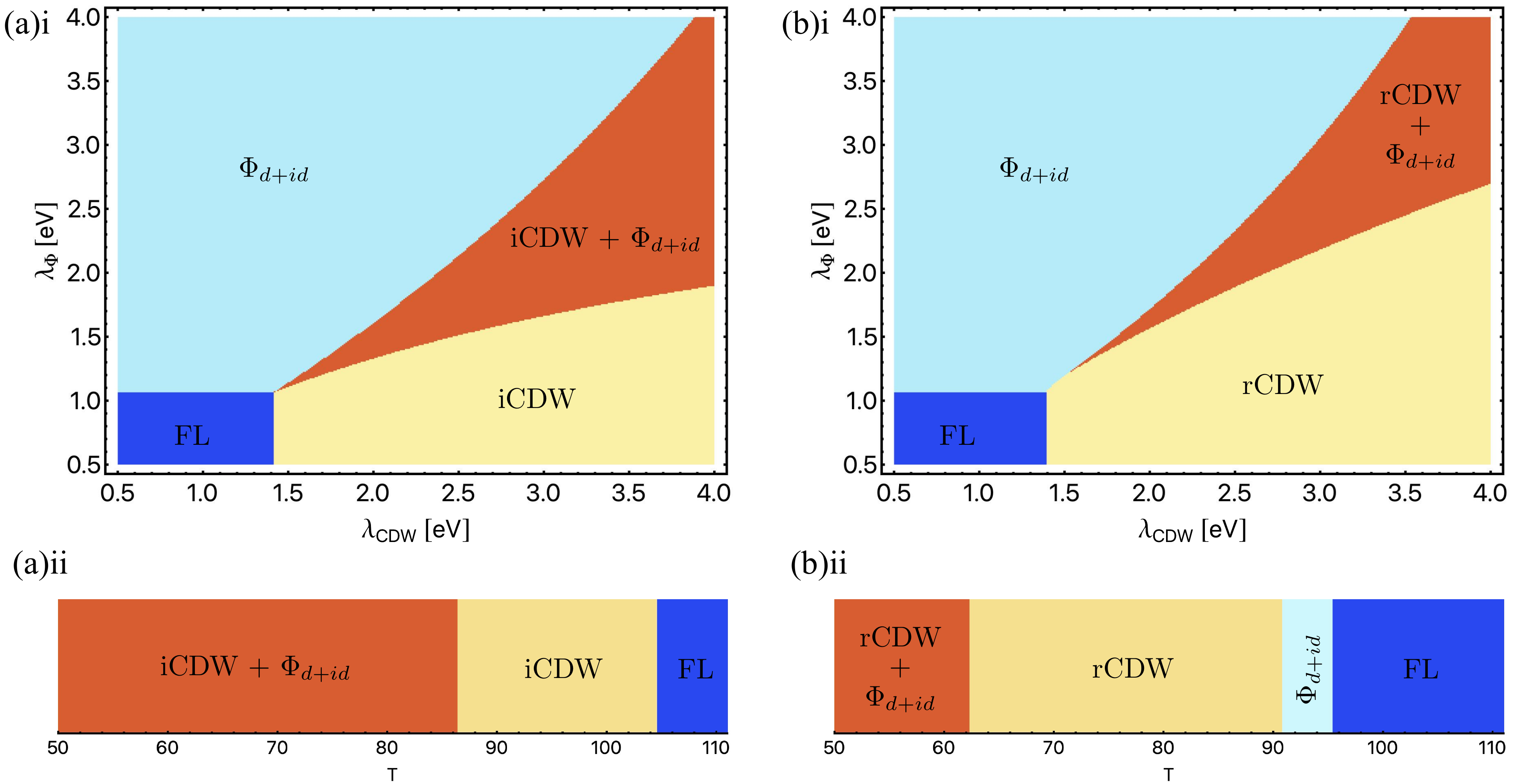}
  \caption{Phase diagrams. Everywhere we have taken $\kappa=2$; i/rCDW + $\Phi_{d+i d}$ represents coexistence, and FL the Fermi liquid metallic state. (a)i The iCDW + $\Phi_{d+i d}$ and (b)i the rCDW + $\Phi_{d+i d}$ phase diagrams, using the eigenvalues $\lambda_\Phi, \lambda_\text{CDW}$ as free parameters. In (a)ii and (b)ii the eigenvalues $\lambda_\Phi, \lambda_\text{CDW}$ are computed explicitly by solution of the RG equations, resulting in a phase diagram as a function of $T$. The two different phase diagrams result from taking (a)i $g_{1,c}< 0, h_1<0$, and (b)ii $g_{1,c}<0, h_1<0$ with $g_{3,d}=0$; precise values of the initial couplings are given in the Supplementary Material. }
\label{f:phase}
\end{figure*}

In addition to broken TRS, the chiral order parameter winds twice along the Fermi surface and vanishes away from it, thereby exhibiting a non-trivial topology with Chern number $|C|=2$. In order to illustrate this, we diagonalise a mean-field Hamiltonian (Methods \eqref{HMF}) defined on a lattice -- we consider a two-orbital kagome lattice model in an infinite ribbon geometry with zigzag edges. The 1D dispersion of the ribbon is plotted in Fig. \ref{f:edge} for the $d+id$ phase, which exhibits two chiral edge modes, with the left/right-movers propagating along the top/bottom of the ribbon. Full details of the lattice model are provided in the Supplementary Material. The nontrivial topological invariant implies a quantised anomalous Hall conductivity $\sigma_{xy} = Ce^2/(2\pi)$ which is carried by two chiral edge modes. We note that this value of $\sigma_{xy} =Ce^2/(2\pi)$ accurately accounts for the intrinsic contribution to the anomalous Hall effect seen in AV$_3$Sb$_5$ \cite{Yang2021}.

\  

\noindent {\bf Coexistence of chiral excitons and charge order.}
The RG procedure determines which phases are dominant, but is not sufficient to determine the actual ground state when two phases have comparable $T_c$. Motivated by experiments on AV$_3$Sb$_5$, we examine the ground state when CDW and chiral excitons are proximate instabilities. To carry out the analysis, we consider the Landau-Ginzburg  free energy for chiral excitons and CDW, written as $\mathcal{F} = \mathcal{F}_\Phi  +\mathcal{F}_C+\mathcal{F}_{\Phi,C}$, with $\mathcal{F}_\Phi$ from Eq. \eqref{F_Phi} and
%%%%%%%%%%%%%
\begin{gather}
\notag \mathcal{F}_C = \!\!\!\!\!\!\!\sum_{\alpha\neq\beta; \nu=c,d} \!\!\!\!\!\!\left\{ (a_\nu\delta_{\nu\nu'}+\tfrac{1}{2}\mathcal{V}^{-1}_{\nu\nu'})C_{\alpha\beta\nu}C^*_{\alpha\beta\nu'} +\tfrac{1}{2}c_{1\nu}  |C_{\alpha\beta\nu}|^4\right\}\\
\notag +  \sum_{\nu} b_\nu \left\{ C_{12\nu}C_{23\nu}C_{31\nu}  + \text{c.c.}\right\} + \!\!\!\!\!\!\sum_{\alpha\neq \beta;\nu\neq\nu'}  \!\!\!\!\!c_{2\nu}  |C_{\alpha\beta\nu}|^2|C_{\alpha\beta\nu'}|^2\\
\notag \mathcal{F}_{\Phi,C}= \!\!\!\!\!\!\sum_{\alpha\neq\beta; \nu\neq \nu'} \!\!\!\! c_{3\nu}\{ C_{\alpha\beta\nu} C^*_{\alpha\beta\nu'} \Phi_{\alpha}\Phi_{\beta}^*+ \text{c.c.}\}\\
+ \sum_{\alpha\neq \beta,\nu} c_{4\nu} |C_{\alpha\beta\nu}|^2(|\Phi_{\alpha}|^2+|\Phi_{\beta}|^2). \label{F_C_Phi}
\end{gather}
%%%%%%%%
The expansion coefficients $a_\nu,b_\nu,c_{i\nu}$ depend on temperature and are computed in the Supplementary Material. Here $\mathcal{V}_{\nu\nu'}$ is the CDW gap equation matrix \eqref{gap_eqn}. Henceforth, we denote the largest eigenvalue of $\mathcal{V}_{\nu\nu'}$ by $\lambda_\text{CDW}$.

In the free energy \eqref{F_C_Phi}, there are six complex numbers, $C_{\alpha\beta\nu}$, describing CDW order. Physically the $C_{\alpha\beta\nu}$ correspond to the magnitude of the order for the three distinct vectors $\bm Q_{\alpha\beta}\in\{\bm Q_{12}, \bm Q_{23}, \bm Q_{31}\}$, on the two distinct Fermi surface flavours. From the gap equation \eqref{gap_eqn} we find that the leading CDW order has $C_{\alpha\beta c}=\rho  C_{\alpha\beta d}$, where $\rho$ is a real number. In particular, $|\rho|=1$ in the particle-hole symmetric limit of $\kappa=1$.
Moreover, the gap equation distinguishes real charge density order (rCDW) whereby $C_{\alpha\beta\nu}^*=C_{\beta\alpha\nu}$ and purely imaginary order (iCDW) whereby $C_{\alpha\beta\nu}^*=-C_{\beta\alpha\nu}$. We separately considered parameter regimes in which rCDW and iCDW were the leading CDW phase.

We turn now to the phase diagram predicted by \eqref{F_C_Phi}. To construct the phase diagrams of Fig. \ref{f:phase}(a)i and \ref{f:phase}(b)i we allow the eigenvalues $\lambda_\text{CDW}$ and $\lambda_\Phi$ to be free variables. {To illustrate the property of coexistence and our phenomenological proposal for these AV$_3$Sb$_5$,} we take a realistic range of coupling eigenvalues, consistent with DFT calculations \cite{Wu2021}, and fix $T=80$ K (which enters via the coefficients $a_\nu,b_\nu,c_{i\nu}$). We find three distinct phases: (i) chiral excitons, (ii) CDW, and (iii) coexistence of chiral excitons and CDW. In the rCDW/iCDW phases, the $3\bm Q$ state is favoured, i.e. CDW order is nonzero for all three nesting vectors ${\bm Q}_{\alpha\beta}$, and corresponds to $C_{12\nu}=C_{23\nu}=C_{31\nu}\neq0$. In the excitonic phases, the chiral (TRS breaking) $d+i d$ state is favoured. In the region of coexistence, chiral excitons and the $3{\bm Q}$ CDW are favoured. We point out that $d$-wave excitons coupled to a nematic CDW (e.g. $C_{12\nu}>C_{23\nu}=C_{31\nu}$) was observed as a local minima, but did not appear as the global minimum over the parameter range searched. 

Experiment indicates that TRS breaking and CDW coexist in AV$_3$Sb$_5$, and set in at $T^*\approx 100$ K. We propose that the coexistence phase demonstrated by our Landau-Ginzburg analysis provides a phenomenological explanation of the physics of kagome metals at $T\lesssim T^*$. 

%Experiment indicates that TRS breaking and CDW coexist in AV$_3$Sb$_5$, and set in at $T^*\approx 100$ K. We propose that the coexistence phase predicted by our Landau-Ginzburg analysis provides a phenomenological explanation of the physics of kagome metals at $T\lesssim T^*$. 

\  

\noindent {\bf Truncated RG flow and phase diagram for AV$_3$Sb$_5$.} To complement the analysis leading to Fig. \ref{f:phase}(a)i and \ref{f:phase}(b)i, we now directly compute the eigenvalues $\lambda_\text{CDW}$ and $\lambda_\Phi$ from the RG procedure. Unlike for the fixed ray analysis, here we must provide initial conditions for the RG flow. Once initialised, we perform the RG flow down from a UV scale of $\Lambda \approx 0.5$ eV to an infrared scale set by $T$. We use the resulting renormalised couplings as input to the free energy, minimising to obtain the resulting ground state. This procedure generates the phase diagrams of Fig. \ref{f:phase}(a)ii and (b)ii.

We discuss now the choice of initial couplings that lead to Figs. \ref{f:phase}(a)ii and (b)ii. 
Given that several of the couplings in Table \ref{couplings_estimate} vanish, we allow for the situation where these couplings take negative values. By inspection of the gap equation \eqref{gap_eqn}, we see that an initial value of $g_{1,i}<0$ promotes CDW (this was first noted in \cite{Park2021} for the problem of a single vHS), while $h_1<0$ promotes chiral excitons. To this end, we first allow for both $g_{1,c}, h_1<0$, and subsequently arrive at the phase diagram of Fig. \ref{f:phase}(a)ii. In addition, we allow for $g_{3,d}=0$, and arrive at Fig. \ref{f:phase}(b)ii. The phase diagram is qualitatively the same for $g_{3,d}<0$.  {Next we mention that the magnitudes and ratios have been estimated from \textit{ab initio} calculations (Table \ref{couplings_estimate}). A more accurate treatment would account for the renormalisation of the couplings in going from lattice to the patch UV cutoff $\Lambda$. We have not included these effects in our analysis. We stress that our use of the values in Table \ref{couplings_estimate} is to illustrate that there exist physically reasonable bare couplings which produce the desired phenomenology.}

%%%%%%%%%%%%%%%%% 

%To complement the analysis leading to Fig. \ref{f:phase}(a)i and \ref{f:phase}(b)i, we now explicitly compute the eigenvalues $\lambda_\text{CDW}$ and $\lambda_\Phi$ directly from the RG procedure. The input for this computation are the initial values of the couplings, listed in Table \ref{couplings_estimate}, taking $U=1.25$ eV. In this way, all parameters have been fixed via the DFT mapping.  Our procedure is to: (i) run the RG computation on the couplings \eqref{intRG}, which provides their $T$-dependence; (ii) diagonalise the gap equation \eqref{gap_eqn} to find the eigenvalues $\lambda_i(T)$ of the order parameters; and finally (iii) examine the appropriate free energy of the leading (two) instabilities, determining whether they compete or coexist. This provides a phase diagram as a function of $T$.

\section{Discussion}
\label{disc}

We introduced and analysed a minimal model to describe interacting fermions near a twofold van Hove singularity (TvHS) -- two opposite concavity vHS near the Fermi level. We found the opposite concavities of the two vHS crucially affect the possible many-body instabilities, relative to the single vHS case.  In particular, excitonic order contends as a possible instability and generically results in a chiral $d$-wave excitonic phase in hexagonal systems such as honeycomb and kagome lattices. We contrast our scenario with topological excitonic states which have been previously  explored theoretically \cite{Yu2014,YuenZhou2014, Efimkin2013, Gon2017}; in our case, the topology of the $d+id$ condensate is not inherited from the Berry curvature at the $K$-points or from spin-orbit coupling, but appears at the $M$-point intrinsically due to interaction driven, spontaneous time-reversal symmetry breaking. These findings suggest a new class of candidate materials for topological excitonic ground states.

TvHS were recently seen experimentally in AV$_3$Sb$_5$ \cite{Kang2021}. We now discuss key features of experiment and the extent to which the TvHS minimal model explains them: First,  signatures of time-reversal symmetry breaking, including a significant anomalous Hall effect, are observed at temperatures near to $T_*$ despite the lack of magnetic ordering \cite{Xu2022,Kenney2021}. The presence of chiral excitonic order would offer an appealing interpretation of the broken time-reversal symmetry and anomalous Hall effect. Second, experiments also report the breaking of threefold rotational symmetry and onset of nematic order around $T_c\lesssim 50$ K. Coupling between excitons and CDW naturally results in a phase consisting of nodal $d$-wave excitons and a nematic CDW, however our analysis of the free energy found this phase was only ever a local minimum in our model. Coupling to phonons may promote this phase to the dominant ground state, and we leave further examination of this scenario to future work.  Third, superconductivity emerges generically as an instability of the TvHS minimal model. However, superconductivity is seen at a much lower temperature scale ($T_c\approx 3.5$ K) \cite{Yang2021, Jiang2021, Li2021b, Zhao2021, Li2021c, Shumiya2021, Mielke2021b} than CDW ($T_*\approx 100$ K). At these temperatures the correct starting point for a description of superconductivity requires incorporating the CDW into the ground state.

Time-reversal symmetry (TRS) breaking and the anomalous Hall conductivity has also been proposed to arise due to a complex CDW state \cite{Park2021,Denner2021, lin2021complex}. Our analysis shows that it is possible for CDW and chiral excitons to coexist, but a key difference between these two states is that chiral excitons break TRS but not translational symmetry, allowing experiment to disentangle the two. To this end, it has been shown that the application of strain and pressure can selectively move the two opposite concavity Fermi surfaces closer or further away from the Fermi level \cite{Consiglio2021}. Moving the Fermi level away from one of the vHS creates a mismatch in the Fermi surface areas, which disfavours the excitonic phase, whereas moving the Fermi level away from the two vHS in a way that keeps the Fermi surface areas roughly equal disfavours charge order, while retaining the tendency to excitonic order. Experimental probes of TRS breaking could be applied in the presence of strain and pressure to disentangle the two phases. Additionally, we suggest that the existence of an exciton condensate should lead to Andreev-like interband tunneling, an effect which has been used to probe excitonic order in bilayer systems \cite{Finck2011}.

Beyond the vanadium metals AV$_3$Sb$_5$, another possible scenario for topological excitonic condensation is to engineer TvHS in van der Waals heterostructures made from materials with hexagonal symmetry such as graphene or transition metal dichalcogenides (TMDCs) \cite{Tritsaris2020,Zhu2020}. In moir\' e systems, the reduced bandwidth of the bands near charge neutrality brings the opposite concavity vHS of the valence and conduction bands closer in energy, so that an bias field could feasibly result in a TvHS.  The valence and conduction bands can be further aligned through spin splitting the bands via a TMDC layer \cite{Zaletel2020, siriviboon2021abundance}, the effect of which can be tuned via twist angle \cite{David2019}. Finally, valley
polarisation is observed in twisted layered systems \cite{LiuDaiReview, lin2021zerofield, scammell2021theory}, which could also be exploited to align the valence and conduction bands, e.g. through methods discussed in \cite{Ying2021,Li2020c}. In the context of layered van der Waals materials, a possible experimental probe would be the enhanced tunneling between layers induced by excitons, e.g. \cite{BurgPRL2018, EfimkinPRB2020}.

\clearpage

\section*{METHODS}

\  

\noindent {\bf Leading instabilities.}
Our discussion of the leading ordered states follows from the solution of the gap equations for the order parameter vertices $\mathcal{O}_i=\{\mathcal S_{\alpha i}, \mathcal C_{\alpha i}, \Delta_{\alpha i}, {\cal P}_{\alpha \pm},\Phi^{C}_{\alpha \pm},\Phi^{S}_{\alpha \pm}\}$. We find the mean field gap equations to be
\begin{align}
\label{gap_eqn_methods}
\notag\frac{\partial}{\partial t} {\Phi}^C_{\alpha+}&=d_4\sum_{\beta\neq\alpha}\Big\{(h_4-2l_4) {\Phi}^C_{\alpha+}   -j_4 {\Phi}^C_{\alpha-}\\
\notag & + (h_1-2l_2) {\Phi}^C_{\beta+} +  (j_1-2j_2) {\Phi}^C_{\beta-}\Big\}\\
\notag\frac{\partial}{\partial t} {\Phi}^S_{\alpha+}&=d_4\sum_{\beta\neq\alpha}\Big\{h_4 {\Phi}^S_{\alpha+} +  j_4 {\Phi}^S_{\alpha-} + h_1 {\Phi}^S_{\beta+} +  j_1 {\Phi}^S_{\beta-}\Big\}\\
\notag \frac{\partial}{\partial t}{\cal P}_{\alpha,+}&=-d_1\Big\{h_2{\cal P}_{\alpha,+} +  h_1 {\cal P}_{\alpha,-} +l_1{\cal P}_{\bar\alpha,+} + l_2{\cal P}_{\bar\alpha,-}\Big\}\\
\
\notag \frac{\partial}{\partial t} {\cal C}_{\alpha,\nu} &= d_{2\nu}(g_{2,c}-2g_{1,c}) {\cal C}_{\alpha,c}   -d_{2\nu} g_{3,c} {\cal C}_{\bar{\alpha},c}\\
\notag & + d_{2\bar{\nu}}(l_2-2h_1) {\cal C}_{\alpha,d} + d_{2\bar{\nu}} (l_3-2h_3) {\cal C}_{\bar{\alpha},d} \\
\notag \frac{\partial}{\partial t} {\cal S}_{\alpha,\nu} &=  d_{2\nu} g_{2,c}{\cal S}_{\alpha,c}   + d_{2\nu} g_{3,c} {\cal S}_{\bar{\alpha},c}  + d_{2\bar{\nu}} l_2 {\cal S}_{\alpha,d}\\
\notag & + d_{2\bar{\nu}} l_3 {\cal S}_{\bar{\alpha},d}\\
\frac{\partial}{\partial t} {\Delta}_{\alpha,\nu}&=-\sum_{\beta\neq\alpha} \Big\{d_{0\bar{\nu}}g_{4,c}{\Delta}_{\alpha,c} + d_{0\bar{\nu}}g_{3,c} {\Delta}_{\beta,c}\\
\notag & + d_{0\nu} j_4{\Delta}_{\alpha,d} + d_{0\nu} j_3{\Delta}_{\beta,d} \Big\}
\end{align}
with indices as defined previously: $c$, $d$, $\pm$ referring to flavour, $\alpha$ to patch, and $\bar{\alpha}$ denoting the patch connected to $\alpha$ by a nesting vector. To make the equations compact, we have introduced $\nu=\{c,d\}$ with $\bar{\nu}=\{d,c\}$. The $d$-factors are nesting coefficients which characterise the relative strength of the particle-particle and particle-hole divergences, and are defined in the Supplementary Material -- we have used notation so that $d_{0c}=1, d_{0d}=d_0, d_{2c}=d_2, d_{2d}=d_3$.  The couplings entering the gap equations are understood to inherit scale-dependence from the RG equations for the couplings \eqref{intRG}. The eigenvectors for this linear system of gap equations give the possible order parameter structures, and those with the largest eigenvalue are the leading instabilities. The set of Feynman diagrams which generate these flow equations are given in the Supplementary Material.

\  

\noindent {\bf Landau-Ginzburg analysis.} 
The susceptibility gap equations \eqref{gap_eqn_methods} are insufficient to determine whether order parameters compete or can form a ground state in which multiple orders coexist. Given a set of degenerate or nearly-degenerate solutions to the gap equations, we determine which combination of these solutions is the favoured ground state by calculating the Landau-Ginzburg free energy. We employ the mean-field decomposition of the fermions coupled to a combination of order parameter matrices, and integrate out the fermionic degrees of freedom, arriving at the free energy

\begin{align}
\mathcal{F} = \tfrac{1}{2\lambda_\Phi}\sum_i |\Phi_i|^2 +\tfrac{1}{2}\!\!\!\!\!\!\sum_{\alpha\neq\beta; \nu=c,d} \!\!\!\!\!\!\mathcal{V}^{-1}_{\nu\nu'}\,C_{\alpha\beta\nu}C^*_{\alpha\beta\nu'}  -\text{Tr} \log \mathcal{G}^{-1}. 
\end{align}
Here the full Green's function
\begin{align}
\mathcal{G}^{-1}(i\omega_n,\bm{q}) = \mathcal{G}_0^{-1}(i\omega_n,\bm{q}) + M,
\end{align}
comprises the order parameter matrix $M=M_\Phi + M_C$,
\begin{align}
M_\Phi &= \begin{pmatrix}
\Phi_1 &0   &0 \\ 
 0&  \Phi_2 &0 \\ 
 0&  0&   \Phi_3
\end{pmatrix} \otimes \sigma_+ + \begin{pmatrix}
\Phi^*_1 &0   &0 \\ 
 0&  \Phi^*_2 &0 \\ 
 0&  0&   \Phi^*_3
\end{pmatrix} \otimes \sigma_-\\
M_C &= \begin{pmatrix}
0 &  C_{12c}   & C_{31c}^* \\ 
C^*_{12c} &  0 & C_{23c} \\ 
C_{31c} &  C^*_{23c} &  0
\end{pmatrix} \otimes \sigma_c \nonumber \\
& \ \ \ \ \ \ \ \ \ \ \ \ \ \ \ + \begin{pmatrix}
0 &  C_{12d}   & C_{31d}^* \\ 
C^*_{12d} &  0 & C_{23d} \\ 
C_{31d} &  C^*_{23d} &  0
\end{pmatrix}  \otimes \sigma_d
\end{align}
and the bare Green's function
\begin{gather}
\mathcal{G}_0^{-1}(i\omega_n,\bm{q}) =\begin{pmatrix}
i\omega_n -\varepsilon_1(\bm q) &0  &0 \\ 
 0&  i\omega_n -\varepsilon_2(\bm q) &0 \\ 
 0&  0& i\omega_n -\varepsilon_3(\bm q) 
\end{pmatrix} \otimes \sigma_c \nonumber \\
+ \begin{pmatrix}
i\omega_n +\varepsilon_1(\bm q) &0  &0 \\ 
 0&  i\omega_n +\varepsilon_2(\bm q) &0 \\ 
 0&  0& i\omega_n +\varepsilon_3(\bm q) 
\end{pmatrix} \otimes \sigma_d.
\end{gather}
The dispersion at each patch is $\varepsilon_1(\bm q) =\tfrac{1}{2}q_x(q_x+\sqrt{3}q_y)$, $\varepsilon_2(\bm q) =\tfrac{1}{4}(-q_x^2+3q_y^2)$ and $\varepsilon_3(\bm q) =\tfrac{1}{2}q_x(q_x+\sqrt{3}q_y)$.  For the two degenerate $d$-wave excitons, parameterised by $\Phi_a$ and $\Phi_b$, we have 
\begin{align}
\Phi_1 &= -\tfrac{1}{\sqrt{2}}\Phi_a -\tfrac{1}{\sqrt{6}}\Phi_b ,\nonumber \\
 \Phi_2 &= \sqrt{\tfrac{2}{3}}\Phi_b, \nonumber \\
  \Phi_3 &= \tfrac{1}{\sqrt{2}}\Phi_a -\tfrac{1}{\sqrt{6}}\Phi_b. 
\end{align}
Rewriting
\begin{align}
\text{Tr} \log \mathcal{G}^{-1} = -\mathcal{F}_0 +\text{Tr} \log \left(1 +\mathcal{G}_0M \right) 
\end{align}
where $\mathcal{F}_0$ is the free energy of a free Fermi gas, and using the expansion
\begin{align}
\text{Tr} \log \left(1 +\mathcal{G}_0M \right)  = \sum_{n=0}^\infty \tfrac{(-1)^n}{n}\text{Tr} (\mathcal{G}_0M)^n
\end{align} 
we evaluate the trace of the first four terms in the expansion, resulting in the free energy stated in the main text. Determining whether the minimum of the free energy can include coexisting $C_{\alpha\beta\nu}$ and $\Phi_a,\Phi_b$ requires knowledge of the coefficients in this expansion; their calculation is detailed in the Supplementary Material.

\  

\noindent {\bf Edge states.} To demonstrate the presence of edge states in the excitonic phase, we employ a simplified model for numerical diagonalisation, describing a kagome lattice with two orbital states $\nu = \pm$,
\begin{gather}
\label{HMF}
H = -\!\!\!\sum_{\langle \bm{r},\bm{r}'\rangle,\nu}{t_\nu c^\dag_{\bm{r}',\nu} c_{\bm{r},\nu}} + \sum_{\bm{r}}{\gamma_0 c^\dag_{\bm{r},1} c_{\bm{r},1}} \nonumber\\ +\sum_{\langle\bm{r}', \bm{r}\rangle}{\Delta(\bm{r}',\bm{r}) c^\dag_{\bm{r'},\nu} c_{\bm{r},\nu'} + \text{h.c.}}
\end{gather}
in which only coupling between nearest neighbours is taken into account. We choose the excitonic pairing function $\Delta(\bm{r},\bm{r}')$ so that the lattice theory possesses an equivalent continuum limit to our field theory description of the three patches surrounding the $M$ points. The gap function is
\begin{gather}
\Delta(\bm{r}',\bm{r} \in \sigma) = \tfrac{1}{\sqrt{6}} \,\Delta_0 \,e^{i(\theta_{\bm{r}'-\bm{r}} - (\ell +1)\varphi_\sigma)}  \ \ , \nonumber \\
(\varphi_A,\varphi_B,\varphi_C) = (0, \tfrac{2\pi}{3}, \tfrac{4\pi}{3}),
\end{gather}
with $\ell = \pm 2$ equal to the phase winding of the excitonic order around the Fermi surface. The results for $\ell = -2$ are plotted in Fig. \ref{f:edge} in the main text with $\gamma_0 = 2t$, $\Delta_0 = 0.5t$, for a ribbon geometry with 60 unit cells.

\let\oldaddcontentsline\addcontentsline
\renewcommand{\addcontentsline}[3]{}

\section*{Acknowledgements}
We thank Brenden Ortiz, Michael Denner, Mingu Kang, Colin Nancarrow and Dmitry Efimkin for discussions and comments. HS and OPS acknowledge funding from ARC Centre of Excellence FLEET. TL acknowledges support from the Deutsche Forschungsgemeinschaft.

\section*{Author Contributions Statement}
H.D.S and J.I conceived of the project idea, performed the RG analysis and calculated the Landau-Ginzburg free energy. T.L. performed edge state calculations. All authors critically discussed the details and contributed to writing.
\newpage
{\color{white}.}
\newpage 
\begin{center}
\widetext
\textbf{\large Supplementary Material}
\end{center}

\setcounter{equation}{0}
\setcounter{table}{0}
\setcounter{section}{0}
\setcounter{figure}{0}
\makeatletter
\renewcommand{\theequation}{S\arabic{equation}}
\renewcommand{\thefigure}{S\arabic{figure}}
\renewcommand{\thesection}{S\arabic{section}}
%\renewcommand{\citenumfont}[1]{S#1}
%\begin{widetext}

\bibliography{refs.bib}

\section{Interactions}

\subsection{Patch model}

In the kagome tight-binding model, the wavefunctions at the pure ($+$) and mixed ($-$) vHS have the following sublattice structure near the $M$-points,
\begin{align}
\label{wavefuncs}
\notag \ket{\bm M_1, +} =\hat{A}, && \ket{\bm M_1, -} =\frac{1}{\sqrt{2}}\left(\hat{B}+\hat{C}\right),\\
\notag \ket{\bm M_2, +} =\hat{B}, && \ket{\bm M_2, -} =\frac{1}{\sqrt{2}}\left(\hat{A}+\hat{C}\right),\\
\ket{\bm M_3, +} =\hat{C}, && \ket{\bm M_3, -} =\frac{1}{\sqrt{2}}\left(\hat{A}+\hat{B}\right).
\end{align}
where $\hat{A}, \hat{B}, \hat{C}$ are basis vectors indicating support on the $A, B, C$ sublattices.  Due to the different sublattice structure of the wavefunctions for the pure and mixed states, the corresponding interactions evaluated on the pure and mixed vHS have different intraflavour interactions, denoted $g_{ic}$ and $g_{id}$, where the flavour $c$ is treated as $p$-type and $d$ as $m$-type. The interflavour vertices do not require this distinction. The most general set of interactions between patches/flavours allowed by momentum conservation then results in a 20 coupling model, 
\begin{align}
\label{Vsupp}
V&= \frac{1}{2}\sum_{\alpha,\beta}\left[V_{g,c} + V_{g,d} + V_h + V_j +V_l\right],\\
\notag V_{g,c}&=  g_{1c} c^\dag_\alpha c^\dag_\beta c_\alpha c_\beta + g_{2c} c^\dag_\alpha c^\dag_\beta c_\beta c_\alpha + g_{3c} c^\dag_\alpha c^\dag_\alpha c_\beta c_\beta +\frac{1}{2} g_{4c} c^\dag_\alpha c^\dag_\alpha c_\alpha c_\alpha, \\
\
\notag V_{g,d}&=  g_{1d} d^\dag_\alpha d^\dag_\beta d_\alpha d_\beta + g_{2d} d^\dag_\alpha d^\dag_\beta d_\beta d_\alpha + g_{3d} d^\dag_\alpha d^\dag_\alpha d_\beta d_\beta +\frac{1}{2} g_{4d} d^\dag_\alpha d^\dag_\alpha d_\alpha d_\alpha, \\
\
\notag V_h&= h_1 c^\dag_\alpha d^\dag_\beta d_\alpha c_\beta +h_2 c^\dag_\alpha d^\dag_\beta d_\beta c_\alpha+h_3 c^\dag_\alpha d^\dag_\alpha d_\beta c_\beta +\frac{1}{2} h_4 c^\dag_\alpha d^\dag_\alpha d_\alpha c_\alpha + \left( c \leftrightarrow d \right),\\
\notag V_j&=  j_1 d^\dag_\alpha d^\dag_\beta c_\alpha c_\beta +j_2 d^\dag_\alpha d^\dag_\beta c_\beta c_\alpha + j_3 d^\dag_\alpha d^\dag_\alpha c_\beta c_\beta +\frac{1}{2} j_4 d^\dag_\alpha d^\dag_\alpha c_\alpha c_\alpha + \left( c \leftrightarrow d \right),\\
\notag V_l&= l_1 d^\dag_\alpha c^\dag_\beta d_\alpha c_\beta + l_2 d^\dag_\alpha c^\dag_\beta d_\beta c_\alpha + l_3 d^\dag_\alpha c^\dag_\alpha d_\beta c_\beta +\frac{1}{2} l_4 d^\dag_\alpha c^\dag_\alpha d_\alpha c_\alpha + \left( c \leftrightarrow d \right).
\end{align}
For honeycomb systems $g_{ic}=g_{id}$, reducing the number of independent couplings to 16. The corresponding Feynman diagrams are presented in Figure 1(d) [of the main text]. Due to the large density of states near the TvHS, the Coulomb repulsion is expected to be strongly screened, and we therefore model these interactions as momentum independent.

\subsection{Tight binding model}
To estimate the bare interactions, we start from the output of DFT results and directly compute the interaction vertices. The interacting tight binding model of \cite{Wu2021} gives
\begin{gather}
\label{TB}
H_{int}= U \sum_{i,\sigma,\mu}n_{i\sigma\mu \uparrow}n_{i\sigma\mu \downarrow} + U'\!\!\!\!\!\!\! \sum_{i,\sigma,\mu<\mu',s,s'}\!\!\!\!\!\!\!\! n_{i\sigma\mu s}n_{i\sigma\mu' s'} + V \!\!\!\!\sum_{\braket{ij}\sigma \sigma' \mu\mu'}\!\!\!\!\!\!n_{i\sigma\mu s}n_{j\sigma'\mu' s'}+ J\!\!\!\!\!\! \sum_{i\sigma \mu<\mu', s,s'}\!\!\!\!\!c^\dag_{i\sigma\mu s}c^\dag_{i\sigma\mu' s'}c_{i\sigma\mu s'}c_{i\sigma\mu' s} + J'\!\! \sum_{i\sigma \mu\neq\mu'}\!\!\!\!c^\dag_{i\sigma\mu \uparrow}c^\dag_{i\sigma\mu \downarrow}c_{i\sigma\mu' \downarrow}c_{i\sigma\mu' \uparrow}
\end{gather}
where $i$ enumerates unit cells, $\sigma=A,B,C$ enumerates sublattice, $\mu=\pm $ enumerates the orbital index, and $s$ is spin. Ref. \cite{Wu2021} finds $U\sim 1 -2$ eV, $U'=0.8 U$, $V\approx0.3U$ and $J=J'=0.1 U$.

\subsection{Patch Model -- initial coupling estimates}

Here we list the estimates for the bare couplings (which serve as initial conditions for RG flow). We evaluate with the spin structure $s,s ; -s,-s$. Before evaluating the orbital form factors, we note the orbital selection rules (i.e. conditions on index $i$ in \eqref{TB}) imply
\begin{align}
g_{i\nu} \propto U, V, && h_i \propto U', V, && j_i \propto J', && l_i \propto J.
\end{align}
Including form factors, evaluated at the $M$-points, we arrive at the initial conditions of Table I [of the main text], repeated here as Table \ref{couplings_estimate}.  The initial conditions taken to arrive at the phase diagram of Figure 4(a)ii and 4(b)ii [of the main text] are presented in Table \ref{initial_conds_1} and \ref{initial_conds_2}, respectively. 

\begin{table}
\begin{center}
\caption{ Estimates of the bare coupling values in the patch model \ref{Vsupp} for AV$_3$Sb$_5$. Projecting the pure and mixed sublattice form factors onto the cRPA results of Ref. \cite{Wu2021} results in the below values, where the intra-orbital, inter-orbital, Hund's, pair hopping, and nearest neighbour repulsions are $U=1$-$2$ eV, with $U'=0.8U$, $J=J'=0.1U$ and $V=0.3U$.} 
\label{couplings_estimate}
\vspace{0.1cm}
\begin{ruledtabular}
 \begin{tabular} {lllllllllllllllllllll} \\[-3.5mm]
 & $g_{i,c}$  & $g_{i,d}$ & $h_{i}$ & $j_{i}$ & $l_i$ \\[1mm] \hline \vspace{-0.2cm} \\ \vspace{0.2cm} 
$i=1$ & $0$ & $\tfrac{1}{4}(U+V)$ & $0$ & $0$ & $\tfrac{1}{2} J$ \\   \vspace{0.2cm}
$i=2$ & $V$ & $\tfrac{1}{4}U+V$ & $\tfrac{1}{2}U'+V$ & $0$ & $0$ \\   \vspace{0.2cm}
$i=3$ & $0$ & $\tfrac{1}{4}(U+V)$ & $0$ & $\tfrac{1}{2}J'$ & $0$ \\   \vspace{0.2cm}
$i=4$ & $U+V$ & $\tfrac{1}{2}U+V$ & $V$ & $0$ & $0$ \\[-1mm]
 \end{tabular}
\end{ruledtabular}
\end{center}
\end{table}
\begin{table}
\begin{center}
\caption{Initial conditions leading to the phase diagram of Figure 4(a)ii [of the main text]. Taking $U=1.3$ eV, with ratios fixed at $U'=0.8U$, $J=J'=0.1U$ and $V=0.3U$ as per Ref. \cite{Wu2021} and Table \ref{couplings_estimate}. Blue entries differ from the initial conditions of Table \ref{couplings_estimate}.} 
\label{initial_conds_1}
\vspace{0.1cm}
\begin{ruledtabular}
 \begin{tabular} {lllllllllllllllllllll} \\[-3.5mm]
 & $g_{i,c}$  & $g_{i,d}$ & $h_{i}$ & $j_{i}$ & $l_i$ \\[1mm] \hline \vspace{-0.2cm} \\ \vspace{0.2cm} 
$i=1$ & {\color{blue}$-\frac{2}{3}V$} & $\tfrac{1}{4}(U+V)$ &  {\color{blue}$-\frac{7}{4}V$} & $0$ & $\tfrac{1}{2} J$ \\   \vspace{0.2cm}
$i=2$ & $V$ & $\tfrac{1}{4}U+V$ & $\tfrac{1}{2}U'+V$ & $0$ & $0$ \\   \vspace{0.2cm}
$i=3$ & $0$ & $\tfrac{1}{4}(U+V)$ & $0$ & $\tfrac{1}{2}J'$ & $0$ \\   \vspace{0.2cm}
$i=4$ & $U+V$ & $\tfrac{1}{2}U+V$ & $V$ & $0$ & $0$ \\[-1mm]
 \end{tabular}
\end{ruledtabular}
\end{center}
\end{table}

\begin{table}[h]
\begin{center}
\caption{Initial conditions leading to the phase diagram of Figure 4(b)ii [of the main text]. Again taking $U=1.3$ eV, with ratios fixed at $U'=0.8U$, $J=J'=0.1U$ and $V=0.3U$. Blue entries differ from the initial conditions of Table \ref{couplings_estimate}}. 
\label{initial_conds_2}
\vspace{0.1cm}
\begin{ruledtabular}
 \begin{tabular} {lllllllllllllllllllll} \\[-3.5mm]
 & $g_{i,c}$  & $g_{i,d}$ & $h_{i}$ & $j_{i}$ & $l_i$ \\[1mm] \hline \vspace{-0.2cm} \\ \vspace{0.2cm} 
$i=1$ & {\color{blue}$-\frac{2}{3}V$} & $\tfrac{1}{4}(U+V)$ &  {\color{blue}$-\frac{7}{4}V$} & $0$ & $\tfrac{1}{2} J$ \\   \vspace{0.2cm}
$i=2$ & $V$ & $\tfrac{1}{4}U+V$ & $\tfrac{1}{2}U'+V$ & $0$ & $0$ \\   \vspace{0.2cm}
$i=3$ & $0$ & {\color{blue} $0$} & $0$ & $\tfrac{1}{2}J'$ & $0$ \\   \vspace{0.2cm}
$i=4$ & $U+V$ & $\tfrac{1}{2}U+V$ & $V$ & $0$ & $0$ \\[-1mm]
 \end{tabular}
\end{ruledtabular}
\end{center}
\end{table}

 %%%%%%%%%%%%%%%%%
\begin{figure}[t]
\vspace{-1cm}
{\includegraphics[width=0.365\textwidth,clip]{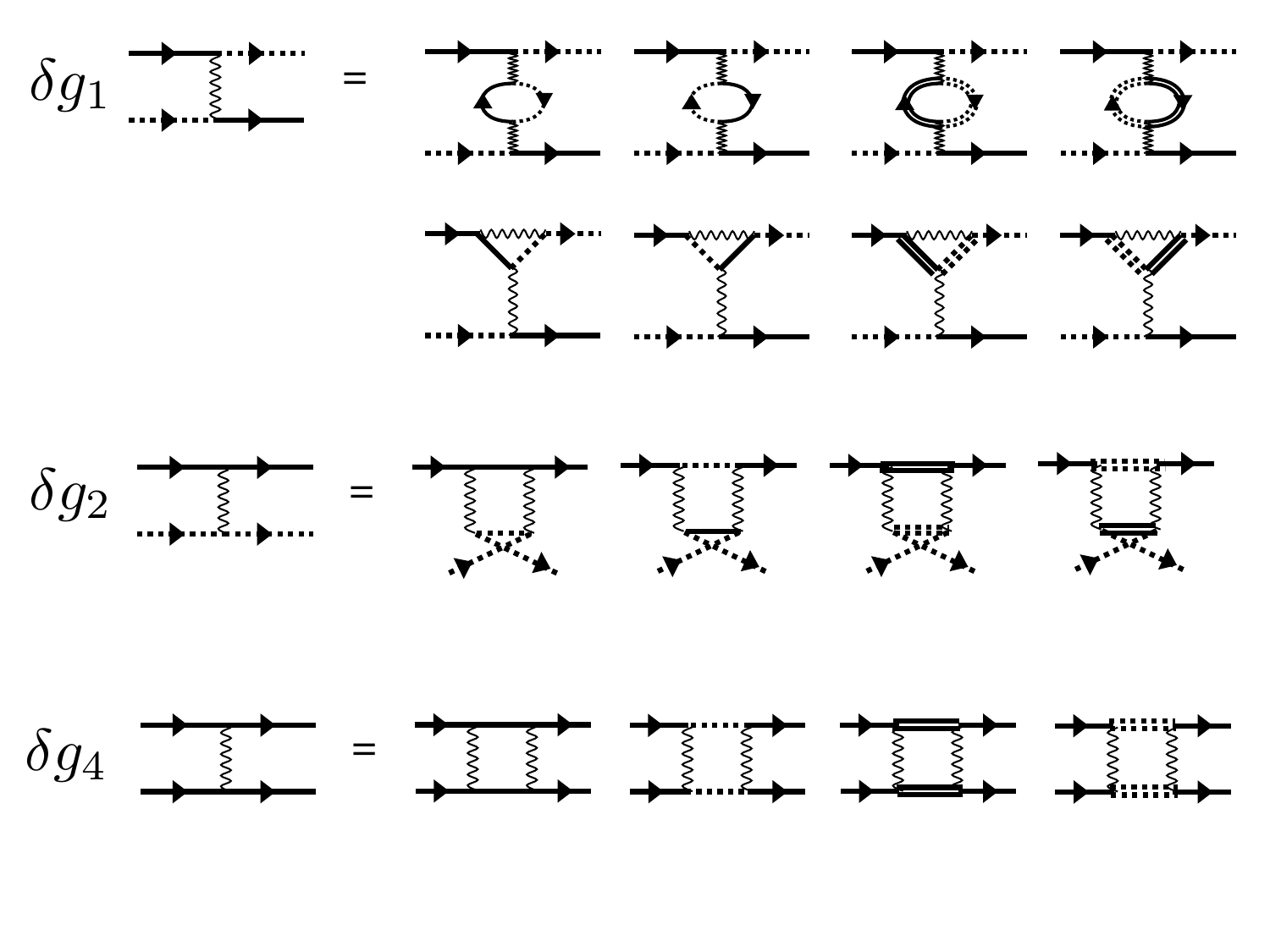}} \hspace{1cm}
{\includegraphics[width=0.365\textwidth,clip]{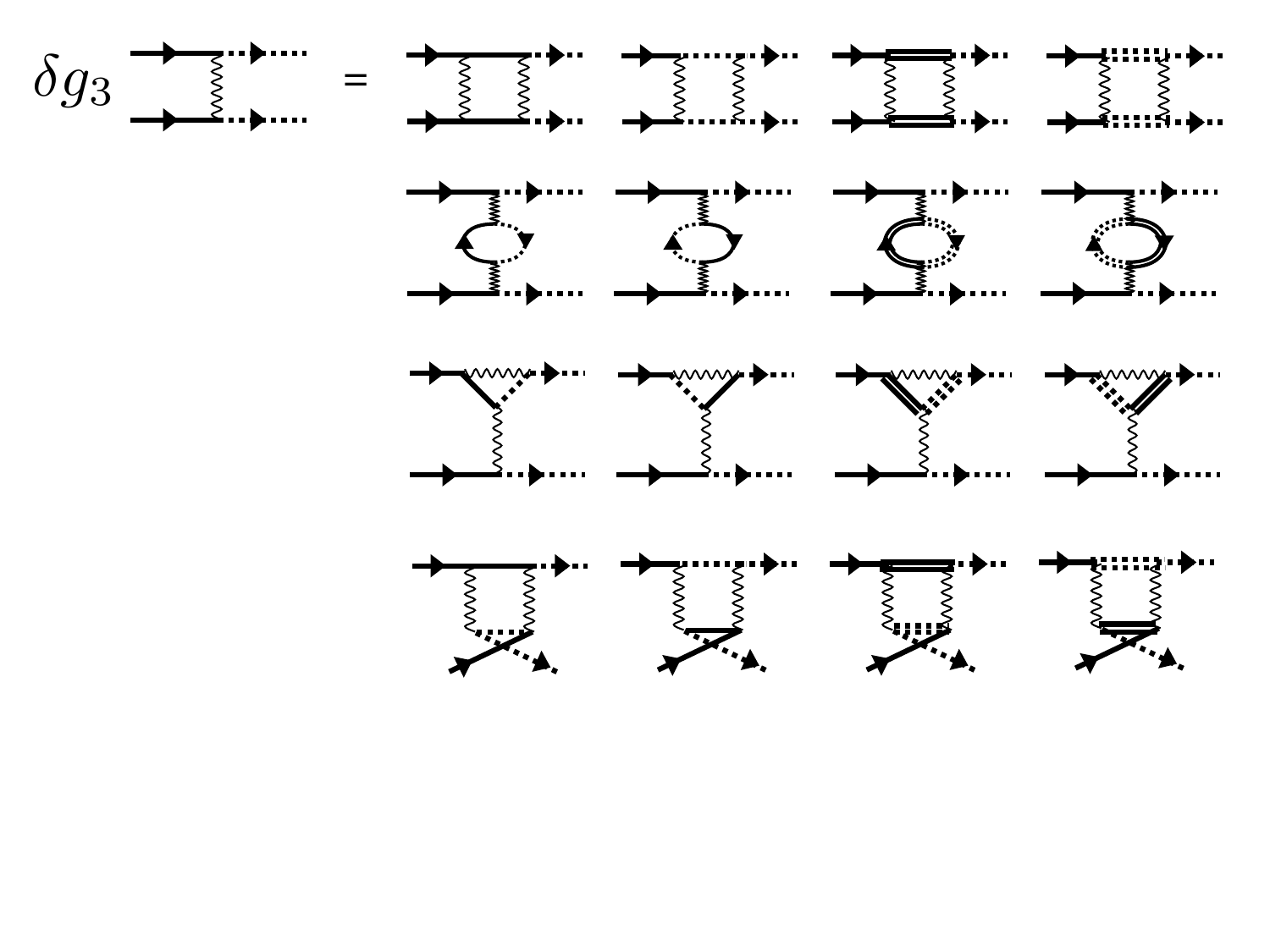}}
{\includegraphics[width=0.365\textwidth,clip]{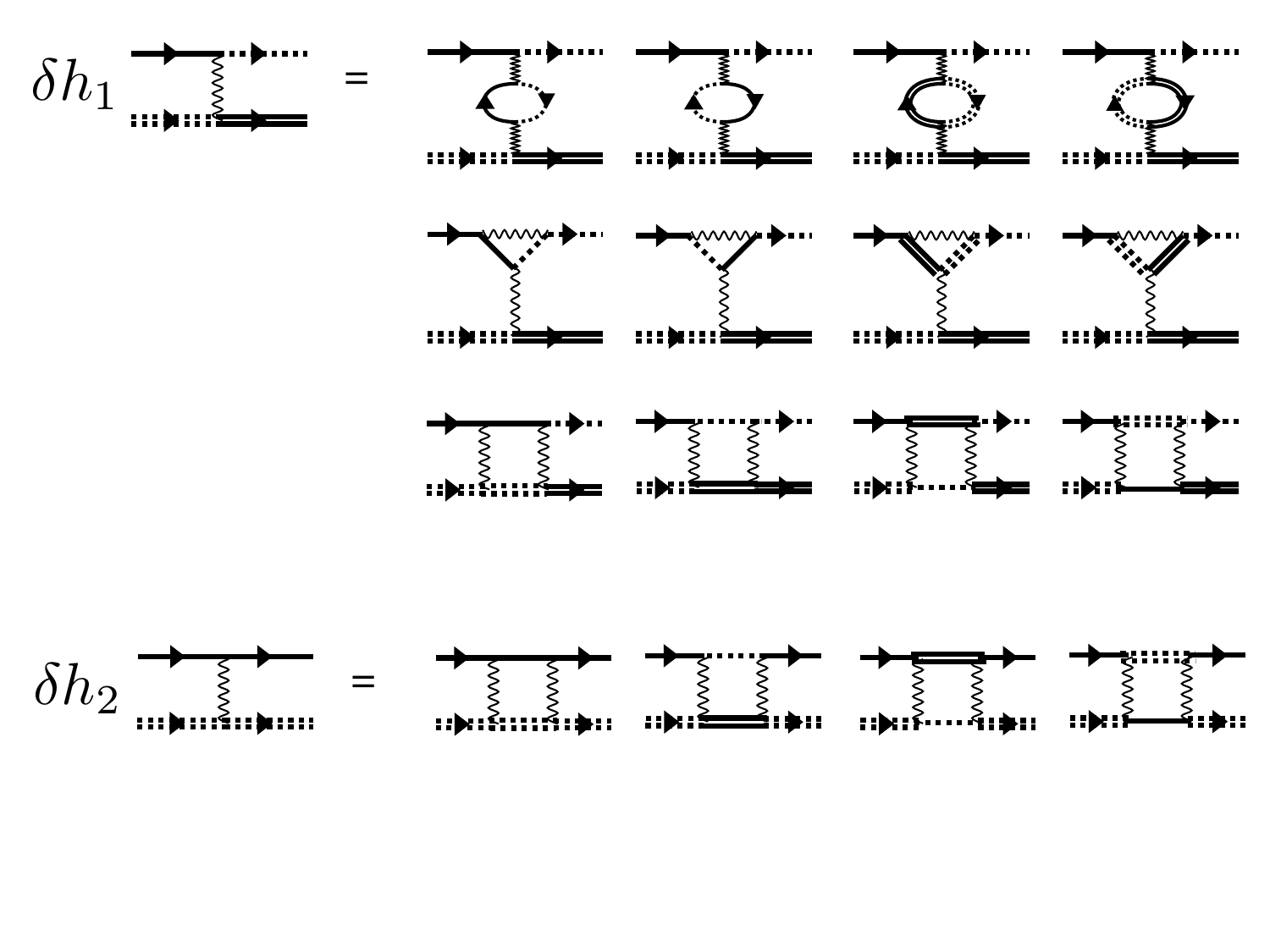}}\hspace{1cm}
{\includegraphics[width=0.365\textwidth,clip]{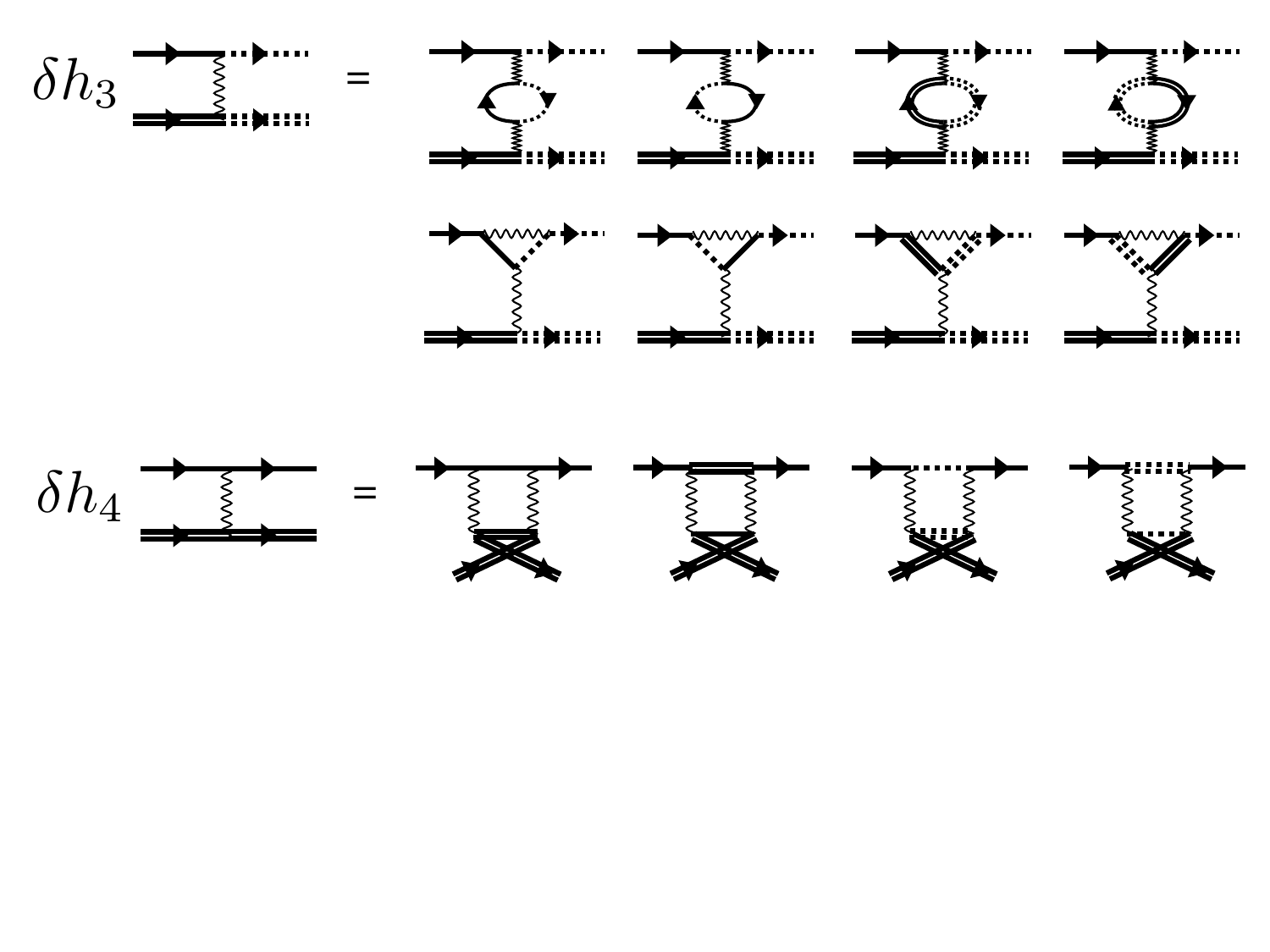}}
{\includegraphics[width=0.365\textwidth,clip]{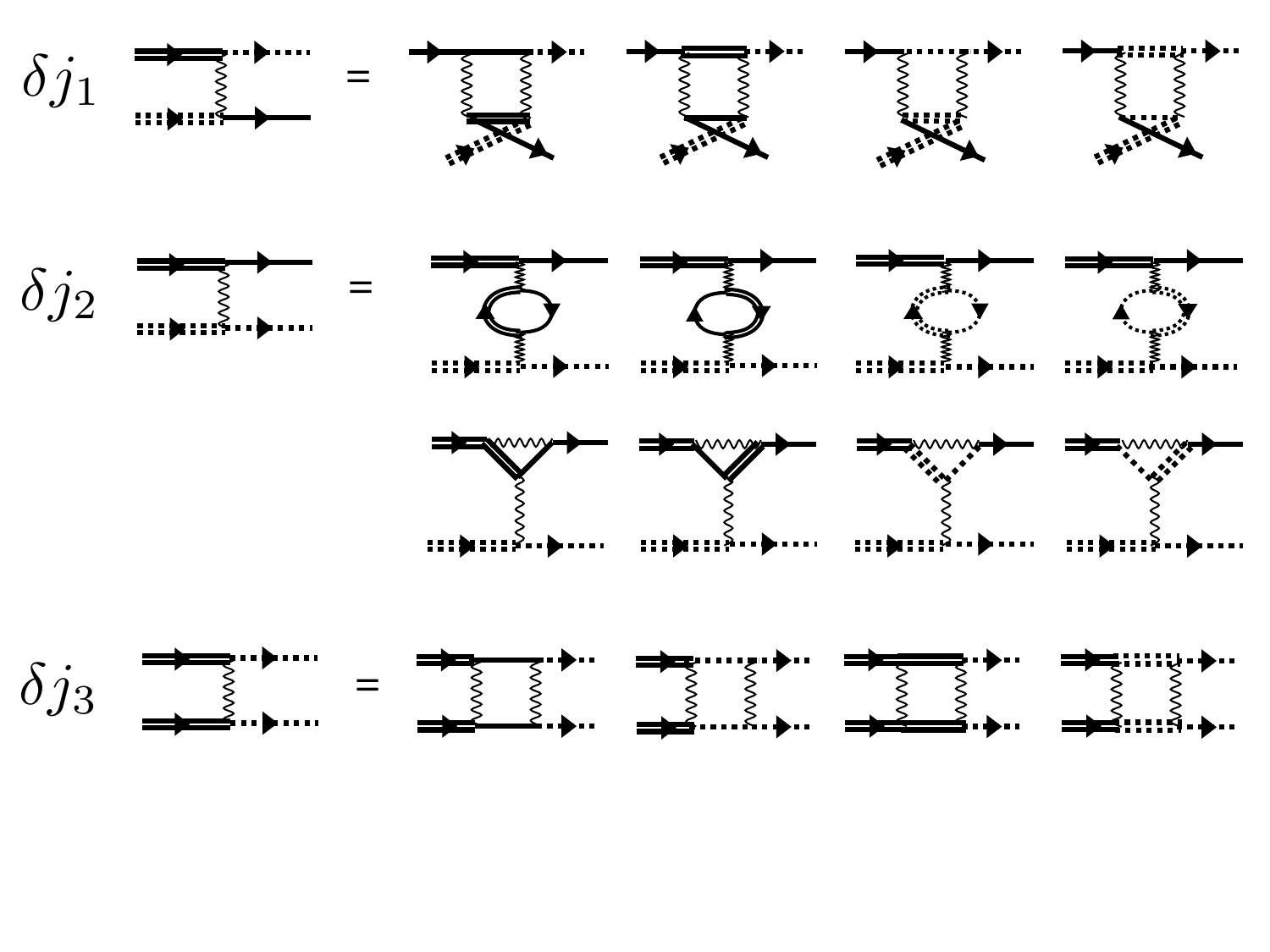}}\hspace{1cm}
{\includegraphics[width=0.365\textwidth,clip]{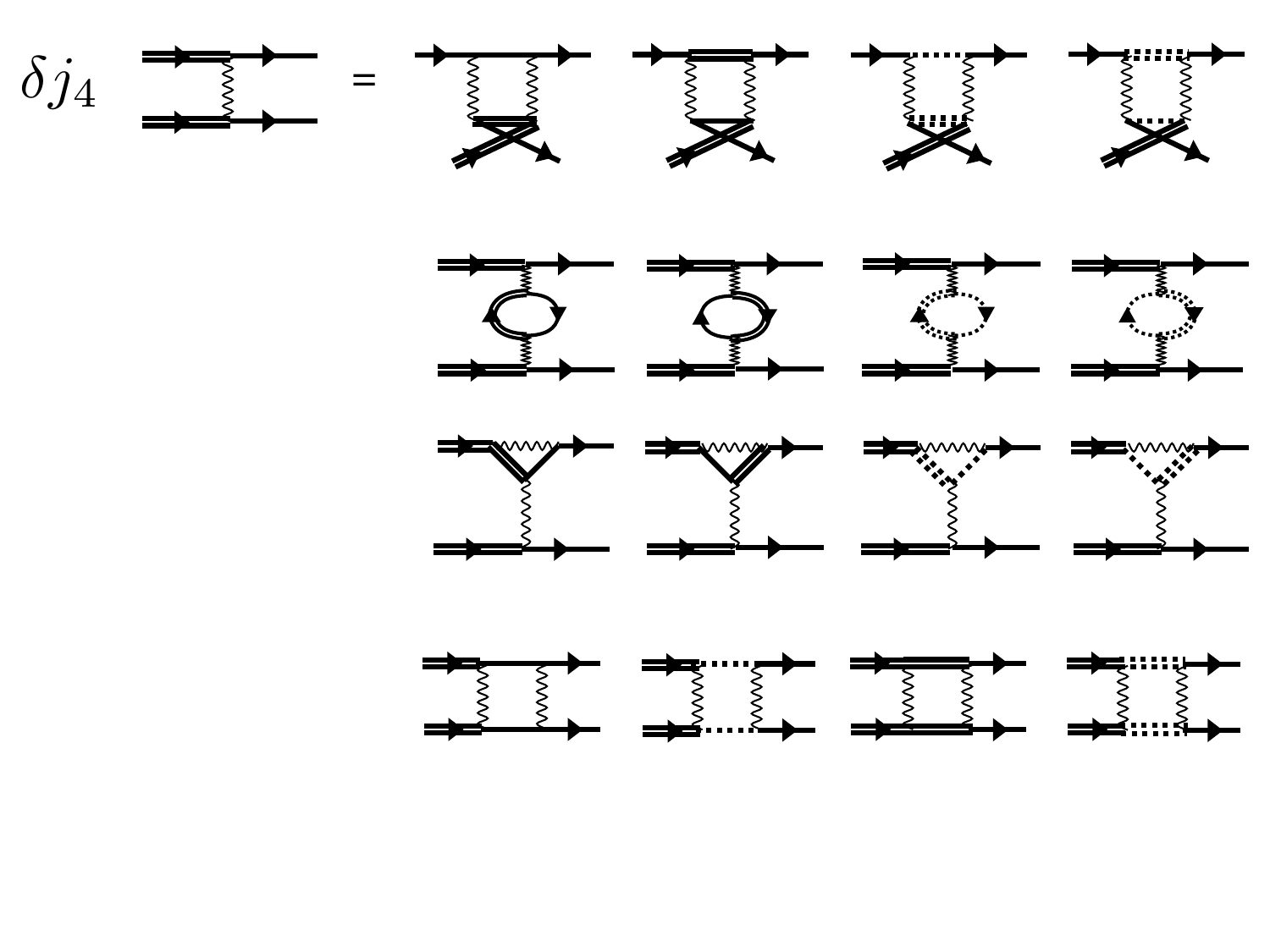}}
{\includegraphics[width=0.365\textwidth,clip]{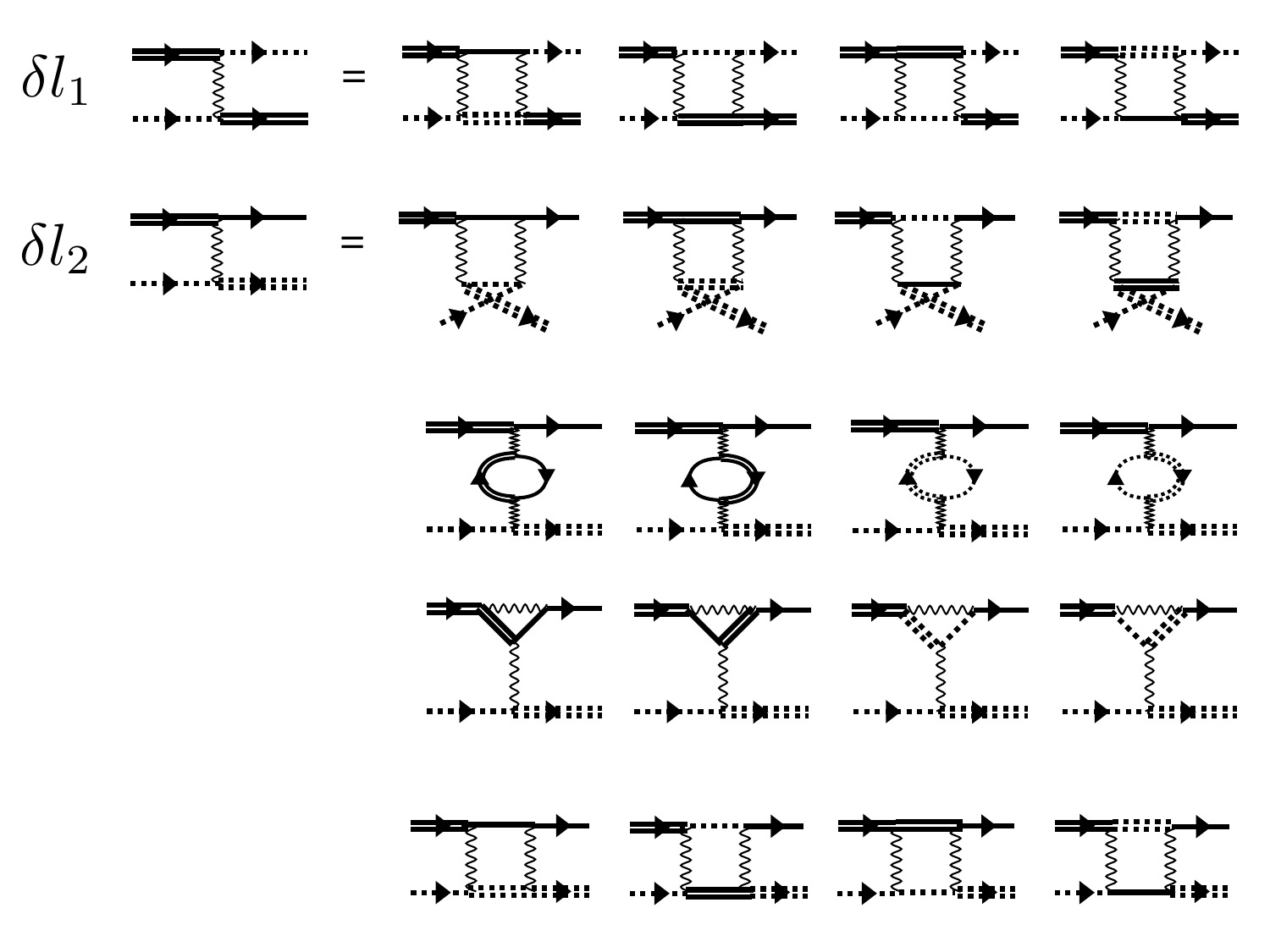}}\hspace{1cm}
{\includegraphics[width=0.365\textwidth,clip]{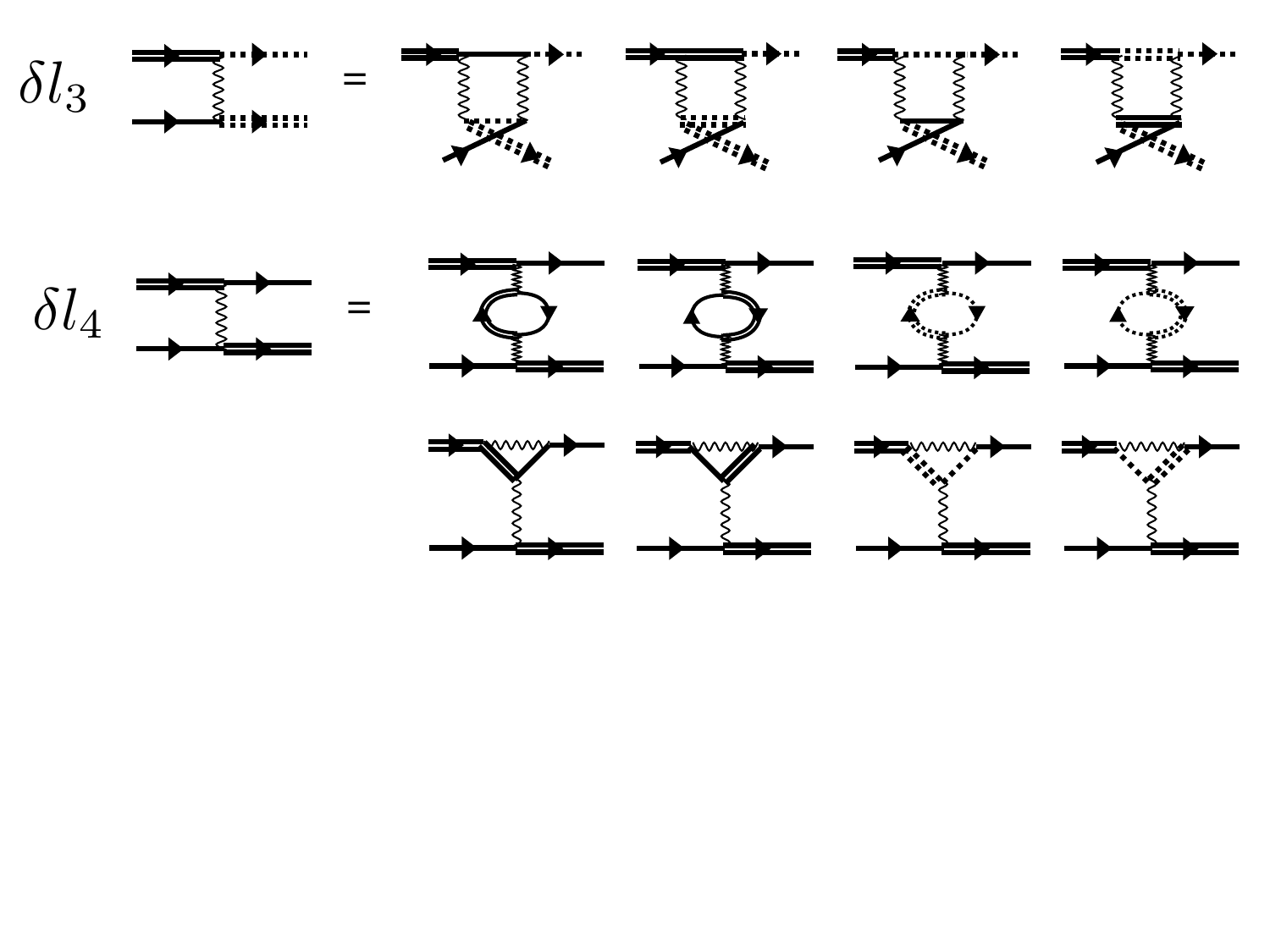}}
%{\includegraphics[width=0.45\textwidth,height=0.225\textwidth,clip]{RG_Blank}}
 \caption{\textbf{Flow equations for the couplings}: The distinct Feynman diagrams representing leading logarithmic ($\log^2$) corrections to the couplings (\ref{Vsupp}). Flavour indices on  $g_{ic}$ and $g_{id}$ are suppressed. The explicit flow equations for the couplings $g_{i\nu}, h_i, j_i, l_i$ are given by (\ref{RG}).}
\label{f:RG}
\end{figure}
%%%%%%%%%%%%%%%%% 

\section{Flow equations for the couplings}\label{s:RG}
\begin{figure}[t]
{\includegraphics[width=0.345\textwidth,clip]{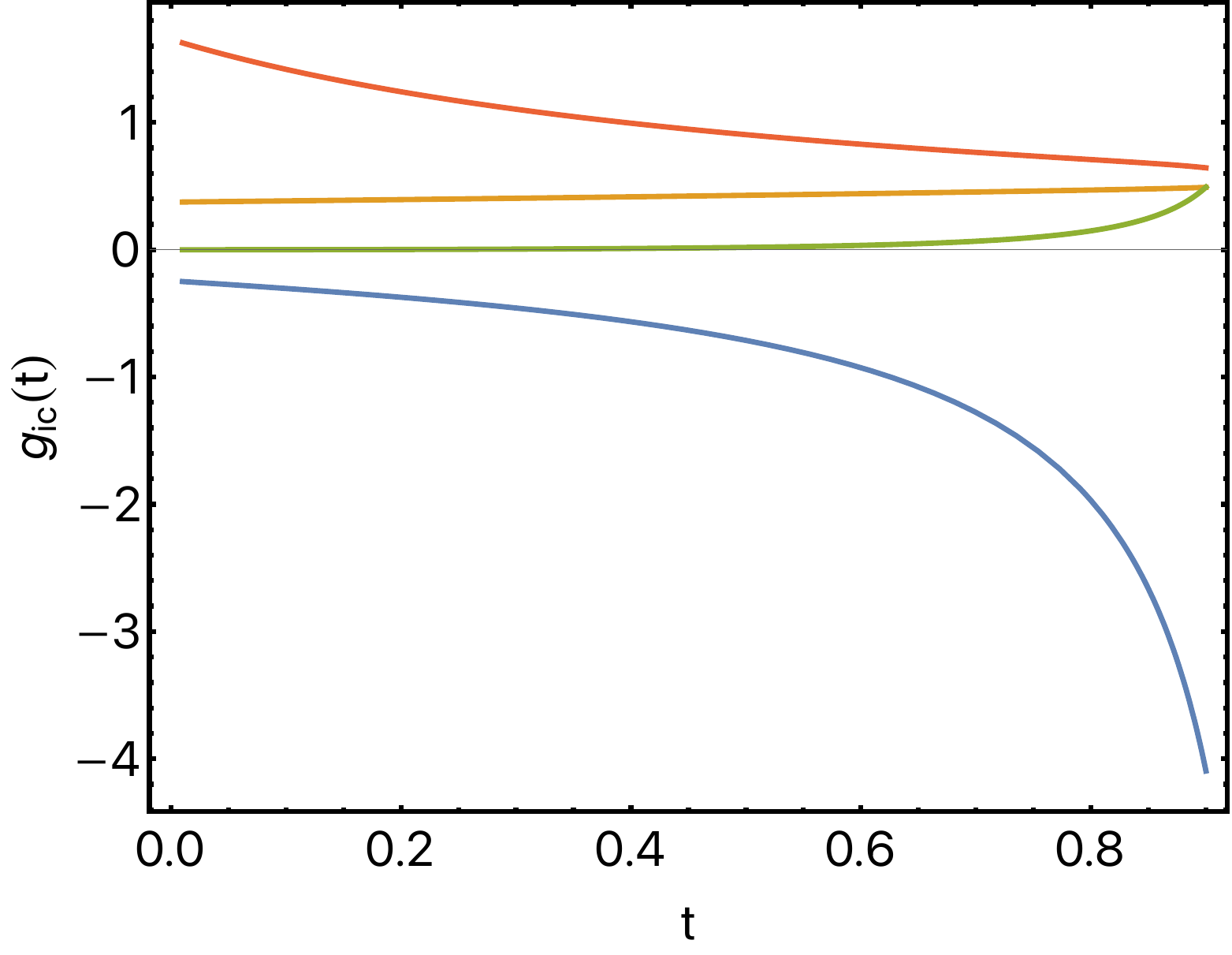}}
{\includegraphics[width=0.355\textwidth,clip]{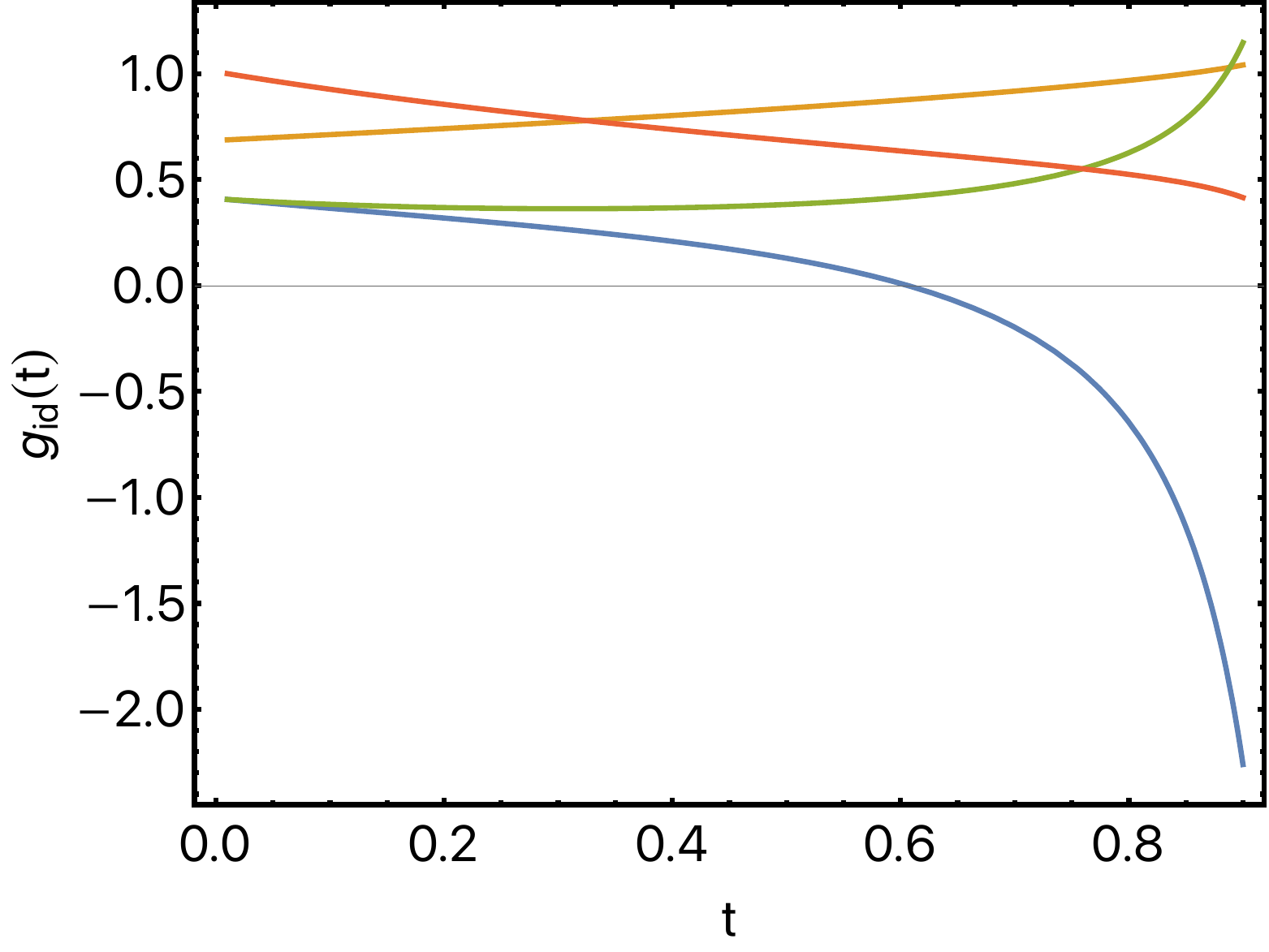}}
{\includegraphics[width=0.345\textwidth,clip]{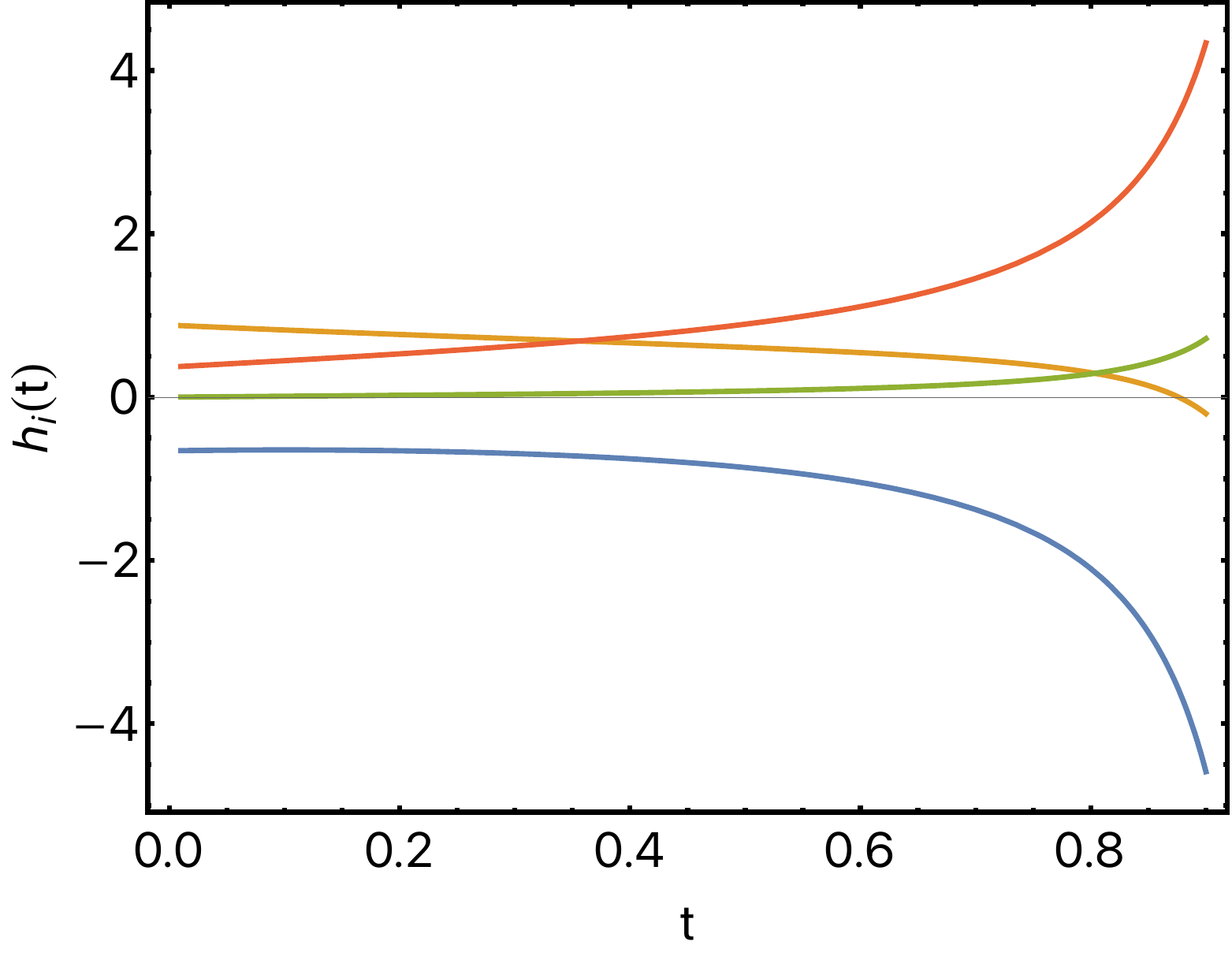}}
{\includegraphics[width=0.355\textwidth,clip]{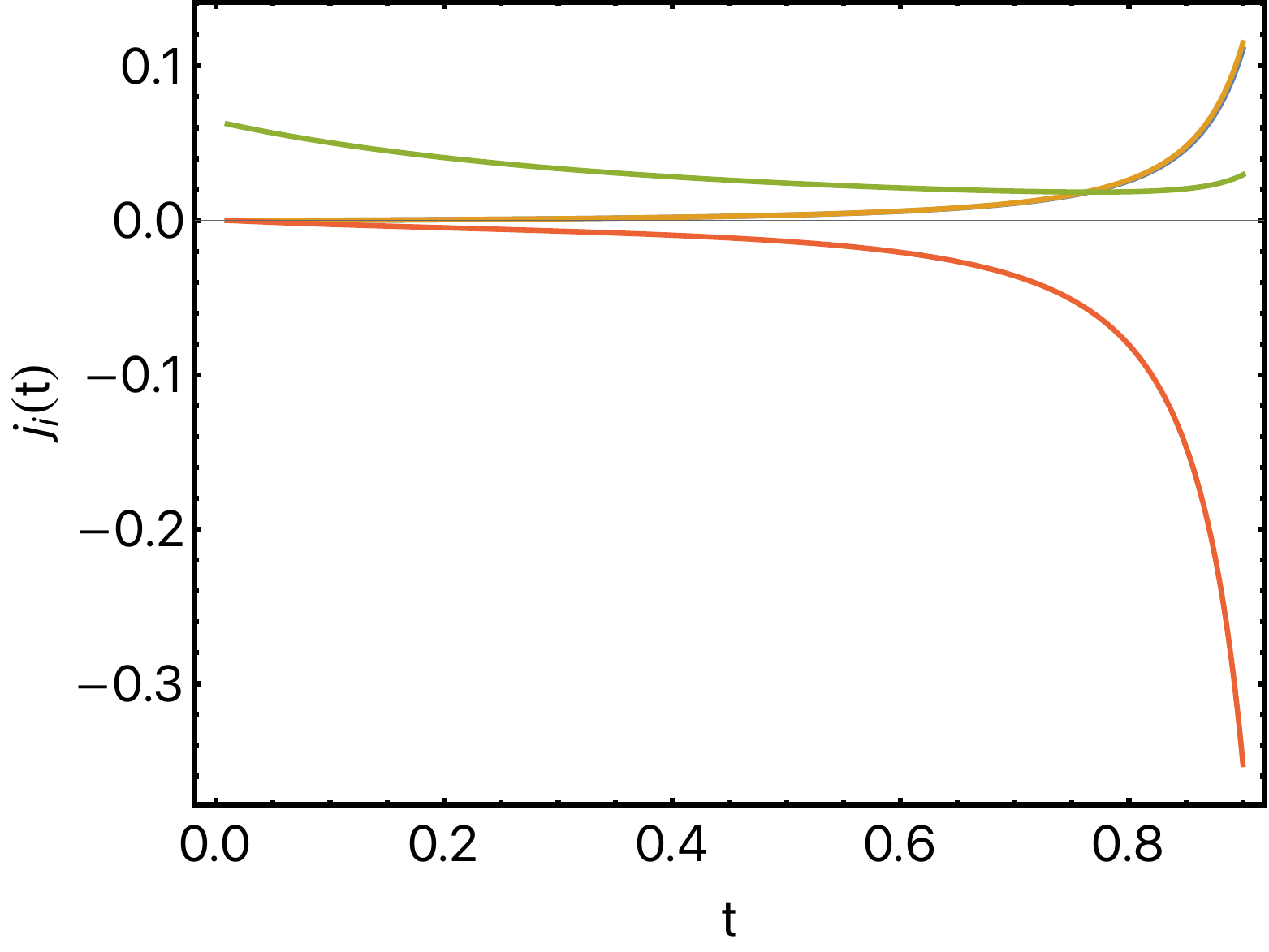}}
{\includegraphics[width=0.345\textwidth,clip]{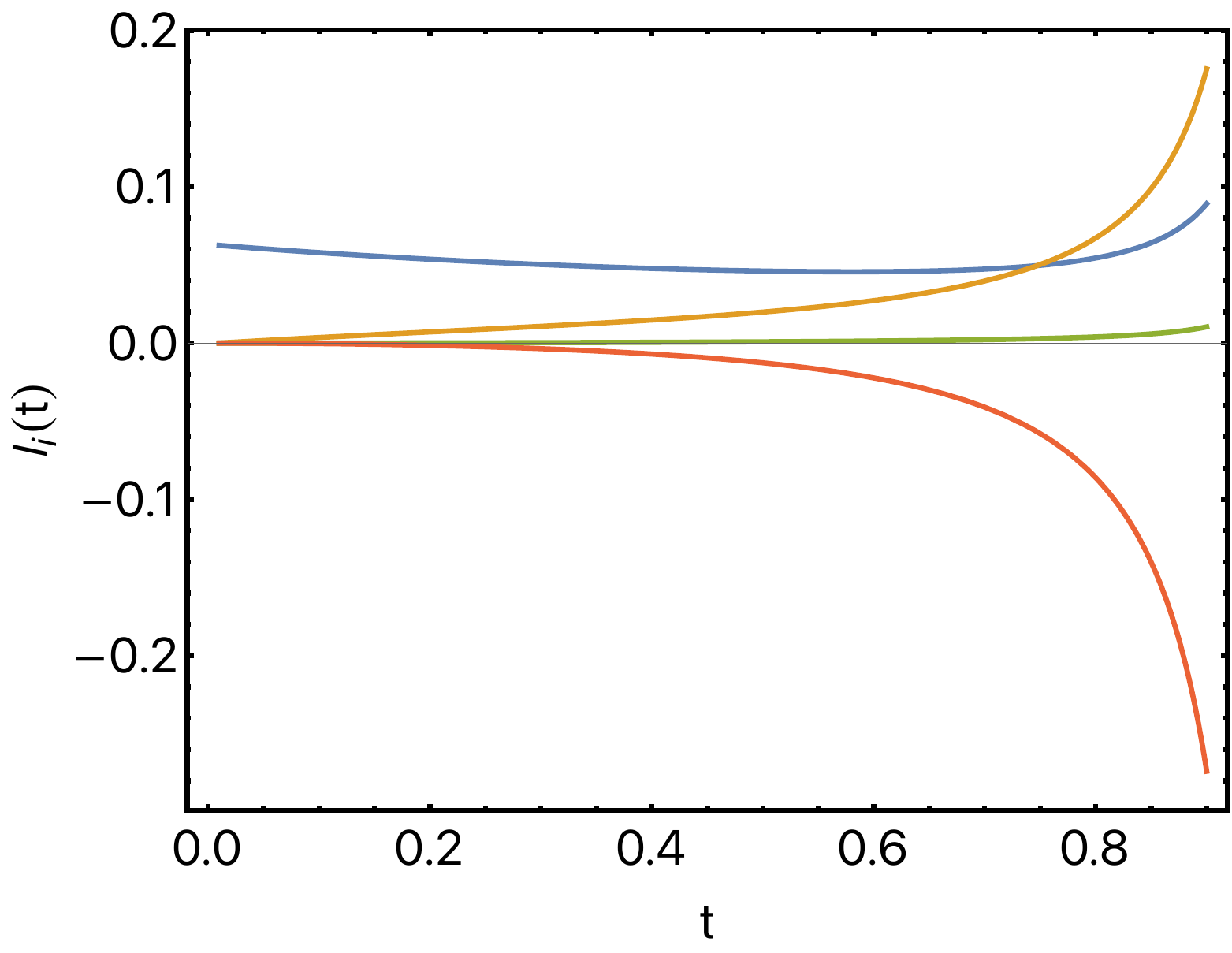}}
{\includegraphics[width=0.345\textwidth,clip]{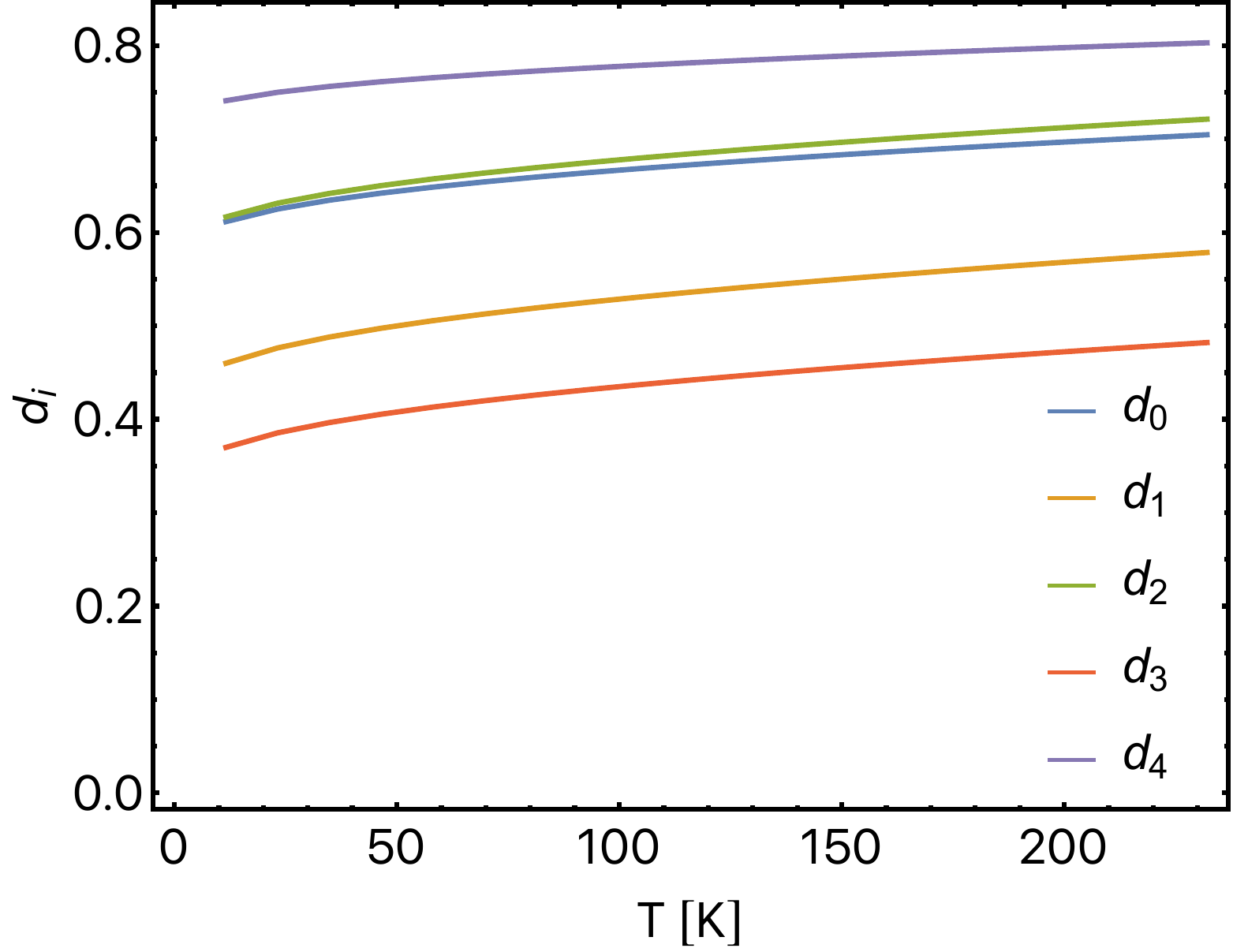}}
\caption{RG flow of the couplings obtained by integration of the flow equations \eqref{RG}, with initial conditions taken from Table \ref{initial_conds_1}. Here \{blue, orange, green, red\} corresponds to $i=\{1,2,3,4\}$. Also included are the $d_i$-factors which appear in the RG analysis \eqref{RG} and \eqref{gap_eqn}, plotted as a function of $T$.}
\label{f:RGflow}
\end{figure}

To $\log^2$ accuracy, the topological distinct Feynman diagrams that contribute to the $\beta$ functions are presented in Figure \ref{f:RG}. Owing to the momentum independence of the contact interactions, the bubble, vertex, and box corrections depicted in Figure \ref{f:RG} all reduce to simple two-particle susceptibilities. 
We define the particle-particle and particle-hole susceptibilities,
\begin{align}
    \Pi_{pp}^{\nu\nu'}(\bm q)&=-i \int \Tr\  G_\nu(\omega, \bm p + \bm q)G_{\nu'}(-\omega, -\bm p) \ \tfrac{d\omega d^2 \bm p}{(2\pi)^3}, 
    &&\Pi_{ph}^{\nu\nu'}(\bm q)= -i\int \Tr\  G_\nu(\omega, \bm p + \bm q)G_{\nu'}(\omega, \bm p)\ \tfrac{d\omega d^2 \bm p}{(2\pi)^3}
\end{align}where $G_\nu(\omega, \bm p)$ is the fermion Green's function for the flavour $\nu$, and $pp$ or $ph$ denote particle-particle or particle-hole. The RG time is given by $t= \Pi_{pp}^{cc}(\bm 0)$, and the $d_i$ nesting factors are defined as
\begin{align}
d_0 &= \frac{d}{dt} \Pi_{pp}^{dd}(\bm 0), \ \ \ \ \ d_1 = \frac{d}{dt} \Pi_{pp}^{cd}(\bm Q_\alpha), \ \ \ \ \ d_2 = \frac{d}{dt} \Pi_{ph}^{cc}(\bm Q_\alpha), \ \ \ \ \ d_3 = \frac{d}{dt} \Pi_{ph}^{dd}(\bm Q_\alpha), \ \ \  \ \ d_4 = \frac{d}{dt} \Pi_{ph}^{cd}(\bm 0).
\end{align}
They are plotted in Figure \ref{f:RGflow} against the infrared scale given by the temperature $T$. 

The explicit $\beta$ functions are,
\begin{align}
\label{RG}
 \notag \dot{g}_{1c}&=2 d_2 g_{1c} \left(g_{2c}-g_{1c}\right)-2 d_3
   \left(-h_1 l_2+h_3
   \left(h_3-l_3\right)+h_1^2\right), \\
   \notag \dot{g}_{2c}&=d_2 \left(g_{2c}^2+g_{3c}^2\right)+d_3
   \left(l_2^2+l_3^2\right),\\
   \notag \dot{g}_{3c}&=-2 d_2 \left(g_{1c}-2 g_{2c}\right) g_{3c}+2 d_3
   h_3 l_2+2 d_3 h_1 l_3-4 d_3 h_1 h_3-d_0 j_3^2-2
   d_0 j_3 j_4+2 d_3 l_2 l_3-g_{3c}^2-2 g_{4c}
   g_{3c},\\
   \notag \dot{g}_{4c}&=-d_0 \left(2 j_3^2+j_4^2\right)-2
   g_{3c}^2-g_{4c}^2,\\
   \
   \
   \notag \dot{g}_{1d}&=2 d_3 g_{1d} \left(g_{2d}-g_{1d}\right)-2 d_2
   \left(-h_1 l_2+h_3
   \left(h_3-l_3\right)+h_1^2\right),\\
   \notag \dot{g}_{2d}&=d_3 \left(g_{2d}^2+g_{3d}^2\right)+d_2
   \left(l_2^2+l_3^2\right),\\
   \notag \dot{g}_{3d}&=-d_0 g_{3d} \left(g_{3d}+2
   g_{4d}\right)-2 d_3 \left(g_{1d}-2
   g_{2d}\right) g_{3d}+2 d_2 h_3 l_2+2 d_2 h_1
   l_3-4 d_2 h_1 h_3+2 d_2 l_2 l_3-j_3^2-2 j_4 j_3,\\
   \notag \dot{g}_{4d}&=-d_0 \left(2 g_{3d}^2+g_{4d}^2\right)-2
   j_3^2-j_4^2,\\
   \
   \
   \notag \dot{h}_1&=d_2 \left(g_{1c} \left(l_2-2 h_1\right)+g_{3c}
   \left(l_3-h_3\right)+h_1 g_{2c}\right)+d_3
   \left(g_{1d} \left(l_2-2 h_1\right)+g_{3d}
   \left(l_3-h_3\right)+h_1 g_{2d}\right)+d_4
   h_1^2-2 d_1 h_2 h_1+2 d_4 h_4 h_1+d_4 j_1^2+2 d_4
   j_1 j_4-2 d_1 l_1 l_2,\\
   \
   \
   \
      \notag \dot{h}_1&=d_2 \left(g_{1c} \left(l_2-2 h_1\right)+g_{3c}
   \left(l_3-h_3\right)+h_1 g_{2c}\right)+d_3
   \left(g_{1d} \left(l_2-2 h_1\right)+g_{3d}
   \left(l_3-h_3\right)+h_1 g_{2d}\right)+d_4(
   h_1^2 +2  h_4 h_1+ j_1^2+2 
   j_1 j_4)-2 d_1 (h_2 h_1 + l_1 l_2),\\
   \
   \
   \
   \notag \dot{h}_2&=-d_1 \left(h_1^2+h_2^2+l_1^2+l_2^2\right),\\
   \notag \dot{h}_3&=d_2 \left(g_{3c} \left(l_2-h_1\right)+g_{1c}
   \left(l_3-2 h_3\right)+h_3 g_{2c}\right)+d_3
   \left(g_{3d} \left(l_2-h_1\right)+g_{1d}
   \left(l_3-2 h_3\right)+h_3 g_{2d}\right),\\
   \notag \dot{h}_4&=d_4 \left(2 h_1^2+h_4^2+2 j_1^2+j_4^2\right),\\
   \
   \
   \notag \dot{j}_1&=2 d_4 \left(h_4 j_1+h_1 \left(j_1+j_4\right)\right),\\
   \notag \dot{j}_2&=2 d_4 \left(h_4 j_2+h_1 \left(j_2+j_4\right)-2 j_2
   l_2-2 j_2 l_4+j_1 l_2-j_4 l_2+j_1 l_4\right),\\
   \notag \dot{j}_3&=-d_0 \left(\left(j_3+j_4\right) g_{3d}+j_3
   g_{4d}\right)+j_3
   \left(-g_{4c}\right)-\left(j_3+j_4\right)
   g_{3c},\\
   \notag \dot{j}_4&=-d_0 \left(2 j_3 g_{3d}+j_4
   g_{4d}\right)+d_4 \left(4 h_1
   \left(j_1+j_2\right)+4 h_4 j_4+4 j_1 l_2-8 j_2
   l_2-2 j_4 l_4\right)-2 j_3 g_{3c}-j_4 g_{4c},\\
   \
   \
   \notag \dot{l}_1&=-2 d_1 \left(h_2 l_1+h_1 l_2\right),\\
   \notag \dot{l}_2&=d_2 l_2 g_{2c}+d_2 l_3 g_{3c}+d_3 l_2 g_{2d}+d_3
   l_3 g_{3d}+2 d_4 \left(h_1 l_2+h_4 l_2+h_1
   l_4-j_2^2-j_4 j_2+j_1
   \left(j_2+j_4\right)-l_2^2-2 l_2 l_4\right)-2 d_1
   \left(h_1 l_1+h_2 l_2\right),\\
   \notag \dot{l}_3&=d_2 \left(l_2 g_{3c}+l_3 g_{2c}\right)+d_3
   \left(l_2 g_{3d}+l_3 g_{2d}\right),\\
     \dot{l}_4&=2 d_4 \left(2 h_1 l_2+h_4 l_4-2 j_2^2+2 j_1 j_2-2
   l_2^2-l_4^2\right).
\end{align}
Identifying $g_{ic}=g_{id}$, as appropriate for a TvHS in a honeycomb system, the $\beta$ functions reduce to 16 independent functions, which may be obtained straightforwardly from the above expressions.

The RG flow of the couplings, i.e. solutions to \eqref{RG}, are presented in Figure \ref{f:RGflow}. For illustration, the initial conditions are taken from Table \ref{initial_conds_1}.

\clearpage

 %%%%%%%%%%%%%%%%%
%%%%%%%%%%%%%%%%% 

\clearpage

\section{Order parameter gap equations}
\begin{figure}[t!]
{\includegraphics[width=0.35\textwidth,clip]{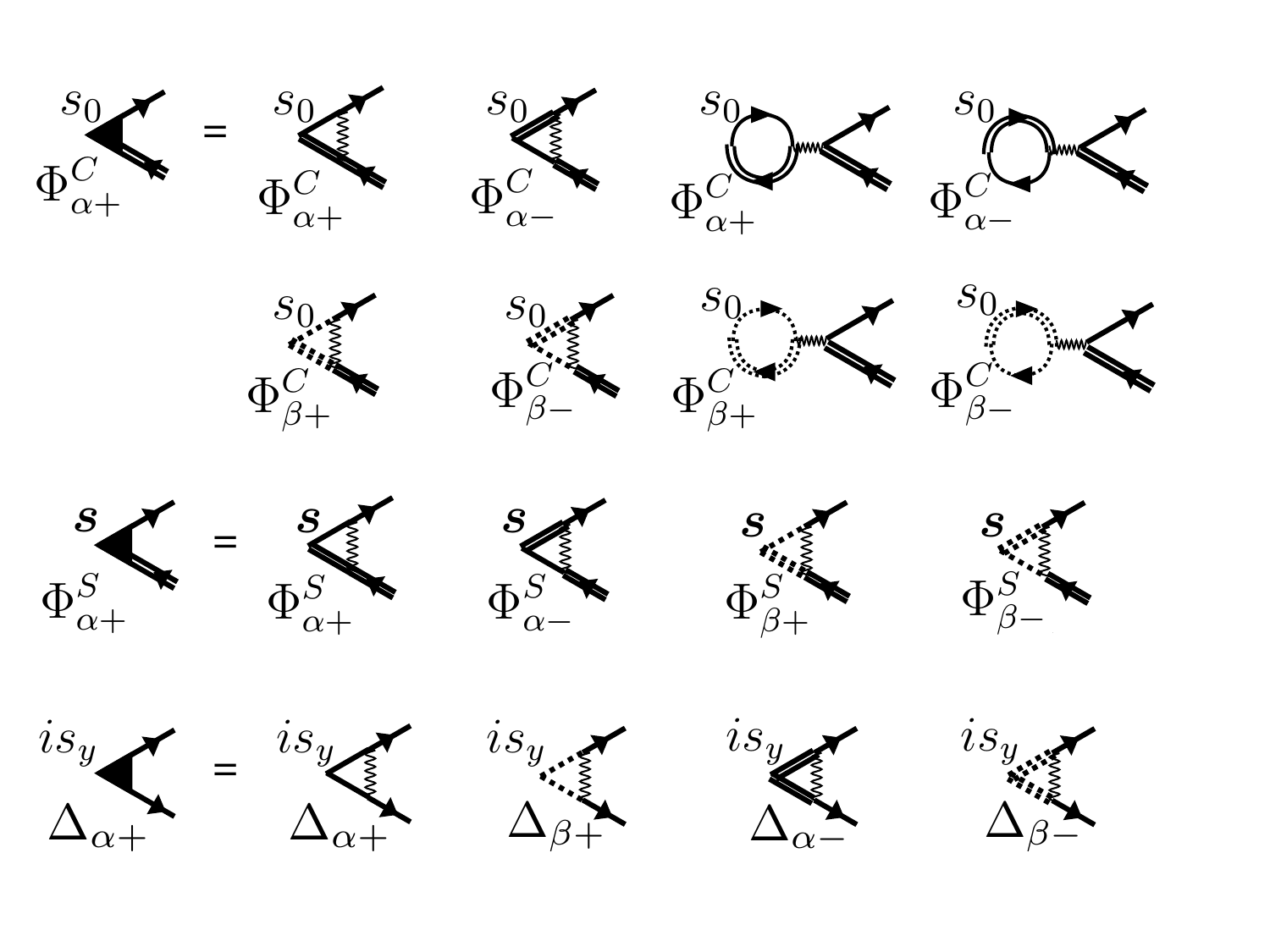}}\vspace{0.25cm}
{\includegraphics[width=0.35\textwidth,clip]{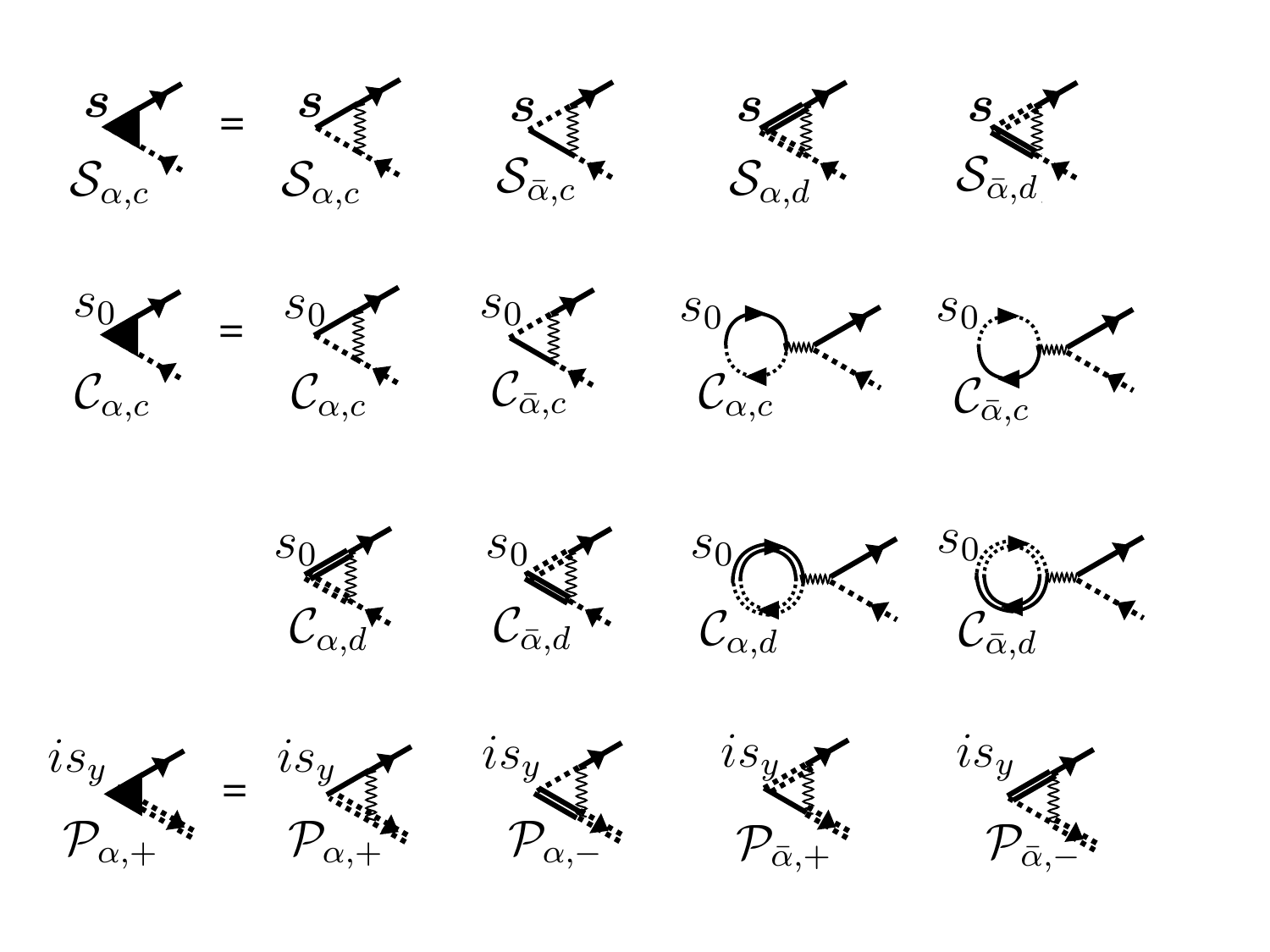}}
\caption{\textbf{Gap equations}: The Feynman diagrams representing the leading logarithmic corrections to the order parameter fields are shown. The result is a set of linear equations for the susceptibilities, which give the {\it gap equations} for the various possible order parameters, in Eq. \eqref{gap_eqn}.}
\label{f:Susc}
\end{figure}

The flow equations for the order parameter vertices $\mathcal{O}_i=\{\mathcal S_{\alpha i}, \mathcal C_{\alpha i}, \Delta_{\alpha i}, {\cal P}_{\alpha \pm},\Phi^{C}_{\alpha \pm},\Phi^{S}_{\alpha \pm}\}$ referenced in the main text are represented diagrammatically in Fig. \ref{f:Susc} and are given explicitly by (repeated here for convenience)
\begin{align}
\label{gap_eqn}
\notag\frac{\partial}{\partial t} {\Phi}^C_{\alpha+}&=d_4\sum_{\beta\neq\alpha}\Big\{(h_4-2l_4) {\Phi}^C_{\alpha+}   -j_4 {\Phi}^C_{\alpha-} + (h_1-2l_2) {\Phi}^C_{\beta+} +  (j_1-2j_2) {\Phi}^C_{\beta-}\Big\}\\
\notag\frac{\partial}{\partial t} {\Phi}^S_{\alpha+}&=d_4\sum_{\beta\neq\alpha}\Big\{h_4 {\Phi}^S_{\alpha+} +  j_4 {\Phi}^S_{\alpha-} + h_1 {\Phi}^S_{\beta+} +  j_1 {\Phi}^S_{\beta-}\Big\}\\
\notag \frac{\partial}{\partial t}{\cal P}_{\alpha,+}&=-d_1\Big\{h_2{\cal P}_{\alpha,+} +  h_1 {\cal P}_{\alpha,-} +l_1{\cal P}_{\bar\alpha,+} + l_2{\cal P}_{\bar\alpha,-}\Big\}\\
\
\notag \frac{\partial}{\partial t} {\cal C}_{\alpha,\nu} &= d_{2\nu}(g_{2,\nu}-2g_{1,\nu}) {\cal C}_{\alpha,\nu}   -d_{2\nu} g_{3,\nu} {\cal C}_{\bar{\alpha},\nu}  + d_{2\bar{\nu}}(l_2-2h_1) {\cal C}_{\alpha,\bar{\nu}} + d_{2\bar{\nu}} (l_3-2h_3) {\cal C}_{\bar{\alpha},\bar{\nu}} \\
\notag \frac{\partial}{\partial t} {\cal S}_{\alpha,\nu} &=  d_{2\nu} g_{2,c}{\cal S}_{\alpha,\nu}   + d_{2\nu} g_{3,c} {\cal S}_{\bar{\alpha},\nu}  + d_{2\bar{\nu}} l_2 {\cal S}_{\alpha,\bar{\nu}} + d_{2\bar{\nu}} l_3 {\cal S}_{\bar{\alpha},\bar{\nu}}\\
\frac{\partial}{\partial t} {\Delta}_{\alpha,\nu}&=-\sum_{\beta\neq\alpha} \Big\{d_{0\bar{\nu}}g_{4,c}{\Delta}_{\alpha,c} + d_{0\bar{\nu}}g_{3,c} {\Delta}_{\beta,c} +d_{0\nu} j_4{\Delta}_{\alpha,d} + d_{0\nu} j_3{\Delta}_{\beta,d} \Big\}
\end{align}
with indices as defined previously---$c$, $d$, $\pm$ referring to flavour, $\alpha$ to patch---and $\bar{\alpha}$ denoting the patch connected to $\alpha$ by a nesting vector. To make the equations compact, we have introduced $\nu=\{c,d\}$ with $\bar{\nu}=\{d,c\}$, and combined the $d$-factors such that $d_{0c}=1, d_{0d}=d_0, d_{2c}=d_2, d_{2d}=d_3$, and we approximate $\partial_t d_i=0$, as stated in the Methods section. The couplings entering the gap equations are understood to inherit scale-dependence from the coupling RG equations \eqref{RG}. The eigenvectors for this linear system of gap equations give the possible order parameter structures, and those with the largest eigenvalue are the leading instabilities. The eigenvalues of \eqref{gap_eqn} are lengthy, unenlightening expressions so we have chosen not to present them. 

\begin{figure}[h!]
{\includegraphics[width=0.35\textwidth,clip]{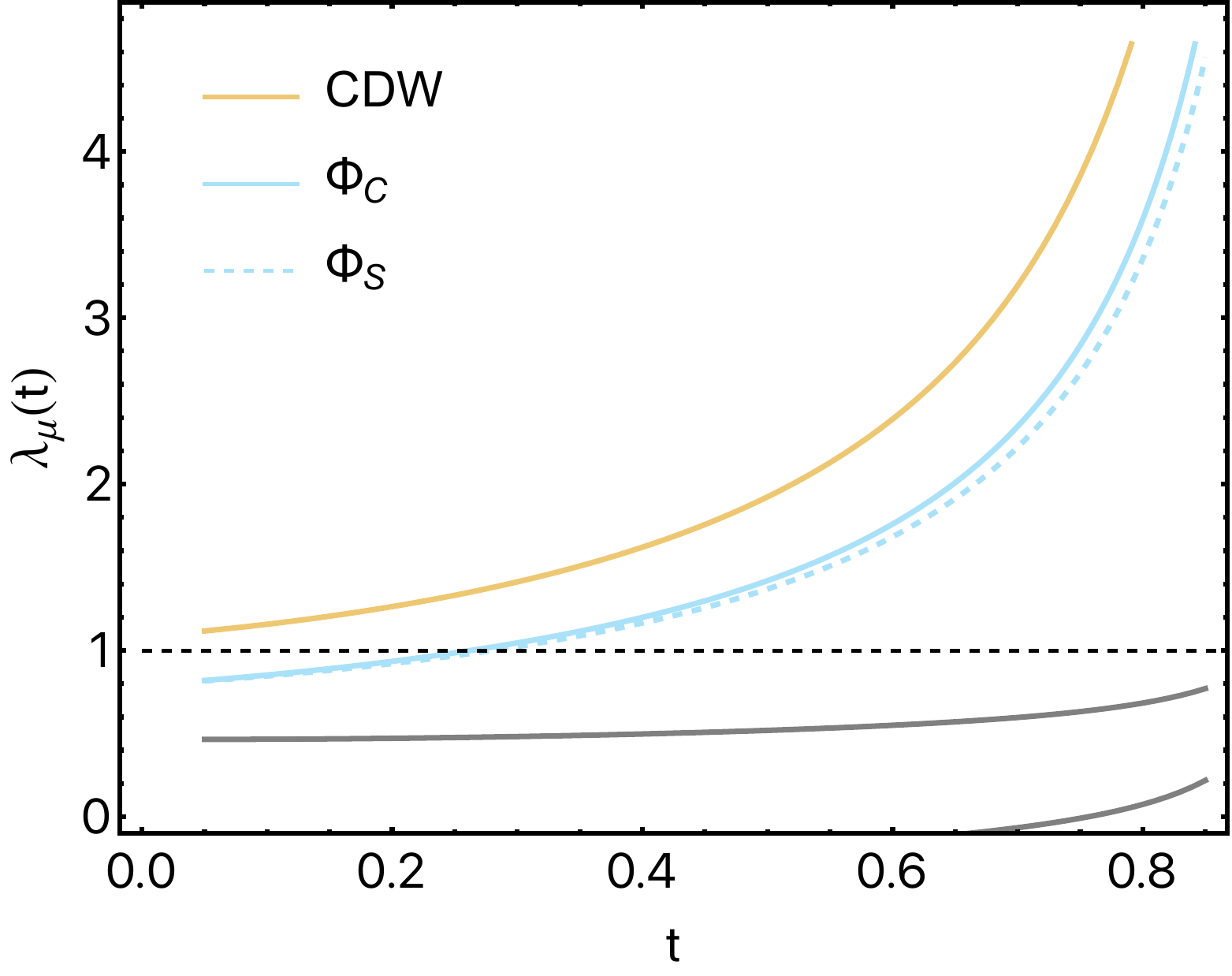}}
{\includegraphics[width=0.35\textwidth,clip]{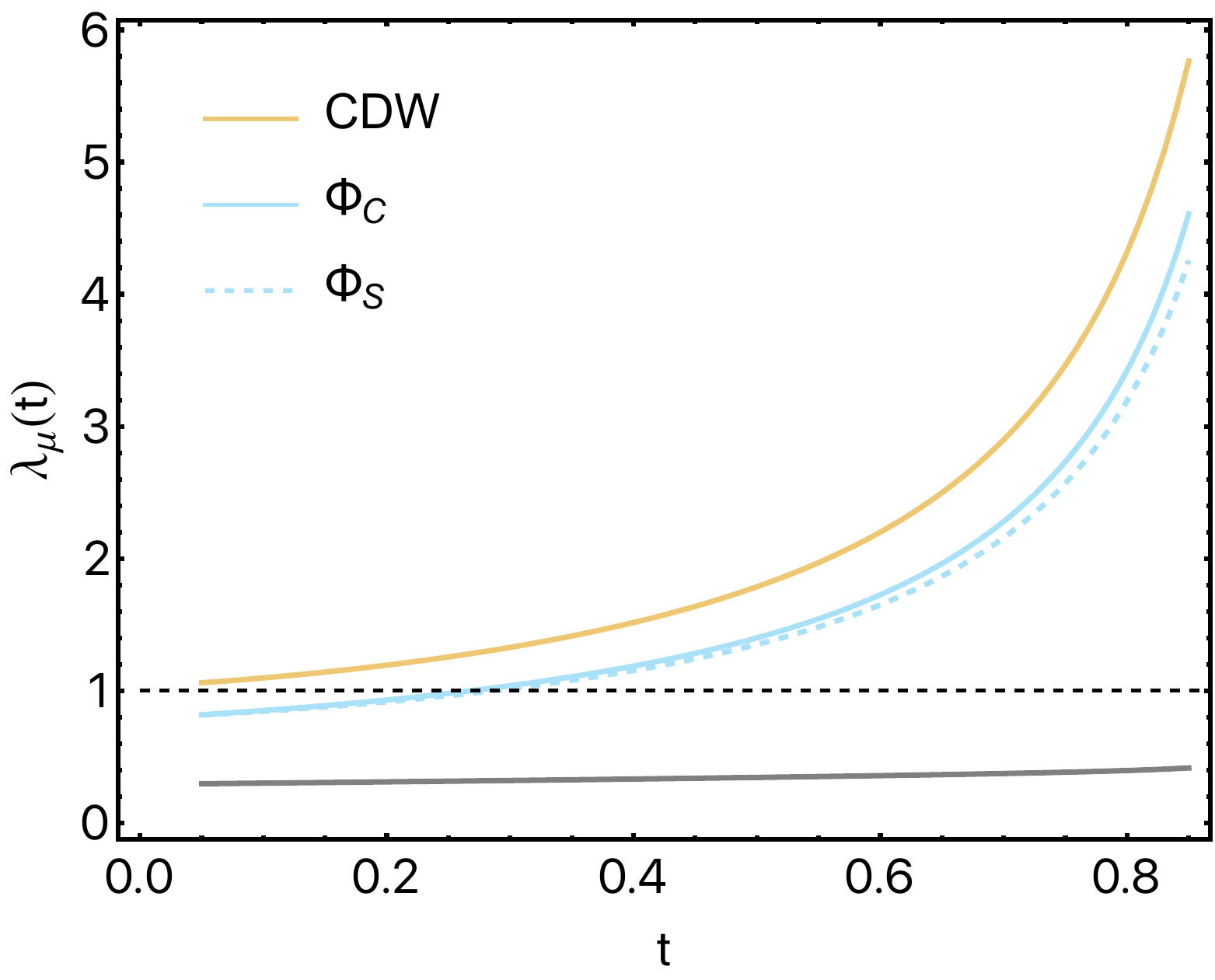}}
\caption{Order parameter eigenvalue flow evaluated from the gap equations \eqref{gap_eqn}, with initial conditions for (a) taken from Table \ref{initial_conds_1}, and for (b) from \ref{initial_conds_2}. The CDW and $\Phi_{C,S}$ eigenvalues are marked, all other eigenvalues are subleading, and plotted in grey.}
\label{f:eigflow}
\end{figure}

The RG flow of the solutions to \eqref{gap_eqn} are present in Figure \ref{f:eigflow}, subject to the initial conditions of Table \ref{initial_conds_1} and \ref{initial_conds_2}.

 %%%%%%%%%%%%%%%%%
%%%%%%%%%%%%%%%%% 
%%%%%%%%%%%%%%%%%
%%%%%%%%%%%%%%%%% 

\section{RG fixed rays}
Integrating the flow equations, one finds that the couplings generically diverge, resulting in an instability. For a given set of initial conditions, the couplings approach fixed ratios of each other at long RG times -- referred to as an RG `fixed ray', `fixed trajectory', or sometimes more loosely refered to as a `fixed point'.

To obtain the RG fixed rays, we insert the scaling form $\{g_{i\nu}, h_i, j_i, l_i\} = \{G_{i\nu}, H_i, J_i, L_i\} \mathtt{S}$, with $\mathtt{S} =1/(t_c-t)$, into the RG equations which at long RG times $t\rightarrow t_c$ allows the differential flow equations \eqref{RG} to be reduced to nonlinear algebraic equations for the $t$-independent coefficients $\{G_{i\nu}, H_i, J_i, L_i\}$. 
We summarise the procedure, which has been discussed in detail elsewhere, c.f Ref. \cite{Park2021}: 

\begin{enumerate}
    \item Let us denote the set of running couplings as $\gamma_i=\{g_{i\nu}, h_i, j_i, l_i\}$ and the set of scaling  coefficients $\Gamma_i= \{G_{i\nu}, H_i, J_i, L_i\}$. 

\item The fixed rays are found via $\dot{\Gamma}_i = \left[\dot{\gamma}_i - \Gamma_i \dot{\mathtt{S}}\right]/\mathtt{S}=\left[\beta_i[\{\Gamma\}] - \Gamma_i\right] \mathtt{S}=0$.  

\item To analyse the stability of the fixed points, we examine the matrix
\begin{align}
T_{ij} = \frac{\partial}{\partial \gamma_j} \left(\beta_i[\{\Gamma\}] - \Gamma_i\right).
\end{align}
Evaluating $T_{ij}$ at the fixed points (i.e. at the solutions to $\Gamma_i = \beta_i[\Gamma_j]$), we discard those fixed points with greater than one positive eigenvalues. The stable fixed rays satisfy this condition.
\end{enumerate}
We have employed a slightly different approach to Ref. \cite{Park2021}, which we found to be more efficient for the present problem.  Ref. \cite{Park2021} eliminate one coupling $g_i$ by using it as a proxy for the RG time. In our case we do not eliminate any $g_i$ and retain all fixed points with just one unstable direction, which may similarly be taken as a proxy for the RG time; our approach reproduces the results of \cite{Park2021}, i.e. in the limit of no interflavour couplings. 
 
 We note that the fixed rays featuring CDW order found in that work are in fact unstable to the addition of interflavour coupling, as evidenced by the fact our model has no fixed rays with charge order. Charge ordering is therefore not a dominant weak coupling instability, as discussed in the main text, and must set in at shorter RG times before fixed ray behaviour sets in, requiring the bare couplings to be adequately large. Below we compute the stable fixed rays; we make use of the result that true weak-coupling instabilities require the associated order parameter eigenvalue coefficient $\Lambda_i\geq1/2$ \cite{Maiti2013}, where the eigenvalue is $\lambda_i  = \Lambda_i {\mathtt S}$.
 
 \clearpage

\subsection{Model: Kagome $\kappa=2$}
In this model, there are eight unique stable fixed rays. Including only the non-zero terms for brevity, the fixed ray scaling coefficients (i.e. $\{g_{i\nu}, h_i, j_i, l_i\} = \{G_{i\nu}, H_i, J_i, L_i\} \mathtt{S}$) are
\begin{align}
   \notag &{\bm 1.}  \ \  \big\{G_{2 c}= 0.00247,G_{3 c}= 0.06035,G_{4 c}=
   -0.18739,G_{2 d}= 0.00356,G_{3 d}= 0.09047,G_{4 d}=
   -0.28107, \\
  \notag  & \hspace{1cm}  H_1= -0.00326,H_3= 0.00026,H_4=
   0.18837,J_1= 0.00324,J_2= 0.00324,J_3=
   0.07354,J_4= -0.45461\big\}\\
  \notag     &{\bm 2.}  \ \   \big\{G_{2 c}= 0.00694,G_{4 c}= -0.01131,G_{2 d}=
   0.01082,G_{4 d}= -0.01697,\\
   \notag &  \hspace{0.5cm}  H_2= -0.00843,H_4=
   0.01318,J_2= 0.12316,J_4= 0.12953,L_2=
   -0.12606,L_4= -0.12248\big\}\\
  \notag     &{\bm 3.}  \ \   \big\{G_{2 c}= 0.00694,G_{4 c}= -0.01131,G_{2 d}=
   0.01082,G_{4 d}= -0.01697,\\
  \notag  &  \hspace{0.5cm}  H_2= -0.00843,H_4=
   0.01318,J_2= -0.12316,J_4= -0.12953,L_2=
   -0.12606,L_4= -0.12248\big\}\\
    \notag      &{\bm 4.}  \ \   \big\{G_{1c}= -0.01244,G_{2 c}= 0.00303,G_{4 c}=
   -0.05957,G_{1d}= -0.01939,G_{2 d}= 0.00472,G_{4 d}=
   -0.08937, H_1= 0.16721, \\ 
   \notag &  \hspace{0.5cm}  H_2= -0.01876,  H_4=
   0.17096,J_1= -0.15909,J_2= -0.07964,J_4=
   -0.2899,L_1= -0.01504,L_2= 0.08341,L_4=
   0.05094\big\}\\
\notag &{\bm 5.}  \ \   \big\{G_{2 d}= 0.12733,G_{3 d}= 0.52591,G_{4 d}=
   -0.84684\big\}\\
\notag   &{\bm 6.}  \ \   \big\{G_{2 c}= 0.08906,G_{3 c}= 0.3514,G_{4 c}=
   -0.55499\big\}\\
 \notag   &{\bm 7.}  \ \   \big\{G_{2 c}= 0.06738,G_{3 c}= -0.30805,G_{4 c}=
   -0.25463\big\}\\
   &{\bm 8.}  \ \   \big\{G_{2 d}= 0.09823,G_{3 d}= -0.465,G_{4 d}=
   -0.38935\big\}.
   \end{align}
At these fixed rays the following ordered phases are supported, respectively:
\begin{align}
{\bm 1.} \ \Delta_d+\Phi^C_d+\Phi^S_d,  \ \ {\bm 2.} \  \Phi^C_s, \ \  {\bm 3.}  \ \Phi^C_s, \ \ {\bm 4.}  \ \Phi^S_s, \ \  {\bm 5.}  \ \Delta_d, \ \ {\bm 6.}  \ \Delta_d,  \ \ {\bm 7.}  \ \Delta_d, \ \ {\bm 8.}  \ \Delta_d.
   \end{align}

\subsection{Model: Kagome $\kappa=1$}
In this model, there are seven unique stable fixed rays. The corresponding nonzero couplings are
\begin{align}
   \notag &{\bm 1.}  \ \   \big\{G_{1c}= -0.00077,G_{2 c}= 0.00038,G_{3 c}= 0.02784,G_{4 c}=
   -0.16242,G_{1d}= -0.00077,G_{2 d}= 0.00038,G_{3 d}= 0.02784, \\
   \notag & \hspace{0.5cm} G_{4 d}=
   -0.16242, H_1= -0.02784,H_2= -0.00038,H_3= 0.00077,H_4=
   0.16242,J_1= 0.02778,J_2= 0.02778, \\
   \notag & \hspace{0.5cm}  J_3= 0.02778,J_4=
   -0.36461\big\}\\
      \notag &{\bm 2.}  \ \   \big\{G_{2 c}= 0.00478,G_{4 c}= -0.01034,G_{2 d}= 0.00478,G_{4 d}=
   -0.01034,\\
  \notag  & \hspace{0.5cm} H_2= -0.00478,H_4= 0.01034,J_2= -0.09656,J_4=
   -0.10118,L_2= -0.09773,L_4= -0.09587\big\}\\
\notag &{\bm 3.}  \ \   \big\{G_{1c}= -0.00841,G_{2 c}= 0.00208,G_{4 c}= -0.05558,G_{1d}=
   -0.00841,G_{2 d}= 0.00208,G_{4 d}= -0.05558, H_1= 0.12905,\\
   \notag  & \hspace{0.5cm} H_2=
   -0.0105,H_4= 0.13597,J_1= -0.12584,J_2= -0.06292,J_4=
   -0.22912,L_1= -0.00841,\\
   \notag  & \hspace{0.5cm} L_2= 0.06452, L_4= 0.04019\big\}\\
   \notag &{\bm 4.}  \ \   \left\{G_{2 d}= 0.05258,G_{3 d}= -0.32,G_{4 d}= -0.28741\right\}\\
   \notag &{\bm 5.}  \ \   \left\{G_{2 d}= 0.06129,G_{3 d}= 0.34472,G_{4 d}= -0.61106\right\}\\
   \notag &{\bm 6.}  \ \   \left\{G_{2 c}= 0.05258,G_{3 c}= -0.32,G_{4 c}= -0.28741\right\}\\
   &{\bm 7.}  \ \   \left\{G_{2 c}= 0.06129,G_{3 c}= 0.34472,G_{4 c}= -0.61106\right\}.
   \end{align}
These fixed rays support the following ordered phases:
\begin{align}
\ \ {\bm 1.}  \  \Phi ^C_d+\Delta _d+\Phi ^S_d,  \ \ {\bm 2.}  \  \Phi
   ^C_s,  \ \ {\bm 3.}  \  \Phi ^S_s, \ \ {\bm 4.}  \  \Delta _d, \ \ {\bm 5.}  \  \Delta _d, \ \ {\bm 6.}  \  \Delta _d, \ \ {\bm 7.}  \  \Delta _d.
\end{align}

\subsection{Model: Honeycomb $\kappa=1$}
In this model, there are three unique stable fixed rays. The corresponding nonzero couplings are
\begin{align}
\notag &{\bm 1.}  \ \   \big\{G_2= 0.00316,G_4= -0.13371,H_2= -0.00316,H_4=
   0.13371,J_2= 0.07916,J_4= -0.34034, \\
   \notag  & \hspace{0.5cm} L_2=
   0.07952, L_4= -0.09172\big\}
  \\
  \notag &{\bm 2.}  \ \   \big\{G_2= 0.00479,G_4= -0.01017,H_2= -0.00479,H_4=
   0.01017,J_2= -0.09694,J_4= -0.10033, \\
   \notag  & \hspace{0.5cm} L_2=
   -0.09783,L_4= -0.09645\big\}\\
    &{\bm 3.}  \ \   \left\{G_2= 0.05258,G_3= -0.32,G_4= -0.28741\right\}.
  \end{align}
  At these fixed rays the following ordered phases are supported:
\begin{align}
{\bm 1.}  \ \Phi ^C_d, \ {\bm 2.}  \ \Phi ^C_s, \ \ {\bm 3.}  \ \Delta _d.
\end{align}

\clearpage

\section{Free energy expansion}\label{LGW}
Here we provide the explicit details of the free energy expansion. 
We wish to calculate the free energy
\begin{align}
\label{freeE}
\mathcal{F} = \tfrac{1}{2\lambda_\Phi}\sum_i |\Phi_i|^2 +\tfrac{1}{2}\!\!\!\!\!\!\sum_{\alpha\neq\beta; \nu=c,d} \!\!\!\!\!\!\mathcal{V}^{-1}_{\nu\nu'}\,C_{\alpha\beta\nu}C^*_{\alpha\beta\nu'} +\tfrac{1}{2}\text{Tr} (\mathcal{G}_0M)^2 -\tfrac{1}{3}\text{Tr} (\mathcal{G}_0M)^3+\tfrac{1}{4}\text{Tr} (\mathcal{G}_0M)^4,
\end{align}
with order parameter matrix $M$ and Green's function $\mathcal{G}_0$ as defined in the Methods. Writing the contributions as $\mathcal{F} = \mathcal{F}_0+\mathcal{F}_2+\mathcal{F}_3+\mathcal{F}_4$;
\begin{align}
\label{F2}
\mathcal{F}_2 &= \tfrac{1}{2\lambda_\Phi}\sum_i |\Phi_i|^2 +\tfrac{1}{2}\!\!\!\!\!\!\sum_{\alpha\neq\beta; \nu=c,d} \!\!\!\!\!\!\mathcal{V}^{-1}_{\nu\nu'}\,C_{\alpha\beta\nu}C^*_{\alpha\beta\nu'}+\tfrac{1}{2}\text{Tr} (\mathcal{G}_0M)^2 \\
\label{F3}
\mathcal{F}_3 &= -\tfrac{1}{3}\text{Tr} (\mathcal{G}_0M)^3\\
\label{F4}
\mathcal{F}_4 &= \tfrac{1}{4}\text{Tr} (\mathcal{G}_0M)^4
\end{align}

\

\begin{figure}[h]
{\includegraphics[width=0.305\textwidth,clip]{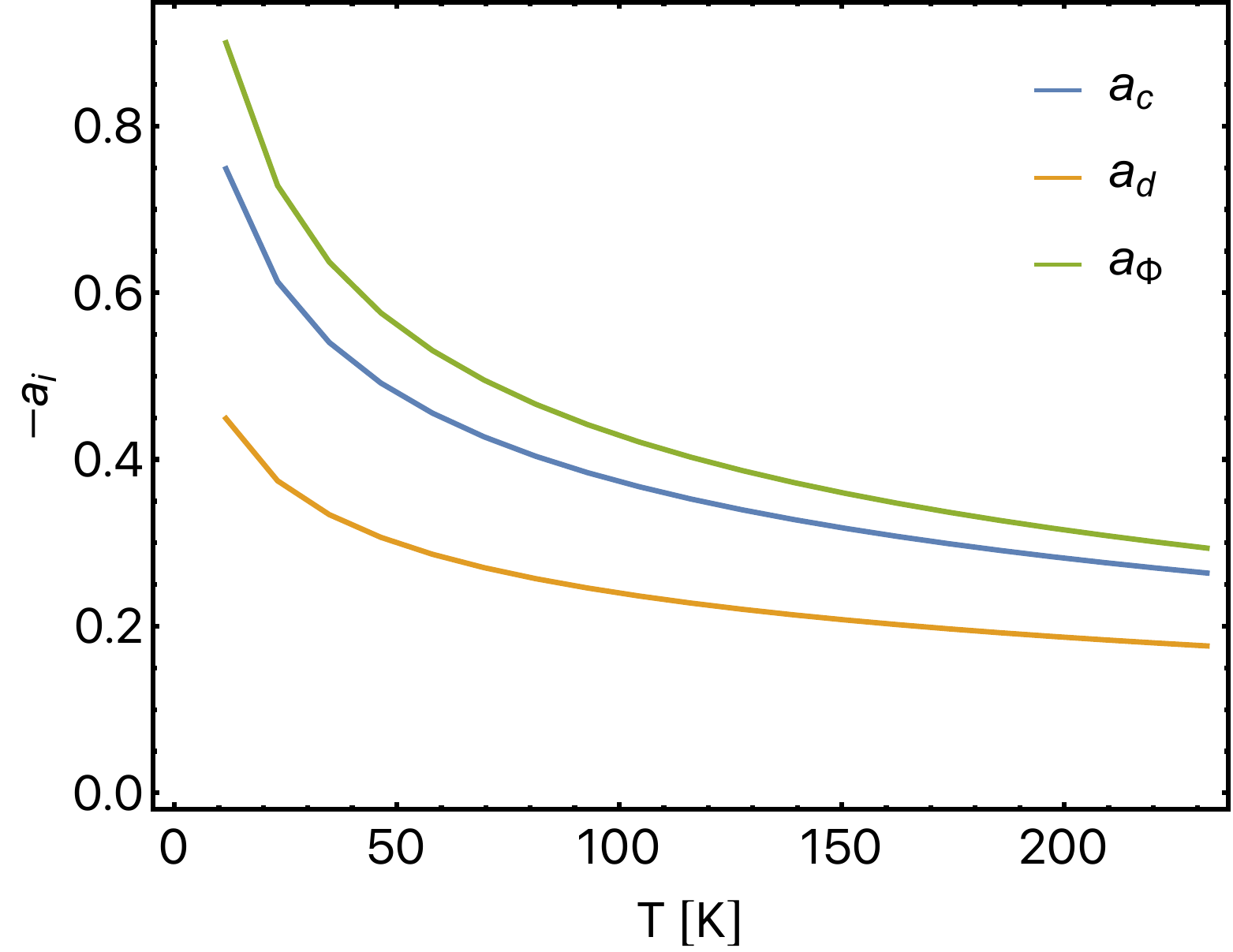}}
{\includegraphics[width=0.31\textwidth,clip]{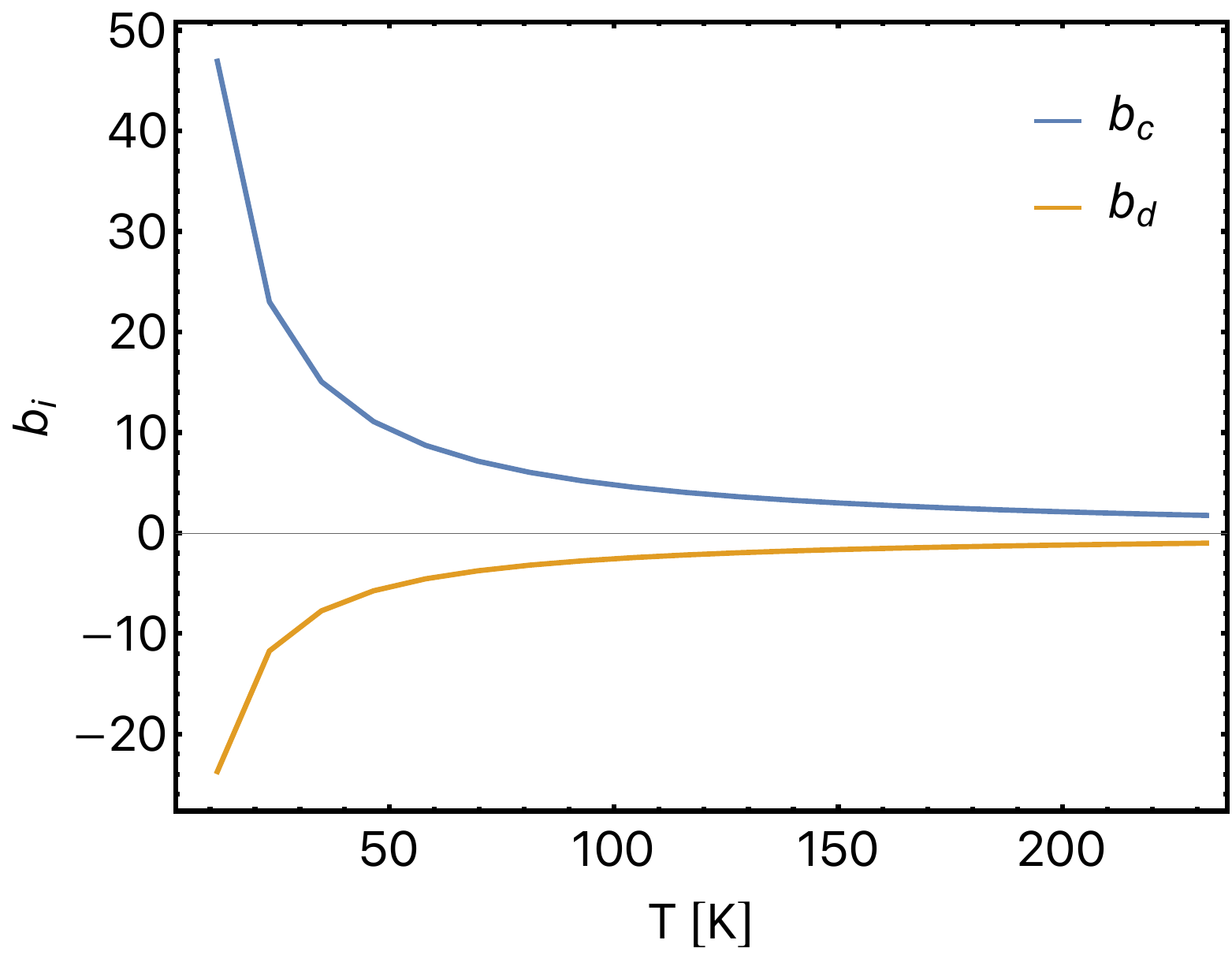}}
{\includegraphics[width=0.32\textwidth,clip]{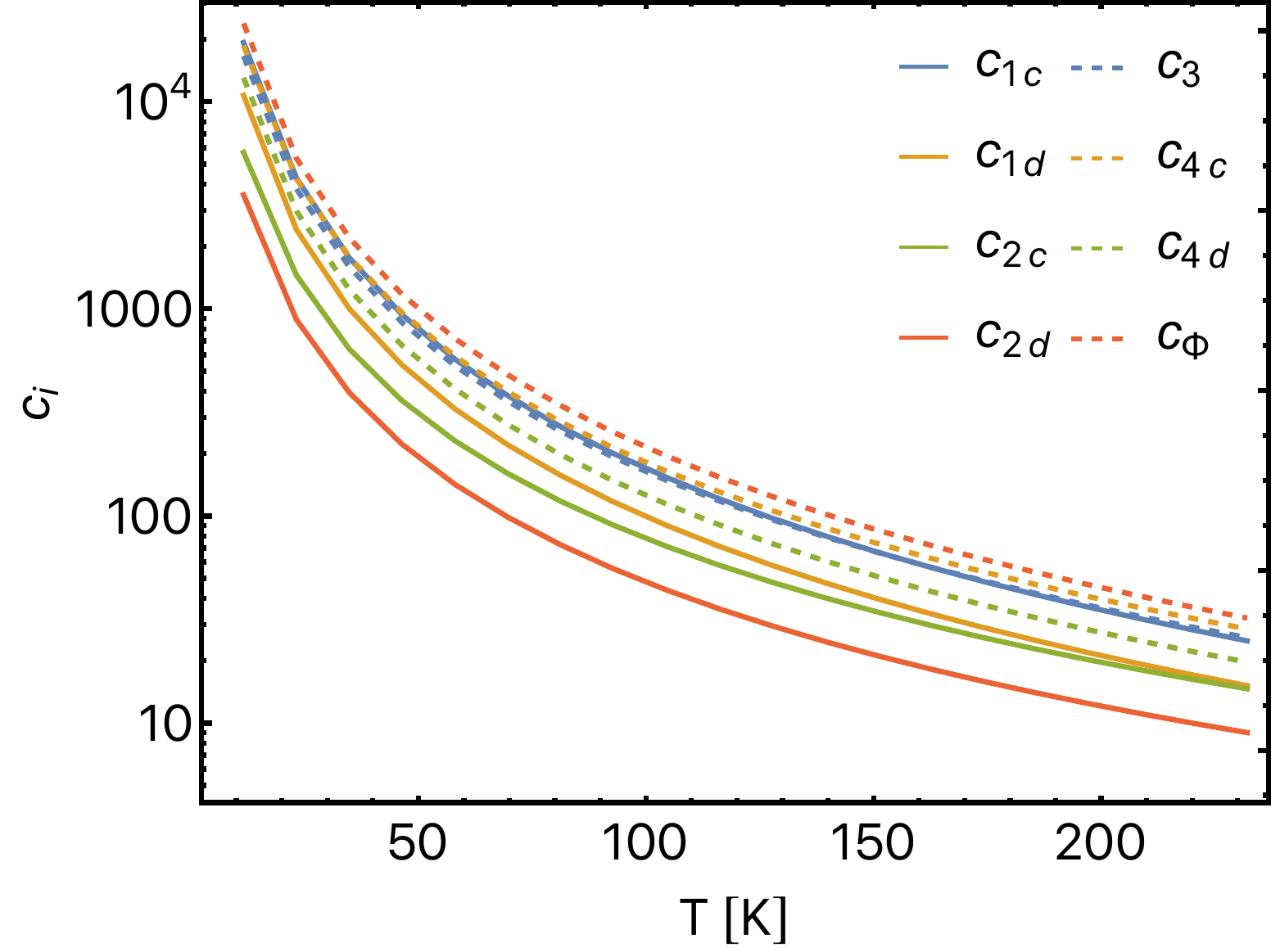}}
%{\includegraphics[width=0.35\textwidth,clip]{Plots/di_plots}}
\caption{Coefficients of the free energy expansion \eqref{freeE} -- used explicitly to obtain the phase diagram of Fig. 4 [of the main text]. Everywhere we take the infrared cut-off as temperature $T$. And have taken $\kappa=2$.}
\label{f:coeff}
\end{figure}

{\bf Quadratic term.} 
Expanding the trace in Eq. \eqref{F2},
\begin{align}
\tfrac{1}{2}\text{Tr} (\mathcal{G}_0M)^2 = a_c\sum_{\alpha\neq\beta} |C_{\alpha\beta c}|^2 + a_d\sum_{\alpha\neq\beta} |C_{\alpha\beta d}|^2 +a_\Phi\left(|\Phi_a|^2+|\Phi_b|^2 \right)
\end{align}
where
\begin{align}
a_c &= \sum_n \int  \mathcal{G}_{ci}(i\omega_n,\bm{q})\mathcal{G}_{cj}(i\omega_n,\bm{q}), && a_d = \sum_n \int  \mathcal{G}_{di}(i\omega_n,\bm{q})\mathcal{G}_{dj}(i\omega_n,\bm{q}), &&a_\Phi = \sum_n \int  \mathcal{G}_{ci}(i\omega_n,\bm{q})\mathcal{G}_{di}(i\omega_n,\bm{q}).
\end{align}
The coefficients $a_i$ are evaluated in Figure \ref{f:coeff}(a). 

\

{\bf Cubic term.}
The cubic term we find from Eq. \eqref{F3} is 
\begin{align}
-\mathcal{F}_3=\tfrac{1}{3}\text{Tr} (\mathcal{G}_0M)^3 = b_c(  C_{12c}C_{23c}C_{31c} +\text{c.c.})+b_d  (  C_{12d}C_{23d}C_{31d} +\text{c.c.})
\end{align}
where
\begin{align}
b_c &= \sum_n \int  \mathcal{G}_{c1}(i\omega_n,\bm{q})\mathcal{G}_{c2}(i\omega_n,\bm{q})\mathcal{G}_{c3}(i\omega_n,\bm{q}), && b_d = \sum_n \int  \mathcal{G}_{d1}(i\omega_n,\bm{q})\mathcal{G}_{d2}(i\omega_n,\bm{q})\mathcal{G}_{d3}(i\omega_n,\bm{q}).
\end{align}
The coefficients $b_\nu$ are evaluated in Figure \ref{f:coeff}(b) for $\kappa =2$. We consider two cases: rCDW and iCDW, corresponding to $\Phi=0,\pi/2$, respectively. For iCDW the cubic terms vanish, while for rCDW it is non-zero. Hence, in general, the cubic term favours rCDW.  

\

{\bf Quartic term.}
Lastly, we arrive at the quartic term from expanding the trace in Eq. \eqref{F4}. We first write the result in terms of $\Phi_{\alpha}$ where $\alpha$ indexes patch, and then convert to the $d$-wave basis $\Phi_{a,b}$:
\begin{gather}
\mathcal{F}_4=\tfrac{1}{4}\text{Tr} (\mathcal{G}_0M)^4 = \tfrac{1}{2}c_{1c} \sum_{\alpha\neq \beta} |C_{\alpha\beta c}|^4 +\tfrac{1}{2}c_{1d} \sum_{\alpha\neq \beta} |C_{\alpha\beta d}|^4 + c_{2c} \!\!\!\!\sum_{\alpha\neq \beta\neq \gamma} |C_{\alpha\beta c}|^2|C_{\alpha\gamma c}|^2 + c_{4d} \!\!\!\sum_{\alpha\neq \beta\neq \gamma} |C_{\alpha\beta d}|^2|C_{\alpha\gamma d}|^2  +\tfrac{1}{2}c_\Phi\sum_{\alpha} |\Phi_\alpha|^4\nonumber \\
+ c_3\sum_{\alpha\neq \beta} (  C_{\alpha\beta d} C^*_{\alpha\beta c} \Phi_{\alpha}\Phi_{\beta}^* + \text{c.c.}) + c_{4c}\sum_{\alpha\neq \beta} |C_{\alpha\beta c}|^2(|\Phi_{\alpha}|^2+|\Phi_{\beta}|^2)+ c_{4d}\sum_{\alpha\neq \beta} |C_{\alpha\beta d}|^2(|\Phi_{\alpha}|^2+|\Phi_{\beta}|^2)
\end{gather}
where (using $\nu\in\{c,d\}$ and $\bar\nu\in\{d,c\}$)%{\color{red} (c2 is c2c and c4 is c2d, c1 is c1c and c3 is c1d c7 and c8 are c4c and c4d, c5 is cphi.)}
\begin{align}
\notag c_\Phi &= \sum_n \int  \mathcal{G}_{ci}^2(i\omega_n,\bm{q})\mathcal{G}_{di}^2(i\omega_n,\bm{q}), \quad c_{1\nu} = \sum_n \int  \mathcal{G}_{\nu i}^2(i\omega_n,\bm{q})\mathcal{G}_{\nu j}^2(i\omega_n,\bm{q}), \quad c_{2\nu} = \sum_n \int  \mathcal{G}_{\nu1}(i\omega_n,\bm{q})\mathcal{G}_{\nu2}(i\omega_n,\bm{q})\mathcal{G}^2_{\nu3}(i\omega_n,\bm{q}), \\
c_3 &= \sum_n \int  \mathcal{G}_{d1}(i\omega_n,\bm{q})\mathcal{G}_{d2}(i\omega_n,\bm{q})\mathcal{G}_{c1}(i\omega_n,\bm{q})\mathcal{G}_{c2}(i\omega_n,\bm{q}), \quad  c_{4\nu} = \sum_n \int  \mathcal{G}_{\bar{\nu}i}(i\omega_n,\bm{q})\mathcal{G}_{\nu j}(i\omega_n,\bm{q})\mathcal{G}^2_{\nu i}(i\omega_n,\bm{q}).
\end{align}
The coefficients $c_{i\nu}$ are evaluated in Figure \ref{f:coeff}(c).

 %%%%%%%%%%%%%%%%%
%%%%%%%%%%%%%%%%% 

\clearpage

\section{Honeycomb-kagome bilayer toy model}
\label{toy}
\begin{figure}[h!]
\hspace{1.5cm}{\includegraphics[width=0.3\textwidth,clip]{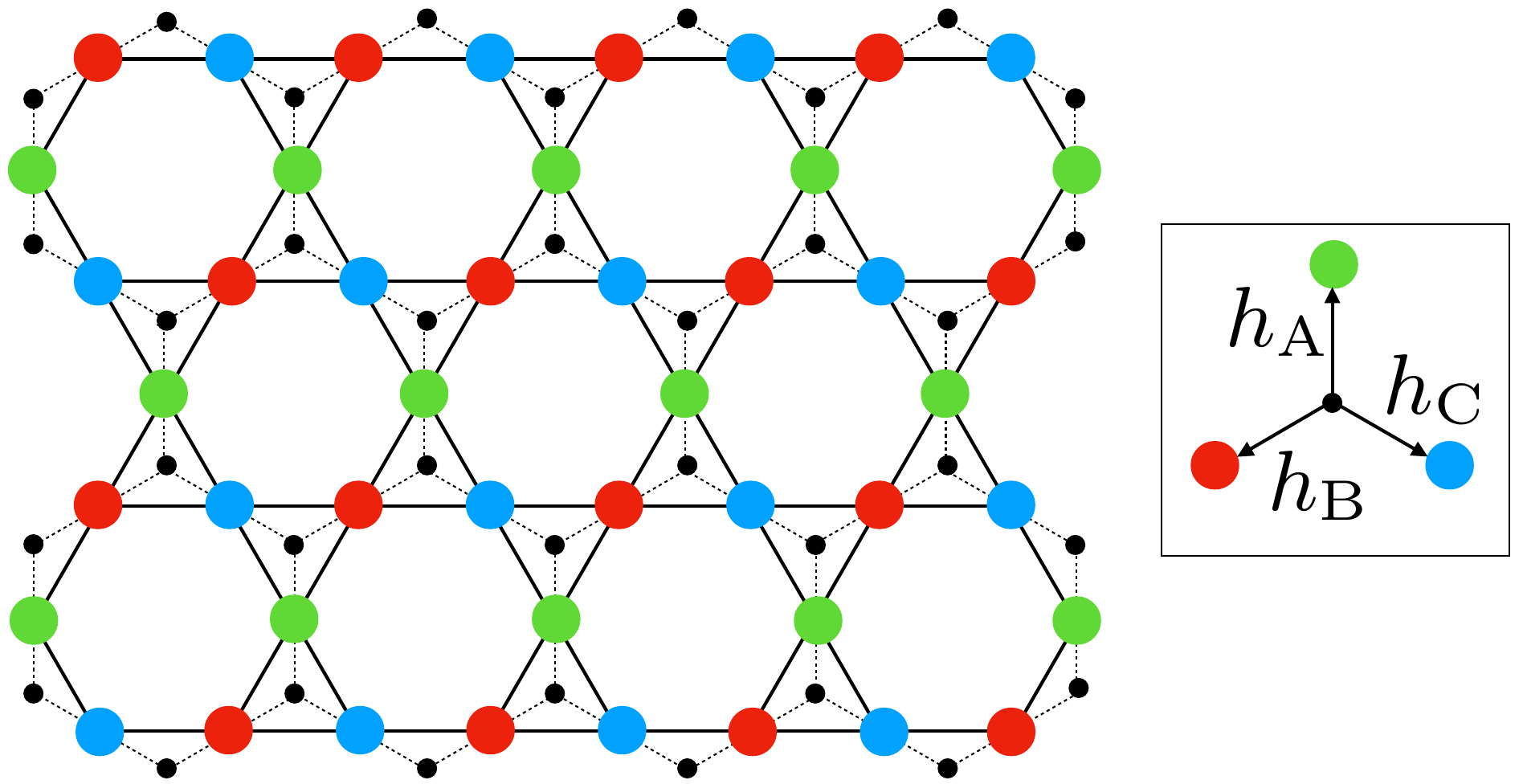}}\vspace{0.5cm}\\
{\includegraphics[width=0.3\textwidth,clip]{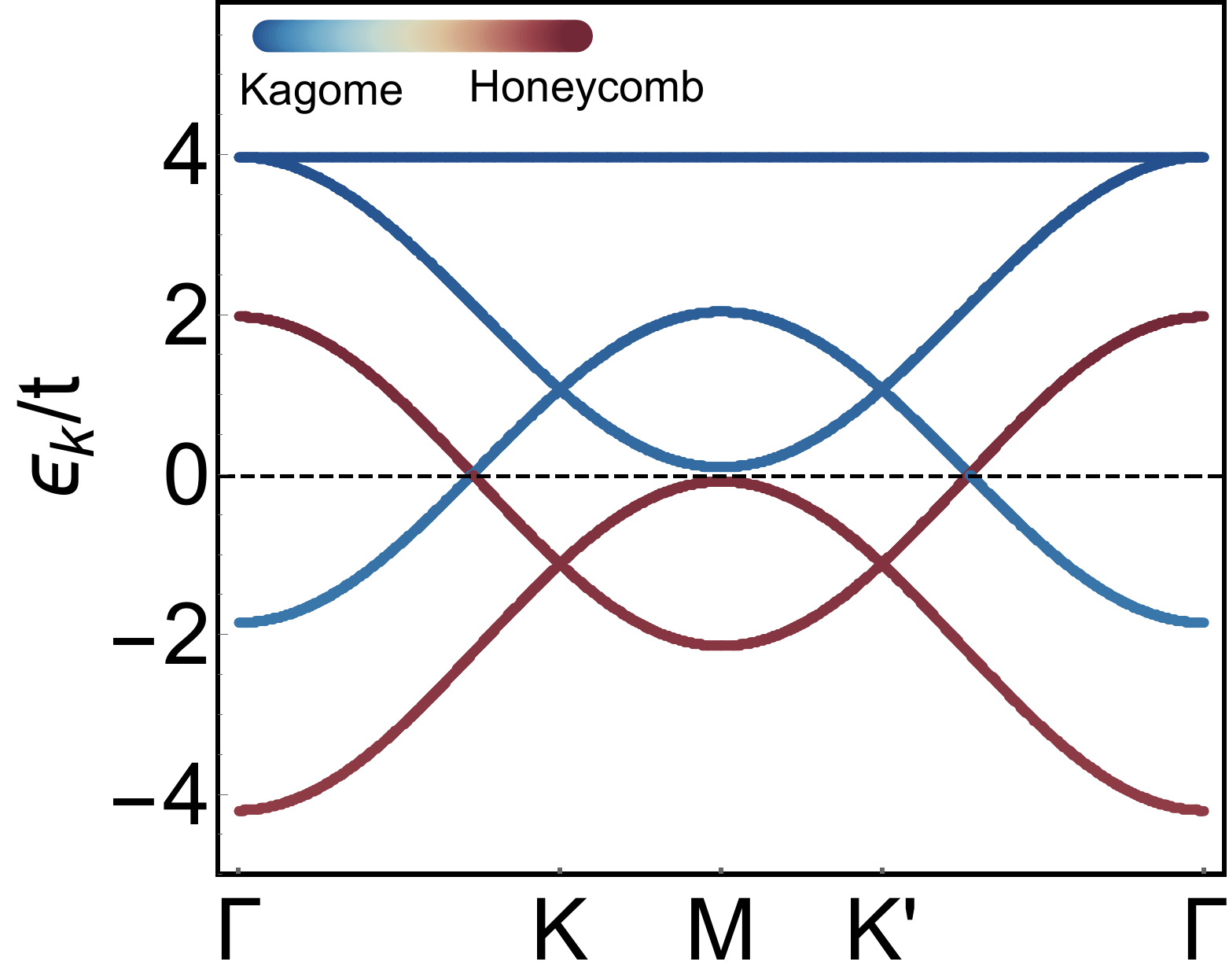}}
\hspace{1cm}
{\includegraphics[width=0.28\textwidth,clip]{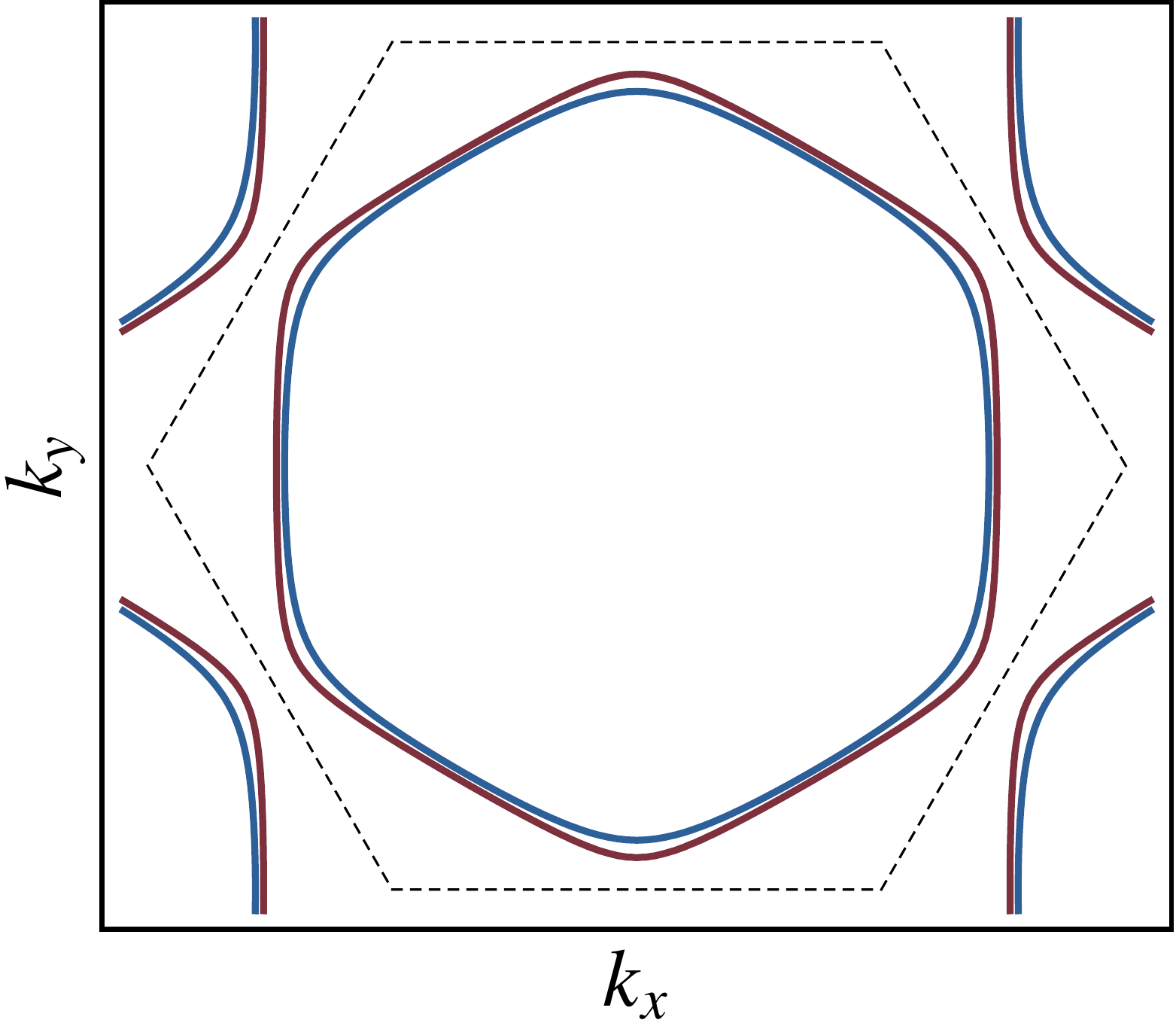}}
\begin{picture}(0,0) 
\put(-250,200){\textbf{(a)}} 
\put(-360,110){ \textbf{(b)}} 
\put(-170,110){\textbf{(c)}} 
\end{picture}
\caption{\textbf{Bilayer tight binding toy model:} (a) Lattice geometry; black dots indicate the honeycomb sites connected by dashed lines, coloured dots indicate kagome sites coloured by sublattice. Inset shows nearest neighbour tunneling vectors $\bm h_\sigma$ connecting honeycomb and kagome sites.  (b)  Bandstructure and (c) Fermi surface for $\gamma_0=3, \gamma_1=0.25$. The colour scaling indicates the wavefunction weight coming from either the kagome or the honeycomb lattice. The chemical potential is shown as the dotted black line in (b), and is chosen to correspond to doping near the $\bm M$-point. Corresponding Fermi surface is plotted in (c), with the Brillouin zone boundary shown in dashed black.}
\label{f:toymodel}
\end{figure}

An illustrative model of a TvHS is the following toy lattice model comprising of a honeycomb-kagome bilayer, with interlayer coupling a site on the honeycomb lattice to the the three nearest sites on the kagome lattice:
\begin{align}
\label{toymodel}
&H_{\text{bilayer}}=H_\text{H} + H_\text{K} + H_\text{T},\\
\notag H_\text{H}&=-\sum_{i} \left\{t_+ \left(a_i^\dag b_i + b_i^\dag a_i\right) + \gamma_0  \left(a_i^\dag a_i + b_i^\dag b_i\right)\right\},\\
\notag H_\text{K}&=-t_-\sum_{\braket{i,j}} \left\{ A_i^\dag B_j + A_i^\dag C_j + B_i^\dag C_j \right\} +\text{h.c.},\\
\notag H_\text{T}&=\gamma_1 \sum_{i} (a_i^\dag+b_i^\dag) (A_i + B_i + C_j) + \text{h.c.}
\end{align}
Here $a_i^\dag, b_i^\dag$ create electrons on the ($a,b$) sublattices of the honeycomb layer at site $i$,  $A_i^\dag, B_i^\dag, C_i^\dag$ create electrons on the kagome ($A,B,C$) sublattices at site $i$, and the sum over $\braket{i,j}$ enumerates nearest neighbours. The interlayer coupling $\gamma_1$ connects a given sublattice site of the honeycomb lattice to the nearest three sublattice sites of the kagome lattice, and $\gamma_0$ is an onsite energy shift which acts to increase the chemical potential in one layer relative to the other, aligning the valence and conduction bands associated to each layer. A schematic of the lattice geometry is depicted in Fig \ref{f:toymodel}(a). Taking $t_+=t_-=1, \gamma_0=3, \gamma_1=0.25$, the bandstructure and Fermi surfaces are shown in Fig. \ref{f:toymodel}(b) \& (c).

 %%%%%%%%%%%%%%%%%
%%%%%%%%%%%%%%%%% 

\section{Properties of the chiral excitonic condensate}\label{ED}

In this section we elaborate on the properties of the chiral excitonic condensate. The excitonic order parameter winds by a phase $\pm 4\pi$ around the Fermi surface and fully gaps the bulk dispersion (c.f. Fig \ref{sfig:kagomebilayerbulk}). As a result, the mean-field Hamiltonian for the excitonic condensate describes a Chern insulator with Chern number $C  =\pm 2$. In order to demonstrate the non-trivial topology, we diagonalise a mean-field Hamiltonian defined on a lattice, which provides a completion of the low-energy description we have so far considered to the entire Brillouin zone. Following on from the toy model of Section \ref{toy}, we shall consider a honeycomb/kagome bilayer in an infinite ribbon geometry with zigzag edges. In addition, we shall present results for a honeycomb-honeycomb bilayers, and two-orbital kagome systems.

\subsection{Honeycomb-kagome bilayer}

\begin{figure*}
\includegraphics[width=0.38\textwidth]{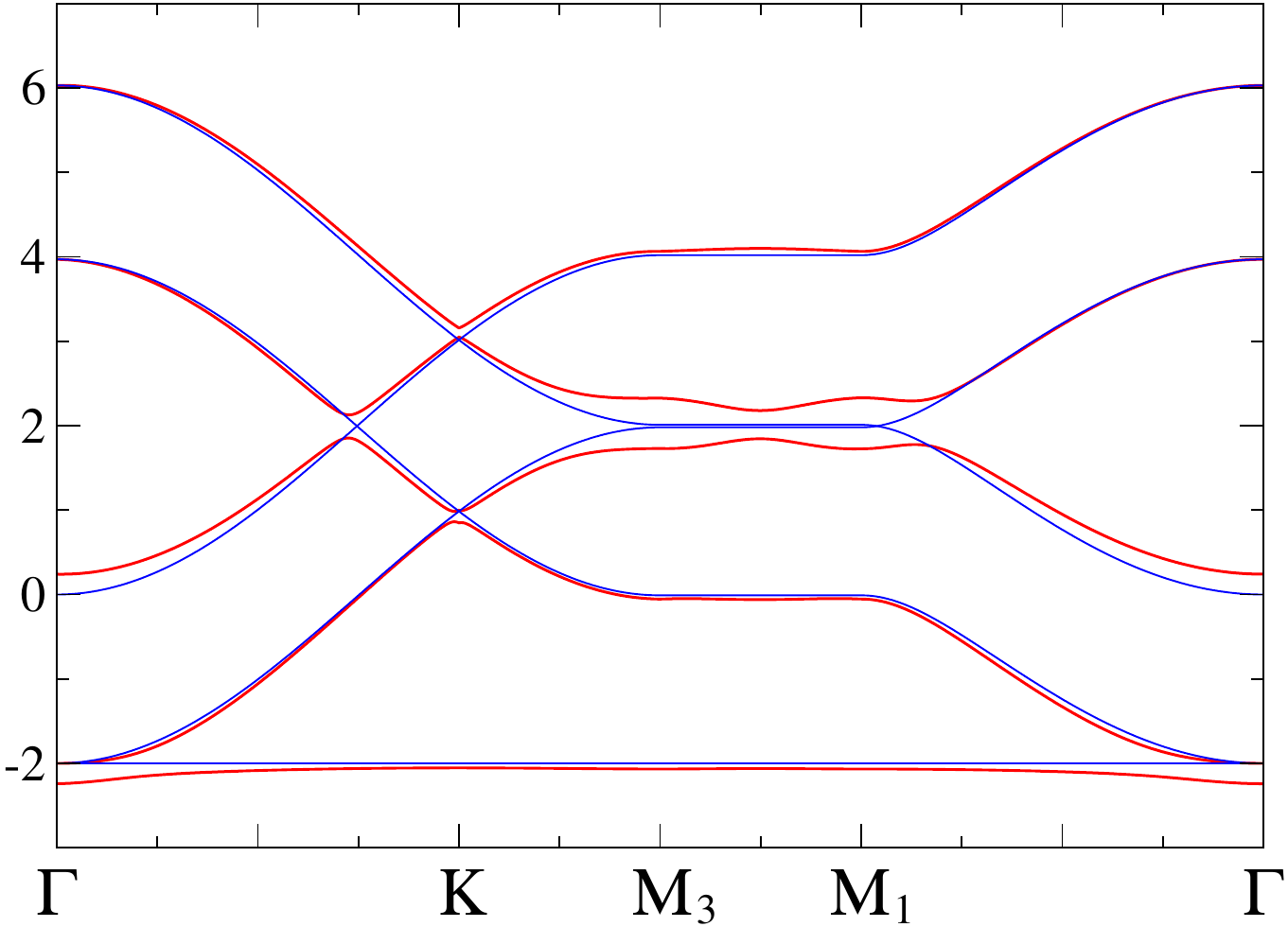} 
\hspace{1cm}
\raisebox{0.8cm}{\includegraphics[width=0.25\textwidth]{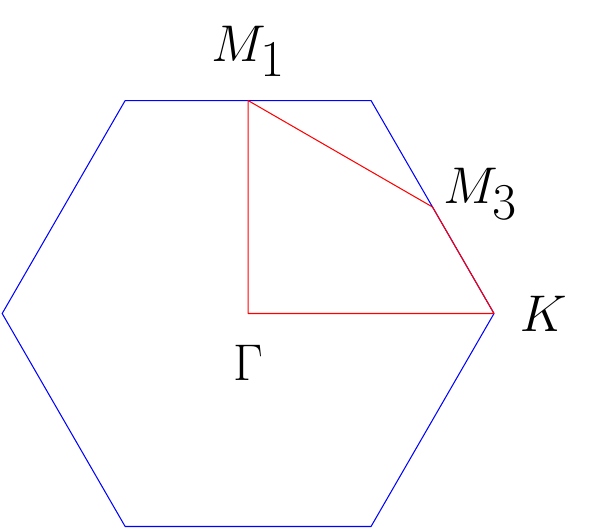}}

\caption{\textbf{Exact diagonalisation results:} Left: the bulk dispersion for the honeycomb/kagome bilayer with parameters $t_+ = t_- = t$, $t_\perp = 0.1t$, and $\Delta = 0.3t$ (red), $\Delta = 0$ (blue). Right: the path in the Brillouin zone along which the dispersion is plotted.}
\label{sfig:kagomebilayerbulk}
\end{figure*}

\begin{figure}
\includegraphics[width=0.3\textwidth]{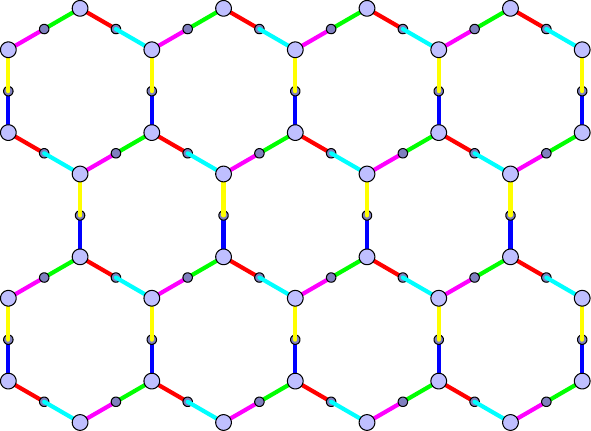}
\caption{The real space pairing Hamiltonian \eqref{realspacepairing} for a $d+id$ excitonic insulator ($\ell = 2$) . The large (small) circles indicate sites in the honeycomb (kagome) layer, while the coloured bonds represent the phases of the excitonic pairing; blue, green, red, yellow, magenta, and cyan bonds corresponding to the phases $e^{-\frac{i\pi}{6}}, e^{-\frac{5\pi i}{6}}, e^{i\pi}, e^{\frac{5i\pi}{6}}, e^{\frac{i\pi}{6}}, 1$.}
\label{sfig:latticebilayer}
\end{figure}

In this section, we shall switch to more compact notation compared to the main text and \ref{toy}, for ease of describing the real space structure of the excitonic order parameter. We introduce the simplified lattice model
\begin{align}
H &= H_+ + H_- + H_\perp + H_\Delta \ \ , \nonumber \\
H_+ &= -t_+\sum_{\langle \bm{r},\bm{r}'\rangle\in +}{ c^\dag_{\bm{r}'} c_{\bm{r}}} + \gamma_0\sum_{\bm{r}\in +}{ c^\dag_{\bm{r}} c_{\bm{r}}} \ \ , \nonumber \\
H_- &= -t_-\sum_{\langle \bm{r},\bm{r}'\rangle\in \nu}{ c^\dag_{\bm{r}'} c_{\bm{r}}}  \ \ , \nonumber \\
H_\perp &= \gamma_1  \sum_{\substack{\bm{r} \in \sigma = \{a,b\}\\ j = \{A,B,C\}}}{c^\dag_{\bm{r}+\bm{h}_{j\sigma}} c_{\bm{r}}} + \text{h.c.}\ \ ,\nonumber \\
H_\Delta &= \sum_{\bm{r}'\in +, \bm{r}\in -}{\Delta(\bm{r}',\bm{r}) c^\dag_{\bm{r'}} c_{\bm{r}}}
\end{align}
consisting of a honeycomb layer stacked on top of a kagome layer, where $\nu = +,-$ denote the honeycomb and kagome layers, $a,b$ denote the sublattices of the honeycomb layer, $A,B,C$ denote the sublattices of the kagome layer, and $\bm{h}_{j\sigma}$ are vectors connecting a site in sublattice $\sigma$ in the honeycomb layer to its nearest neighbors in sublattice $j$ in the kagome layer. The Hamiltonian consists of the bilayer toy model -- with nearest neighbor hopping in the honeycomb and kagome planes with hopping energies $t_\pm$, a relative chemical potential $\gamma_0$ between the two layers, as well as tunneling $\gamma_1$ -- along with an excitonic pairing term $\Delta(\bm{r}',\bm{r})$ from a site in the honeycomb layer to its three nearest neighbors in the kagome layer. We choose the excitonic pairing function $\Delta(\bm{r},\bm{r}')$ so that the lattice theory possesses an equivalent continuum limit to our field theory description of the three patches surrounding the $M$ points. The spatial wavefunctions of the eigenstates $|\bm{k}\approx\bm{M}_j,\nu\rangle$ of $H_\parallel$ in the upper and lower layers ($\nu = +,-$) are given explicitly by
\begin{align}
\psi^\dag_{\bm{k},\nu} =\frac{1}{\sqrt{N}} \sum_{\bm{r}\in \nu}{\varphi_{\bm{k},\nu}(\bm{r}) c^\dag_{\bm{r}}}
\end{align}
where
\begin{align}
\varphi_{\bm{k},+}(\bm{r}) = \tfrac{1}{\sqrt{2}} e^{i\bm{M}_j\cdot\bm{r}}  , \ \ \ \ \varphi_{\bm{k},-}(\bm{r}) = \begin{cases} \tfrac{1}{\sqrt{2}} e^{i\bm{M}_j\cdot\bm{r}} \     & j\neq \sigma_j \\
0 \ \  \ & j = \sigma_j
\end{cases}
\end{align}
where $\bm{k}\approx \bm{M}_j$, and the sublattice index $(\sigma_1,\sigma_2,\sigma_3) = (a,b,c)$ corresponds to the sites in the kagome lattice for which the 2D projection of the bond vector $\bm{h}_{\sigma_jA}$ is parallel to $\bm{M}_j$. Near the $M$ points, the effective Hamiltonian projected onto states near the Fermi surface is given by
\begin{gather}
H_\Delta = \sum_{j, \bm{k}\approx \bm{M}_j}{\Delta_j \psi^\dag_{\bm{k},+} \psi_{\bm{k},-}} + \text{h.c.} , \nonumber \\
\Delta_j =  \tfrac{1}{2}e^{\frac{i\pi}{6}}\sum_{\sigma\neq j}{ \Delta(\bm{r}_A +\bm{h}_{\sigma A},\bm{r}_A) - \Delta(\bm{r}_B+\bm{h}_{\sigma B},\bm{r}_B)}
\end{gather}
where $\bm{r}_A,\bm{r}_B$ are the coordinates of the A and B sites in the first unit cell. The direct tunneling Hamiltonian $H_\perp$ does not appear in the effective Hamiltonian near the $M$ points, since the contributions from hopping processes involving opposite sublattices in the honeycomb layer interfere destructively. A $d\pm id$ order parameter $(\Delta_1,\Delta_2, \Delta_3) = (\Delta, \Delta e^{\frac{2\pi i}{3} \ell}, \Delta e^{\frac{4\pi i}{3}\ell})$ for $\ell = \pm 2$ implies that the real-space pairing functions are given by
\begin{align}
&\Delta(\bm{r}_A+\bm{h}_{aA},\bm{r}_A) = -\Delta(\bm{r}_B+\bm{h}_{aB},\bm{r}_A) = e^{-\frac{i\pi}{6}}\Delta  \nonumber \\
&\Delta(\bm{r}_A+\bm{h}_{bA},\bm{r}_A) = -\Delta(\bm{r}_B+\bm{h}_{bB},\bm{r}_A) = e^{-\frac{i\pi}{6} + \frac{2\pi i}{3} \ell}\Delta  \nonumber \\
&\Delta(\bm{r}_A+\bm{h}_{cA},\bm{r}_A) = -\Delta(\bm{r}_B+\bm{h}_{cB},\bm{r}_A) = e^{-\frac{i\pi}{6} +\frac{4\pi i}{3} \ell}\Delta 
\label{realspacepairing}
\end{align}
which we illustrate in Fig. \ref{sfig:latticebilayer}.

\begin{figure}[t]
\includegraphics[width=0.48\textwidth]{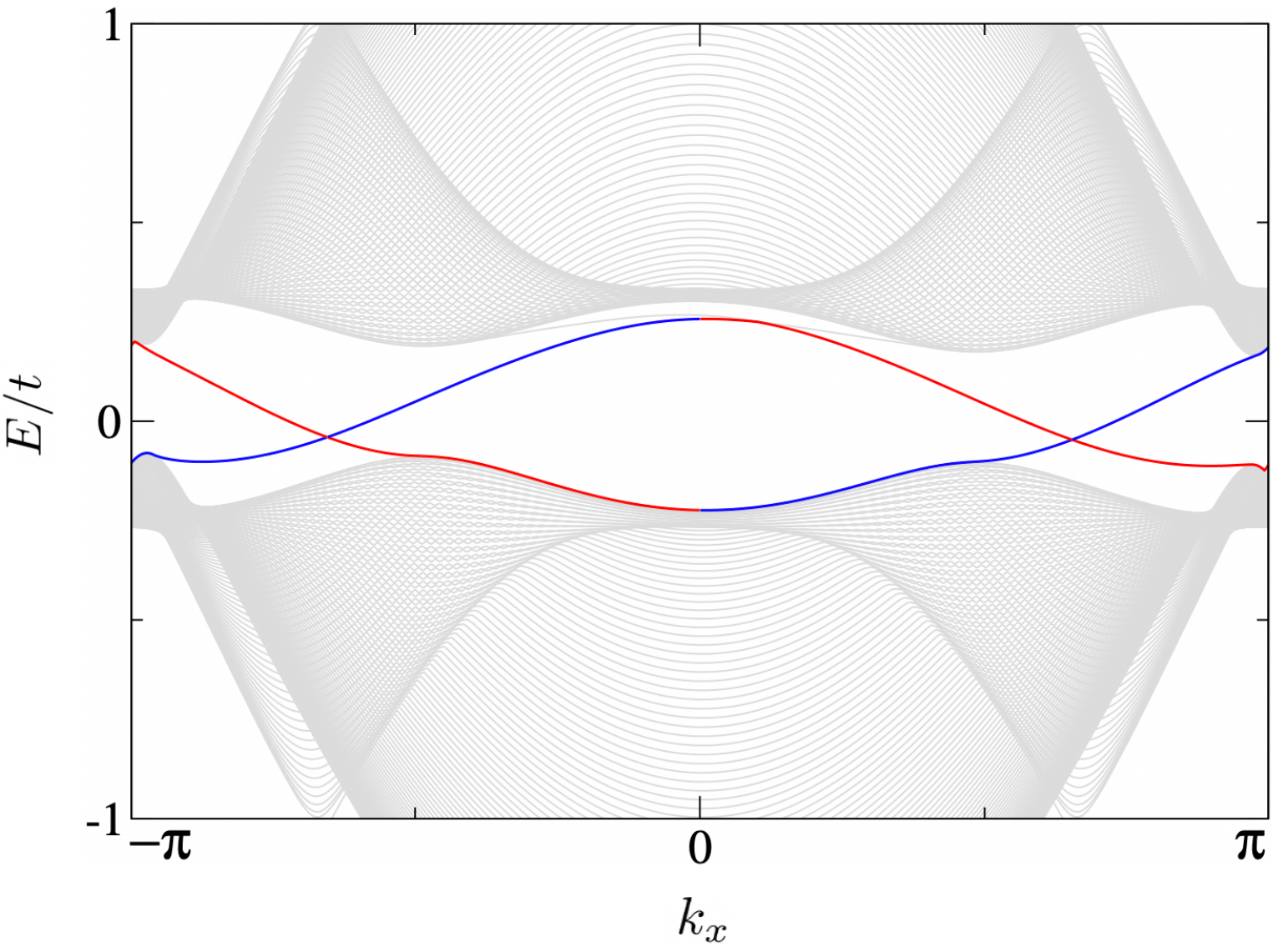}
\includegraphics[width=0.48\textwidth]{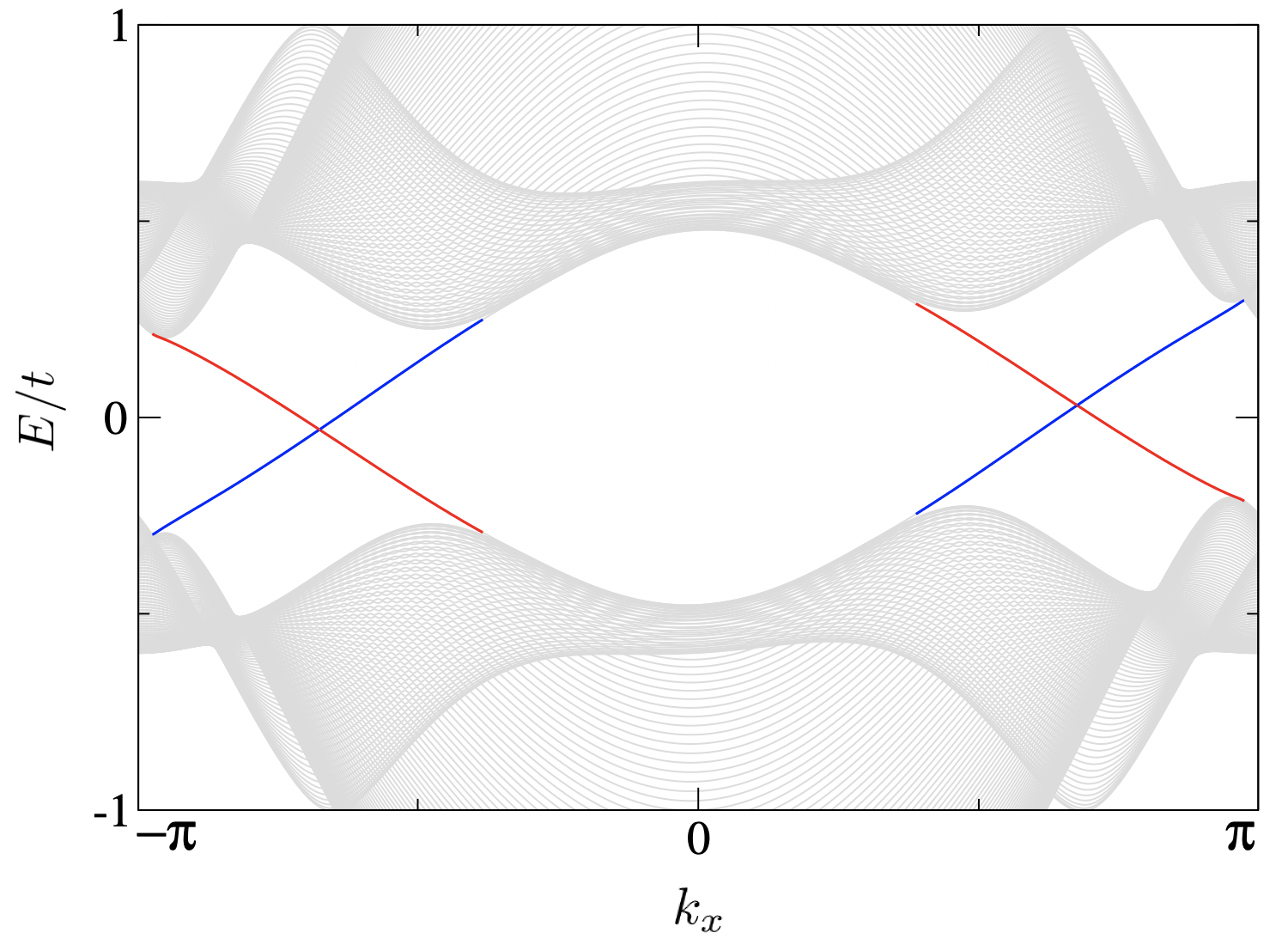}
\caption{Right: The 1D dispersion of an AA stacked honeycomb bilayer for an infinite ribbon with width 120 unit cells, with parameters $t_+ = t_-=t$, $t_\perp = 0.1t$, $\Delta = 0.3t$. The edge states on the top/bottom of the ribbon are plotted in red/blue. Left: kagome}
\label{sfig:honeycombbilayerdisp}
\end{figure}

We diagonalise the mean-field lattice Hamiltonian for an infinite ribbon geometry in the $\ell = 2$ phase with zigzag edges for parameters $t_+ = t_-$, $\gamma = 0.1t$, $\Delta = 0.4t$, and show the 1D dispersion as a function of momentum $k_x$ along the ribbon in Fig. \ref{sfig:honeycombbilayerdisp} in the main text, with energy in units of $t_+$. The dispersion exhibits two chiral edge modes, with the left-(right-)movers propagating along the top (bottom) of the ribbon represented by red (blue).  The bulk dispersion is shown in Fig. \ref{sfig:kagomebilayerbulk}. For comparison, we have also diagonalised a lattice model consisting of two $AA$ stacked honeycomb layers. In this case the chiral $d$-wave order generates pairing between an $A$ site in one layer and its nearest neighboring $B$ sites in the opposite layer, with the pairing function
\begin{align}
&\Delta(\bm{r}_A +\bm{d}_1 +\bm{c},\bm{r}_A) = -\Delta(\bm{r}_B-\bm{d}_1+\bm{c},\bm{r}_B) = \tfrac{1}{2} \Delta \nonumber \\
&\Delta(\bm{r}_A +\bm{d}_1+\bm{c},\bm{r}_A) = -\Delta(\bm{r}_B-\bm{d}_1-\bm{c},\bm{r}_B) = \tfrac{1}{2}  e^{\frac{2\pi i}{3}\ell}   \Delta\nonumber \\
&\Delta(\bm{r}_A +\bm{d}_1-\bm{c},\bm{r}_A) = -\Delta(\bm{r}_B-\bm{d}_1-\bm{c},\bm{r}_B) =\tfrac{1}{2}  e^{\frac{4\pi i}{3} \ell} \Delta
\end{align}
The 1D dispersion for an infinite ribbon in the $\ell = 2$ phase with parameters $t_+ = t_-=t, \gamma_0 = 2t, \gamma_1 = 0.1t, \Delta = 0.3t$ is shown in Fig. \ref{sfig:honeycombbilayerdisp}, with $t_\pm, \gamma_0,\gamma_1$ referring (as in the kagome model) to the intralayer hopping energies, chemical potential shift between the layers and vertical hopping energies respectively.

\subsection{Two-orbital kagome systems}
We now derive the real space description for two-orbital kagome systems -- arising due to a kagome-kagome bilayer or pair of orbitals in a monolayer -- and diagonalise the excitonic mean field Hamiltonian for this case as well.  First, we consider a kagome lattice with two orbitals $\nu = \pm$ described by the Hamiltonian
\begin{gather}
H = -t\sum_{\langle \bm{r}',\bm{r}\rangle} c^\dag_\nu(\bm{r}') c_\nu(\bm{r}) - \frac{1}{2}\sum_{\bm{r}}\nu \mu_\nu c^\dag_\nu(\bm{r}) c_\nu(\bm{r})
\end{gather}
where $\mu_\nu$ is a relative energy shift. We go to the momentum representation
\begin{gather}
c^\dag_{\bm{k},\sigma,\nu} = \tfrac{1}{\sqrt{N}} \sum_{\bm{r}\in\sigma} e^{i\bm{k}\cdot\bm{r}} c^\dag_{\nu}(\bm{r})
\end{gather}
which gives us
\begin{gather}
H(\bm{k}) = \sum_{\bm{k},\sigma,\sigma',\nu} c^\dag_{\bm{k},\sigma',\nu}\mathcal{H}_{\sigma'\sigma;\nu}(\bm{k})c_{\bm{k},\sigma,\nu}  \nonumber\\
\mathcal{H}_{\sigma'\sigma;\nu} = -\tfrac{\nu \mu_\nu}{2} \delta_{\sigma\sigma'} -2t\left(\begin{array}{ccc}
0 & \cos(\bm{k}\cdot \bm{d}_c) & \cos(\bm{k}\cdot\bm{d}_b) \\
\cos(\bm{k}\cdot\bm{d}_c) & 0 & \cos(\bm{k}\cdot \bm{d}_a) \\
\cos(\bm{k}\cdot\bm{d}_b) & \cos(\bm{k}\cdot\bm{d}_a) & 0
\end{array}\right)_{\sigma'\sigma}
\end{gather}
where $\bm{d}_a,\bm{d}_b,\bm{d}_c = \bm{d}_j$ are 2D vectors satisfying $\bm{d}_j^+ = \frac{a}{2} e^{\frac{2\pi i}{3} j}$ with $\{a,b,c\}$ corresponding to $j = \{0,1,2\}$. At the $M$ points $\bm{M}_j$, satisfying $\bm{M}_j^+ = \frac{2\pi}{\sqrt{3}a}  i e^{\frac{2\pi i}{3} j}$, we have
\begin{gather}
\bm{M}_j\cdot\bm{d}_i = \text{Re}\, \bm{M}_j^+ \bm{d}_i^- = \begin{cases} 0 \ \ , \ \ & j = i \\
\mp \tfrac{\pi}{2} \ \ , \ \ & j = i\pm 1 \ \ \text{mod 3}
\end{cases}
\end{gather}
At the $\bm{M}_1$ point we have
\begin{gather}
\mathcal{H}_\nu(\bm{k}\approx \bm{M}_j) = -\tfrac{\nu \mu_\nu}{2} - 2t \left(\begin{array}{ccc}
0 &0 & 0\\
0 & 0 & 1 \\
0 & 1 & 0
\end{array}\right)
\end{gather}
which has a spectrum $E = \{-\tfrac{\nu \mu_\nu}{2} -2t, -\tfrac{\nu \mu_\nu}{2}, -\tfrac{\nu \mu_\nu}{2} + 2t\}$ with corresponding creation operators
\begin{gather}
\psi^\dag_{\bm{k},m,\nu} = \tfrac{1}{\sqrt{2}}(b_{\bm{k},\nu}^\dag + c^\dag_{\bm{k},\nu}) \ \ , \ \ \psi^\dag_{\bm{k},p,\nu} = a^\dag_{\bm{k},\nu} \ \ , \ \ \psi^\dag_{\bm{k},m',\nu} = \tfrac{1}{\sqrt{2}}(b^\dag_{\bm{k},\nu} - c^\dag_{\bm{k},\nu}) \ \ .
\end{gather}

Note that at the $\Gamma$ point, we have
\begin{gather}
\mathcal{H}_\nu(\bm{k}=\Gamma) = -\tfrac{\nu \mu_\nu}{2} - 2t\left(\begin{array}{ccc}
0 & 1 & 1 \\
1 & 0 & 1 \\
1 & 1 & 0
\end{array}
\right)
\end{gather}
which has eigenvalues $E = \{-\tfrac{\nu\mu_\nu}{2} -4t, -\tfrac{\nu \mu_\nu}{2} +2t, -\tfrac{\nu \mu_\nu}{2} + 2t\}$. This allows us to identify the highest band, with energy $E = -\tfrac{\nu \mu_\nu}{2}+2t$, as the flat band, and the lower two bands as those hosting Dirac points. We consider the case where the middle band of the $\nu = +$ orbital is aligned with the lower band of the $\nu = +$ orbital, so $\tfrac{\mu_\nu}{2} - 2t = -\tfrac{\mu_\nu}{2} \rightarrow \mu_\nu = 2t$. Let us now consider an additional excitonic pairing term 
\begin{gather}
H_{\text{ex}} = \sum_{\bm{r},\bm{d}_i,\pm} \Delta(\bm{r}\pm\bm{d}_i,\bm{r}) c^\dag_-(\bm{r}\pm\bm{d}_i) c_+(\bm{r}) + \text{H.c.} \nonumber \\
= \sum_{\bm{k},\sigma,\sigma'} \sum_{\bm{r}_\sigma \pm \bm{d}_i \in \sigma'} \Delta(\bm{r}_\sigma \pm \bm{d}_i,\bm{r}_\sigma) e^{\mp i\bm{k}\cdot\bm{d}_i} c^\dag_{\bm{k},\sigma',-} c_{\bm{k},\sigma,+} + \text{H.c.}
\end{gather}
We project onto the bands that touch at the M points, i.e. $\psi^\dag_{\bm{k},\nu = +} = \psi^\dag_{\bm{k},p,+}$ and $\psi^\dag_{\bm{k},-} = \psi^\dag_{\bm{k},m,-}$, via the relations
\begin{gather}
c^\dag_{\bm{k}\approx \bm{M}_j,\sigma=j,\nu} = \psi^\dag_{\bm{k}\approx \bm{M}_j,p,\nu} \ \ , \nonumber \\
c^\dag_{\bm{k}\approx \bm{M}_j, \sigma = j\pm 1,\nu} = \tfrac{1}{\sqrt{2}}(\psi^\dag_{\bm{k},m,\nu} \pm \psi^\dag_{\bm{k},m',\nu})
\end{gather}
where the relation $\sigma = j \pm 1$ is understood to hold mod 3. We then obtain
\begin{gather}
H_{\text{ex}} =\sum_j \sum_{\bm{k}\approx \bm{M}_j,\sigma =j, \sigma' \neq j} \sum_{\bm{r}_\sigma \pm \bm{d}_i \in \sigma'} \Delta(\bm{r}_\sigma \pm \bm{d}_i,\bm{r}_\sigma) e^{\mp i\bm{M}_j\cdot\bm{d}_i} c^\dag_{\bm{k},\sigma',-} c_{\bm{k},\sigma,+} + \text{H.c.}  
\nonumber \\
= \tfrac{1}{\sqrt{2}}\sum_j\sum_{\bm{k}\approx \bm{M}_j} \sum_{i\neq j} \Delta(\bm{r}_j \pm \bm{d}_i,\bm{r}_j) e^{\mp i\bm{M}_j\cdot\bm{d}_i} \psi^\dag_{\bm{k},-} \psi_{\bm{k},+} \equiv \sum_j \sum_{\bm{k}\approx \bm{M}_j} \Delta_j \psi^\dag_{\bm{k},-} \psi_{\bm{k},+} 
\end{gather}
with
\begin{gather}
\Delta_j = \tfrac{1}{\sqrt{2}}\sum_{i\neq j}\Delta(\bm{r}_j \pm \bm{d}_i,\bm{r}_j) e^{\mp i\bm{M}_j\cdot\bm{d}_i}  \nonumber \\
= \tfrac{1}{\sqrt{2}}(\Delta(\bm{r}_j+\bm{d}_{j+1},\bm{r}_j) e^{-i\bm{M}_j\cdot\bm{d}_{j+1}} + \Delta(\bm{r}_j-\bm{d}_{j+1},\bm{r}_j) e^{i\bm{M}_j\cdot\bm{d}_{j+1}} + \Delta(\bm{r}_j+\bm{d}_{j-1},\bm{r}_j) e^{-i\bm{M}_j\cdot\bm{d}_{j-1}} + \Delta(\bm{r}_j-\bm{d}_{j-1},\bm{r}_j) e^{i\bm{M}_j\cdot\bm{d}_{j-1}})
\nonumber  \\
= \tfrac{1}{\sqrt{2}}(-i\Delta(\bm{r}_j+\bm{d}_{j+1},\bm{r}_j) +i \Delta(\bm{r}_j-\bm{d}_{j+1},\bm{r}_j) + i\Delta(\bm{r}_j+\bm{d}_{j-1},\bm{r}_j) -i \Delta(\bm{r}_j-\bm{d}_{j-1},\bm{r}_j))
\end{gather}
There is freedom in the choice of $\Delta(\bm{r}+\bm{d}_i,\bm{r}_j)$, however note that setting it to a constant value results in zero. One possibility is
\begin{gather}
\Delta(\bm{r}_j \pm \bm{d}_{j'},\bm{r}_j) = \pm\tfrac{1}{\sqrt{6}}\Delta_j e^{\frac{2\pi i}{3} (j'-j)} \ \ ,
\end{gather}
since we have
\begin{gather}
\tfrac{1}{\sqrt{2}}(-i\Delta(\bm{r}_j+\bm{d}_{j+1},\bm{r}_j) +i \Delta(\bm{r}_j-\bm{d}_{j+1},\bm{r}_j) + i\Delta(\bm{r}_j+\bm{d}_{j-1},\bm{r}_j) -i \Delta(\bm{r}_j-\bm{d}_{j-1},\bm{r}_j) )
 \nonumber \\
=-\tfrac{i}{\sqrt{12}}\Delta_j e^{-\frac{2\pi i}{3} j}\left[
2e^{\frac{2\pi i}{3} (j+1)} -2 e^{\frac{2\pi i}{3}(j-1)}\right] = \Delta_j 
\end{gather}%=-\tfrac{i}{\sqrt{12}}=\Delta_j} e^{-\frac{2\pi i}{3} j}\left[2e^{\frac{2\pi i}{3} (j+1)} -2 e^{\frac{2\pi i}{3}(j-1)}\right] = \frac{\Delta_j}{2\sqrt{3}} (4\sin (\frac{2\pi}{3})) =\frac{\Delta_j}{2\sqrt{3}} 4\cdot\frac{\sqrt{3}}{2} = \Delta_j 
Thus for $\Delta_j = \Delta_0 e^{\frac{2\pi i}{3} \ell j}$ we have
\begin{gather}
\Delta(\bm{r}_j \pm \bm{d}_{j'},\bm{r}_j) = \pm \tfrac{1}{\sqrt{6}} \Delta_0 e^{\frac{2\pi i}{3}(j'+ (\ell-1) j)}
\end{gather}
The resulting real space Hamiltonian is diagonalised for a ribbon geometry with 60 unit cells, the result being Fig. 4 of the main text.

\end{document}